# The Returns to Science In the Presence of Technological Risks

Matt Clancy - Research Fellow, Open Philanthropy
matt.clancy@openphilanthropy.org

Disclaimer: The views expressed in this report do not necessarily reflect the views of Open Philanthropy.

# Executive Summary

This report has two goals:

1. Estimate the social impact of spending on science
2. Determine the conditions under which accelerating scientific progress raises or lowers social welfare

In estimating the social impact of spending on science, I incorporate the long-run impact of science on per capita income and health, and integrate it into a utility framework inspired by Open Philanthropy's work in this area. I then compare the benefits of the average dollar of science today to the social impact of simply giving a dollar to an individual earning $50,000 per year. I argue the social returns to science are substantially higher than this alternative use, with my preferred model finding the social ROI of science is roughly 70x the social ROI of my benchmark.

Most of the report, however, concerns the second question: is faster science desirable? The core dilemma is that we have good reason to think science will soon enable a wider range of state and non-state actors to genetically engineer new viruses, leading to more frequent and deadly pandemic events. The report develops a quantitative economic model that lets us compare historical benefits of science to forecasts of these costs.

A variety of forecasts about the potential harms from advanced biotechnology suggest the crux of the issue revolves around civilization-ending catastrophes. Forecasts of other kinds of problems arising from advanced biotechnology are too small to outweigh the historic benefits of science. For example, if the expected increase in annual mortality due to new scientific perils is less than roughly 0.15% per year (and there is no risk of civilization-ending catastrophes from science), then in this report's model, the benefits of science will outweigh the costs. I argue the best available forecasts of this parameter, from a large number of superforecasters and domain experts in dialogue with each other during the recent existential risk persuasion tournament, are substantially smaller than these break-even levels. I show this result is robust to various assumptions about the future course of population growth and the health effects of science, the timing of the new scientific dangers, and the potential for better science to reduce risks (despite accelerating them).

On the other hand, once we consider the more remote but much more serious possibility that faster science could derail advanced civilization, the case for science becomes considerably murkier. In this case, the desirability of accelerating science likely depends on the expected value of the long-run future, as well as whether we think the forecasts of superforecasters or domain experts in the existential risk persuasion tournament are preferred. These forecasts differ substantially: I estimate domain expert forecasts for annual mortality risk are 20x superforecaster estimates, and domain expert forecasts for annual *extinction* risk are 140x superforecaster estimates. The domain expert forecasts are high enough, for example, that if we think the future is "worth" more than 100 years of current social welfare, in one version of my model we would not want to accelerate science, because the health and income benefits would be outweighed by the increases in the remote but extremely bad possibility that new technology leads to the end of human civilization. However, if we accept the much lower forecasts of extinction risks from the superforecasters, then we would need to put very very high values on the long-run future of humanity to be averse to risking it.

When possible, in this report I present what I think is the most compelling evidence for different sets of assumptions. For many important questions though, there is a dearth of relevant evidence. What is the long-run future worth? How bad will technologies that have not yet been invented be? On these issues, I hope a contribution of the report is to help sort through the quantitative consequences of different starting assumptions. For most of the report, on questions for which data is lacking, I aim to adopt a neutral stance on which assumptions are to be preferred. In the final section of the report though, I give my personal assumptions, as well as my interpretation of what we should do in response to the findings in this report.

# Table of Contents

# 1.0 Setting the Stage

Consider the following hypothetical scenario:

A philanthropist makes a grant to an organization whose work makes scientific institutions more effective. Some examples of possible grantees could be:

- A group that organizes randomized control trials in government agencies that fund science, to identify more effective processes and procedures
- A group that helps set up new kinds of research organizations
- A training program for program managers in DARPA-like agencies
- An organization that systematically replicates important new research

Let us suppose these grants are highly effective, and the pace of scientific and technological progress meaningfully accelerates. As a result, people are healthier, richer, and happier, and the philanthropist feels like it's done a good job.

However, the acceleration of science may not be uniformly positive. Let's suppose that these grants bring forward in time an important discovery that allows individuals to engineer powerful new bioweapons. It could be that the damage to life and happiness from these bioweapons is so large that it overwhelms all of the good effects of science. In that case, these grants begin to look like a mistake: if the philanthropist had not done them, these terrible events would have been delayed.

And indeed, it's possible that these events would not only be delayed, but perhaps they could have been forestalled entirely if the philanthropist had not made these grants. Operating in parallel to the normal scientific institutions, a revolution born of artificial intelligence could be getting underway, operating mostly independently of the publicly funded academic research institutions affected by the philanthropist's grants. If artificial intelligence advances sufficiently to lead to a transformation of society, perhaps it will also lead to the development of technologies or strategies that eliminate the threat of bioweapons of mass destruction. In this case, the philanthropist's innovation policy program would have been *really* bad, since it brought forward the offensive capabilities of terrorists before the defensive capabilities (possibly from AI) were available. Many preventable deaths might have ensued.

Is accelerating science net welfare improving or not, in this new era of AI and rapidly advancing biotechnology? At the end of the day, this is a question of magnitudes. There are reasons that science can help us or hurt us, and ultimately to make an informed decision about whether the philanthropist should try to accelerate scientific progress on the whole, we need to try and quantify these costs and benefits in some kind of comparable units. That's what this report tries to do.

# 2.0 The Dilemma is Real

In this section, I present some evidence that the dilemma posed above needs to be taken seriously. We really do appear to be living in an era where we can not take for granted that faster and better science will necessarily be net positive for humanity.

## 2.1 Economics and the Science-Safety Tradeoff

In 1955, John Von Neumann published a prescient article titled "Can We Survive Technology?" His main argument was that new technologies increasingly expanded the geographic scope of intervention, rendering the world increasingly vulnerable to bad uses of technology originating in any country or region.

A natural response to such a fear is to either develop the good technology, but not the bad, or simply to slow down overall scientific progress. Von Neumann is quite skeptical either strategy is feasible. He writes:

> *The crisis will not be resolved by inhibiting this or that apparently particularly obnoxious form of technology. For one thing, the parts of technology, as well as of the underlying sciences, are so intertwined that in the long run nothing less than a total elimination of all technological progress would suffice for inhibition. Also, on a more pedestrian and immediate basis, useful and harmful techniques lie everywhere so close together that it is never possible to separate the lions from the lambs. This is known to all who have so laboriously tried to separate secret, "classified" science or technology (military) from the "open" kind; success is never more—nor intended to be more— than transient, lasting perhaps half a decade. Similarly, a separation into useful and harmful subjects in any technological sphere would probably diffuse into nothing in a decade.*
>
> Von Neumann (1955), pgs 669-670

And later:

> *Finally and, I believe, most importantly, prohibition of technology (invention and development, which are hardly separable from underlying scientific inquiry), is contrary to the whole ethos of the industrial age. It is irreconcilable with a major mode of* *intellectuality as our age understands it. It is hard to imagine such a restraint successfully imposed in our civilization.*

Von Neumann (1955), pg 670

Von Neumann's policy prescription is basically that we have no choice but to muddle through, adapting best we can as we go. But even if Von Neumann's intuitions here are correct, they are not necessarily applicable to the current age. In a world where transformative AI could be soon, several years of differential technological development (which Von Neumann seems to think feasible) could be quite valuable. Moreover, we are not really imagining a program to *slow* the progress of science - merely to direct energies towards goals other than its *acceleration*.

Two more semi-recent papers from economics address the science-safety dilemma from a very high-level perspective. Jones (2016) and Aschenbrenner (2020) each develop a semi-endogenous growth model where inventors can choose to invent conventional technology or safety technology. I have written explainers of these papers (and related others) on New Things Under the Sun here so rather than recapitulate those descriptions, here I merely pull out what I consider to be some important takeaways.

- Under relatively standard assumptions about the curvature of the utility function (linear in survival probability, but declining in the marginal utility of consumption), an increasingly wealthy society will "pivot to safety," trading slower economic growth for greater probability of survival.
- In these models, the magnitude of this shift is strongly influenced by the curvature of the utility function; the more quickly marginal utility from consumption diminishes, the greater the pivot to safety.
- A variety of correlational evidence also suggests a pivot to safety has been underway for decades among rich countries, consistent with the models' predictions.
    - Jones (2016) shows the health share of R&D has been rising, measured in various ways
    - Jones (2016) also shows the health share of overall spending has risen dramatically; this is not only about prices, as real consumption of health services also appears to have risen significantly.
    - Singla (2023) documents a sharp rise in the cost of complying with major health, safety, and environmental regulations since 1970
    - The sharp rise in R&D spending on alternative energy can also be interpreted as a pivot to safety

In short, standard economic models suggest societies will over time begin to confront the safety-growth tradeoff differently, increasingly prioritizing safety over growth and these models seem consistent with data. That suggests we should take seriously the claim that safety considerations deserve greater consideration than accelerating economic growth, all else being equal.

Note though that the more sophisticated model in Jones (2016) does not actually imply growth is bad per se, only that it isn't as good as safety, and competes with safety for innovation resources. Aschenbrenner (2020), in contrast, does envision a world where growth can actually be bad (or more precisely, larger economies pose greater annual risks of existential catastrophe, all else equal). But his model also makes a counter-intuitive case for *accelerating* economic growth, under some scenarios. The basic idea is, we may be living in a world where, if we were sufficiently rich, we would spend a lot more on safety technology. We want to get to that world as quickly as we can, because every year we face some risk of destroying ourselves. Although faster growth leads to bigger annual risks of destruction, until we develop safety sufficiently, this tradeoff may be worth it because we experience fewer years of risk overall by more quickly getting rich enough to pivot hard to safety.

I discuss the case for acceleration a bit more in section 10.0.

## 2.2 The Special Danger of Synthetic Biology

While the preceding applies to technology broadly, we have in mind a very specific technological capability that may be coming soon, namely the potential abuses of synthetic biology. To be very concrete, we are worried that instead of bombings, shootings, and hijackings, the lone wolf terrorist and terrorist organizations of the future will use advanced scientific and technological capabilities to design viruses that are much more deadly than what nature has ever created. The primary scenario I consider in this report is that, at some point in the near future, scientific and technological capabilities advance to a point where there is a discrete jump in the annual risk of a "biocatastrophe" by which I mostly mean new more dangerous forms of bioterrorism relying on man-made viruses (a form of synthetic biology). I call this the time of perils. Note my usage here refers specifically to biological perils, not perils from unaligned AI or other forms of new technology.[1]

[1] One reason this report focuses specifically on biocatastrophes rather than risks from advanced AI is that the rate of progress in AI is currently driven by major labs that operate outside the traditional academic ecosystem where science today is largely performed. Fundamental advances in the life sciences, in contrast, continue to be predominantly made in the academic ecosystem.

Synthetic biology is not the only technology with the potential to destroy humanity - a short list could also include nuclear weapons, nanotechnology, and geoengineering. But synthetic biology appears to be the most salient at the moment. Advances in fundamental biology have been quite rapid over the last decade and there is no reason to believe at present they are hitting a wall.

Worse, a 2014 book on barriers to bioweapons by Sonia Ben Ouagrham-Gormley argues that the primary impediment to bioweapons has been related to challenges in performing biological research, rather than access to materials (as was a primary impediment to nuclear terrorism). Traditionally, a large set of specialist skills has been required to do cutting edge biological work, and this has acted as a barrier to the use of biological agents as weapons. But advances in artificial intelligence have the potential to dramatically lower these barriers.

For example, a 2023 working paper by Soice et al. demonstrated how large language models can be used as a tutor for developing bioweapons. While some have criticized this paper for failing to compare to pre-LLM technologies such as Google search, these models are likely to become more advanced, and it seems unlikely that it will be possible to prevent access to open source versions of these models without strong safety guardrails that prevent their use for nefarious ends. It is not clear how far one can get without tacit knowledge about biological research that is not codified in the text that large language models are trained on, but at a minimum, it seems likely that some barriers to the use of synthetic bioweapons will fall substantially. See Sandbrink (2023) and Montague (2023) for more discussion.

## 2.3 Forecasting a Time of Perils

Another place we can look for evidence about whether our dilemma is genuine is the Existential Risk Persuasion Tournament (XPT). This extraordinary tournament recruited 170 people, half of whom were generalist superforecasters (meaning they had a demonstrated ability to perform well on short-term forecasting tournaments) and half of whom were subject matter experts (typically people who worked in academic departments, research labs, or think tanks on these issues). This group debated 59 questions related to long-run existential risks and submitted forecasts on a range of questions related to existential risk, including questions related to the scenario discussed in this report.

In the tournament, participants initially submitted their own forecasts, before being placed in online groups with others of the same "type": superforecasters were grouped with other superforecasters, and domain experts with other domain experts. After internal discussion with their group, participants could resubmit their forecasts. In the next stage, each superforecaster group was merged with a group of domain experts, and allowed to discuss and debate a second time, before again resubmitting individual forecasts. Finally, these groups were shown the forecasts and discussion of other groups and allowed to once more update their beliefs.

The tournament used two strategies to incentivize participants. First, prizes for forecast accuracy on questions related to 2024 and 2030 will be given after those dates. Second, forecasters were in some cases also asked to forecast what they believed their peers would forecast, and were rewarded on accuracy here as well. In the end, respondents typically spent dozens of hours on the tournament.

In this report, I consider the forecasts from this forecasting tournament the gold standard for estimating the future risks of biocatastrophes. This does not mean these forecasts are necessarily correct - as we will see, there is substantial disagreement between superforecasters and domain experts. But in my view, these are the best forecasts that are feasible to generate, and I think the burden of proof is on others to argue why the forecasts presented there should not form our default.

Though forecasting the future is very challenging, the estimates here are probably the best that currently exist in the world, as they reflect the considered judgment of a large number of people with a track record in forecasting, exposed to relevant expertise, and incentivized to perform well. It is therefore worth asking if these forecasts are consistent with the time of perils framework used in this report. Do we see evidence that forecasts expect new biological risks to increase in the future?

There are two directly relevant questions for assessing this.

*What is the probability that a genetically-engineered pathogen will be the cause of death, within a 5-year period, for more than 1% of humans alive at the beginning of the period...*
*... by the end of 2030?*
*... by the end of 2050?*
*... by the end of 2100?*

And:

*What is the probability that a non-genetically-engineered pathogen will be the cause of death, within a 5-year period, for more than 1% of humans alive at the beginning of the period...*
*... by the end of 2030?*
*... by the end of 2050?*
*... by the end of 2100?*

In the following table, I present the median answers submitted in the final round of the tournament, i.e., after people have had an opportunity to discuss and debate.

| Forecast | Group | Year | | |
|---|---|---|---|---|
| | | 2030 | 2050 | 2100 |
| Genetically-engineered pathogen killing >1% population | Superforecasters | 0.25% | 1.5% | 4% |
| | Domain experts | 1.22% | 8% | 10.25% |
| Non-genetically-engineered pathogen killing >1% population | Superforecasters | 0.5% | 1.69% | 3.62% |
| | Domain experts | 1% | 5% | 8.14% |

Table 1. Forecasts of pandemics. Excerpted from Table 23 in the XPT Report

Note that superforecasters and domain experts predict substantially different pandemic risks. I discuss this in more detail in section 11.1. For now I simply verify that both sets of forecasters foresee something like a time of perils scenario.

To see if these forecasts are consistent with the notion of a time-of-perils, I want to find the implied annual risk of each type of pandemic occurring over time. A time of perils framework would predict the annual risk from genetically engineered pandemics to rise over time, especially against the annual risk of non-genetically engineered (i.e., naturally occurring) pandemics. In technical Appendix A1, I describe how one can create annual estimates of pandemics from the above data.

I find the implied average annual rate of pandemic risk is:

| Forecast | Group | Year | | |
|---|---|---|---|---|
| | | 2023-2030 | 2030-2050 | 2100 |
| Genetically-engineered pathogen killing >1% population | Superforecasters | 0.04% | 0.06% | 0.05% |
| | Domain experts | 0.18% | 0.35% | 0.05% |
| Non-genetically-engineered pathogen killing >1% population | Superforecasters | 0.07% | 0.06% | 0.04% |
| | Domain experts | 0.14% | 0.21% | 0.07% |

Table 2. Implied Annual Risks of Pandemic Events (see Appendix A1)

Most of these changes are much smaller than the standard deviation of median beliefs, so this should be regarded as weak evidence. Note also that the number of biosecurity domain experts is only 14. I discuss some implications of different assumptions about the time of perils in section 7.0.

Those caveats aside, these forecasts are consistent with a time of perils framework. Both superforecasters and domain experts predict annual risks of a genetically engineered pandemic will be higher in 2030-2050 than over 2023-2030. Domain experts predict annual risks from genetically engineered pandemics will rise by 0.17pp per year (nearly doubling), compared to a forecast increase of just 0.07pp (a 50% increase) for naturally occurring pandemics. Superforecasters foresee a rise of 0.02pp (a 50% increase), but actually see the risks of naturally occurring pandemics declining by 0.01pp (a 15% decrease). Note both groups also foresee a long-run decline in the annual risks of genetically engineered pandemics after 2050, a phenomenon whose implications I discuss a bit in sections 10.0 and 11.2.

As noted above, the confidence intervals on these estimates is likely to be pretty wide, so that it is possible that forecasters do not foresee a time of perils at all. To the extent this is true, the case for the desirability of accelerating science will be much more straightforward, as discussed in sections 7.0 and 11.2.

## 2.4 Discussion

Sections 2.1-2.3 suggest the dilemma imagined by the philanthropist of section 1.0 is genuine.

- A minority of economists have long recognized that science could pose dangers, and recent work finds evidence for a rational pivot away from faster growth towards safer growth
- Advances in synthetic biology have the potential to cause great harm, and traditional barriers to their access from the difficulty of performing relevant research are probably falling due to advances in artificial intelligence.
- A recent forecasting tournament also foresees the annual risk of genetically engineered pandemics rising in the future relative to the risks from non-genetically engineered viruses, though these forecasts are too noisy to lean too hard on them.

Taken together, we cannot ignore the possibility that accelerating scientific progress would cause net harm. However, it is not yet clear how these dangers stack up against the potential benefits of science. This is necessary in order to proceed.

# 3.0 Baseline Quantitative Model: More Science

To set the stage, and anticipate some conclusions of the following quantitative models, let's consider a thought experiment. This thought experiment greatly simplifies things, but we will see in the analysis of the following sections that more complex models return similar quantitative results.

Because I assume in this paper that the effects of science happen with long delays, I'll frame this thought experiment as one in which a parent takes actions to make the life of their children and grandchildren go as well as possible. Specifically, let's imagine a parent knows that the birth of their child will happen to coincide with a time when society crosses a dangerous threshold of scientific and technological knowledge. From the day the child is born, small groups, or even individuals will be able to deploy synthetic biology to wreak mass death and destruction on the human race. I'll call this the "time of perils." Once that threshold is crossed, humanity enters a more dangerous world, where the annual probability of dying from one of these biocatastrophes is some fixed probability $d$ ($d$ stands for death and destruction). For example, if $d = 0.1\%$, then in every year after the child's birth, people everywhere have an extra 0.1% chance of dying from a biocatastrophe enabled by future advanced technology.

Our hypothetical parent is powerless to permanently halt the advance of science that will bring about this danger. But let's assume they do have the power to slow down science and delay the onset of the time of perils by a single year. Should they do it?

This hinges on the net value of a year of science. In our model, the cost of one year of science is the additional mortality risk it presents to the child. Let's assume the child's lifetime utility is some $U$. The cost of a year of science is the risk of dying during one extra year spent in the time of perils, and thereby forfeiting $U$. Since we have additional mortality risk $d$, the cost to the child of one year of science is $dU$.

Unfortunately, it is not only the child's life that is placed in peril. If they die, they won't have a child of their own, whose lifetime utility we'll also assume to be $U$. Thus, for a parent who wants the life of their child and grandchild to go well, the cost of a year of science is $2dU$, which is the additional extra risk $d$ of losing **two** lives with utility $U$. A parent probably also cares about great grandchildren, but we'll set that aside for the moment.

That's the cost of science. But science also brings benefits. In this paper, I assume science increases our future utility in two ways: by raising our income a little bit in every period, and by extending our lifespan. In this paper, we conservatively assume science affects income with a considerable lag, so in this thought experiment, we'll imagine only the grandchild benefits from it. Let's suppose $x$ is their proportional increase in lifetime utility due to higher income, and $y$ is their proportional increase due to greater survival. To calculate benefits from doing a year of science we should multiply the grandchild's utility by $(1 + x)(1 + y) - 1$ which is approximately equal to $x + y$ for small values of $x$ and $y$.

If the benefits of science exceed the costs for your child and grandchild, you'll prefer not to pause science. That occurs if:

$$2dU < (x + y)U$$

If these two are equal though, you'll be indifferent. Let $d^*$ denote the break-even level of mortality risk from these biocatastrophes, such that if $d = d^*$ you are indifferent to science happening or not. Then:

$$2d^*U = (x + y)U$$

If we divide both sides of the above equation by $2U$, then we find the breakeven value of $d^*$ is where $d^* = (x + y)/2$. If $d$ is greater than this, it's best if you pause science; if it is less, you want it to proceed.

So what are plausible values of $x$ and $y$?

Let's start with $x$, the proportional increase in lifetime utility due to income effects. A simple way to ballpark this is to assume that utility is the same in every year, compute how much better off a person would be in one year if their income was higher, and then divide the utility with extra income by the utility without it. In this report, I'll follow the lead of Open Philanthropy, and use a utility function increasing in the log of income, and where a year of healthy life today is worth 2 "utils". In other words, without extra income, one year's utility is about 2. In the paper, I assume an extra year of science raises incomes by 0.25% in all future years (discussed in detail in section 4.8). Since $log(1 + 0.0025) \approx 0.0025$ "utils", the proportional increase in utility due to income effects is on the order of $0.0025/2 = 0.00125$, so that $x = 0.125\%$.

Next, let's think about $y$, the proportional increase in lifetime utility due to longer life. In 2019, average life expectancy in the world was 72.8, and had been increasing by a fairly consistent 0.338 years per year for several decades. In the paper, I attribute about 56% of this increase in life expectancy to direct and indirect effects of science (discussed in detail in section 4.8), which implies a year of science normally increases our lifespan by about $72.8/(72.8 - 0.56 \times 0.338) - 1 = 0.0026$. If we assume the same proportional gains are possible during the time of perils, that implies $y = 0.261\%$.

To summarize, in this report I argue a year of science increases lifetime utility by about 0.125% via its impact on income and another 0.261% via its impact on health. Taken together, this implies $d^* = (0.125\% + 0.261\%)/2$, or $d^* = 0.2\%$. In other words, if a year of science by our ancestors imperils our future utility by an average of less than 0.2% per year, then the benefits of science exceed the cost. If the danger is greater than 0.2% per year, we would prefer them to pause science.

There is of course a lot of uncertainty about the values of $x$ and $y$. But I actually think we can be relatively confident about the order of magnitude of them. I think it's unlikely that they are both much smaller than half these levels, or much larger than twice these levels, so that we might want to more properly view $d^*$ as likely falling in a range from 0.1% to 0.4%.

As we will see, this is oversimplified in many ways. But in the following more sophisticated models, it turns out additional complications don't much matter, at least for computing break-even values. In my most preferred model, I obtain break even values of $d^* = 0.12\%$.

How does this stack up against the likely increase in mortality that might occur in the time of perils? In the following, my preference is to quantify the time of perils using forecasts made in the existential risk persuasion tournament. I argue $d = 0.002\text{-}0.04\%$ matches those forecasts best. Notably, these all lie well below the range of values of $d^*$. This implies that if the historical benefits to science are maintained, we should expect them to be significantly larger than the costs anticipated by both superforecasters and biological domain experts.

What if we care about our descendants beyond grandchildren? For example, suppose the parent also factors in the effects of science on their great-grandchildren. In that case, the costs of a year of science increase to $3dU$, but the benefits grow to $2(x + y)U$, and $d^*$ would increase to $2(x + y)/3$. If they also factor in their great-great-grandchildren, then the costs of science rise to $4dU$ and the benefits grow to $3(x + y)U$, and $d^*$ increases again to $3(x + y)/4$. This highlights the importance of choosing a reasonable timeframe for evaluating science. If we believe the benefits of science are persistent (and if $x + y > d$), then the longer the time frame, the greater are the benefits of science. On the other hand, if we expect the benefits of science to be transitory, then the opposite result might hold.

We consider these issues and more in the rest of the report.

## 3.1 Model Setup

My basic approach will be to define a utility function and a way to weigh the value of utility across time. We then compute total utility for everyone on earth, from now until eternity, under two scenarios: a status quo scenario, where health gains and income gains follow historic trends, and a "pause science" scenario, where science is turned off for one year. The difference between the utility in these two scenarios is our estimate for the average value of science today. Implicitly, for these results to be relevant to the question of the desirability of accelerating science, I am also assuming that the average value of science today is broadly the same as the value of "more" science happening in the future (for example, via more efficient use of existing R&D resources).

I will assume a discrete model of time, where everyone who is alive in a given period experiences a utility flow that depends on their health and income.

## 3.2 Discount Rates

An important component of this report is how we trade off utility losses today against gains tomorrow and vice versa. How we weigh the utility of future people versus those born today is therefore an important modeling decision.

In thinking about discount rates, there are two important considerations that pull in different directions. First, there are not many ethical reasons to place less weight on the welfare of future people than those born today. This is a rationale for applying a very low discount rate to the utility of future people (meaning the utility of future people is weighted very similarly to the weight of people today). For example, in Davidson (2022), a report on the social returns to productivity, Davidson uses a discount rate of just 0.2% for this reason.

At the same time, this report is premised on the possibility that the future could be radically different from the present in ways that make historic trends a poor guide to the future. This suggests it may soon become very difficult to *forecast* future utility. To the extent it becomes more challenging to assess the impacts of policies in the future, we may end up discounting the future more for these epistemic reasons, so that the utility of future people counts for considerably less than the utility of people born today.

My approach in this report is to imagine two possible futures which I call epistemic regimes. So long as the current epistemic regime persists, the future looks "enough" like the past for long-run trends to be useful for assessing the impacts of policy. However, if we exit the current epistemic regime, I assume historic trends are no longer a useful guide to policy implications.

I assume there is a constant annual probability $p$ that we remain in the current epistemic regime, where past trends are informative about future trends, and a constant annual probability 1 - $p$ that we exit the current regime.

Exiting the current epistemic regime can mean many things. It could mean that transformative AI leads to explosive economic growth. It could mean that unaligned AI leads to the extinction or disempowerment of the human race. It could mean that nuclear war breaks out, or we pass a climate tipping point, and human civilization falls apart. More broadly, it could simply mean that policy impacts become sufficiently hard to forecast that the net effect on welfare is unknown. The common thread between all these scenarios is that if we enter any of them, trends over the last century no longer apply. I assume the expected utility a representative person obtains, if we exit the current regime, does not depend on whether we accelerate science or not.

To this I add the important assumption that the policy being investigated - a pause to science - has no impact on the annual probability we exit the epistemic regime or what utility will be if we exit the current regime (an assumption I relax in section 8.0). This assumption means we do not have to specify the value of exiting the regime; utility in the next epistemic regime will disappear from our calculations. In practical terms, this will turn out to be equivalent to applying a discount rate $p$, justified on epistemic grounds. We will use a higher discount rate not because people today are worth more than people in the future, but because we are more confident historical trends are informative about people today than in the future.

Let $u(t)$ be the utility of people alive in period $t$, in the current epistemic regime, and $\widehat{u}(t)$ be the utility of people alive in period $t$, if we exit the current epistemic regime. Moreover, let $n(t)$ be the number of people alive in period $t$ in the current epistemic regime, and $\widehat{n}(t)$ be the corresponding number if we exit the current epistemic regime.

In this report, I end up applying only the epistemic discount rate $p$ to future people, so that if we knew for certain the effects of policies on future people, their utility would count the same as those alive today. Normalizing today to $t = 0$, total expected utility for people today and in the future can be written as:

$$V_0 = \sum_{t=0}^{\infty} \left\{ p^t n(t) u(t) + (1 - p^t) \hat{n}(t) \hat{u}(t) \right\} \tag{1}$$

Let's take this equation from left-to-right:

- The sum $\sum_{t=0}^{\infty}$ signifies that we'll be summing utility over all future periods.
- The term $p^t$ corresponds to the probability we are still in the current epistemic range in period $t$. If so the utility obtained is proportional to:
  - $u(t)$, which is the utility obtained in period $t$ in the current epistemic regime.
  - $n(t)$ is the number of people alive in period $t$.
- The term $(1 - p^t) \hat{n}(t) \hat{u}(t)$ corresponds to the utility we expect to obtain in the event we have exited the current epistemic regime. This happens with probability $1 - p^t$, in which case we use $\hat{n}(t)$ and $\hat{u}(t)$ to compute the number of people we expect to be alive and their utility.

Note that if we assume the policy under investigation has no impact on utility outside the current epistemic regime (as I do for now), in practice the term $(1 - p^t) \hat{n}(t) \hat{u}(t)$ will not affect our estimate of the policy impact. But the terms $p^t$, which are attached to utility in the current epistemic regime, will matter, providing a rationale for giving more weight to near-term effects of the policy intervention.

As noted, in this baseline model I am assuming the decision to pause science or not does not affect the annual probability of exiting the current epistemic regime. This assumption is fine if we think the time of perils can cause mass death and destruction, but not have effects so dramatic that it affects the probability we continue in our current epistemic regime (where the past is a good guide to the future). If the decision to pause science can affect the prospects for human extinction or other civilization-killing events, then this framework is no longer appropriate. In section 8.0, I explicitly model this kind of dynamic.

## 3.3 Utility Function

Inside the current epistemic regime, I chose a utility function to capture the following constraints, inspired by Open Philanthropy's moral assumptions:

- Utility increases with the log of income. See, for example, Tom Davidson's report on the social returns of productivity.

- Losing a year of healthy life is worth 2 natural log points. See [Technical Updates to Our Global Health and Wellbeing Cause Prioritization Framework.]

I interpret this as meaning that, today, a year of life is worth 2 natural log points for everyone, regardless of income.

A utility function that satisfies both these constraints is:

$$u(t) = 2 + ln(y_t/y_0) \quad (2)$$

for each year *t* of healthy life, and $u_t = 0$ if a person is dead in year *t*. I assume $t = 0$ corresponds to today, so that $y_0$ corresponds to today's income. This utility function normalizes the value of a DALY today, when $y_t = y_0$, to 2, while simultaneously allowing the flow value of utility to increase with the log of income. In particular, note that if income grows at steady exponential rate 1 + *G*, then $y_t = (1 + G)^t y_0$ and $u(t) = 2 + tln(1 + G)$.[2] In practice, I'll generally be approximating this as $u(t) = 2 + tG$

## 3.4 Income Growth

In the status quo, I assume a constant proportion of income is directed to conducting science, and per capita income grows in each period by a factor of 1 + *G*, so that $y_t = (1 + G)^t y_0$ and $ln(y_t/y_0) \approx tG$. Since the economy is growing, this implies science is "getting harder" since growing absolute levels of science R&D yields a constant rate of growth.

[2] This formulation does have the undesirable implication that saving the lives of higher income individuals is more valuable than saving the lives of lower income individuals. There are a few reasons I chose to model things this way. First, in this particular report, everyone in a given time period is assumed to have the same income, so we are never trading off the health of higher income individuals and lower income individuals within the same cohort. Second, while this framing does imply saving the lives of someone in the future increases utility more than saving the life of someone today (since future people have more income), the report also has a discount rate that is steep enough to ensure we, in practice, never value the lives of future wealthy people more than poorer people living today. Finally, in an earlier version of this report, I did completely separate utility from health and income, so that a healthy year of life was always worth 2, but living or dying did not adversely affect utility from income. However, this complicates the interpretation of utility during the time of perils, because it implicitly assumes the gains from income that accrue to a large healthy population continue to count in the social welfare function, even as the population shrinks due to the increased mortality of the time of perils.

Now let's compare this to the world where we pause science. For an individual born today, when we pause science for a year, I assume that after $T$ years growth slows from $G$ to $g$, where $g < G$, for one year. It then reverts to $G$ in the following year. For concreteness sake:

- $y_T = (1 + G)^T$
- $y_{T+1} = (1 + G)^T(1 + g)$
- $y_{T+t} = (1 + G)^{T+t-1}(1 + g)$ for $t > 1$.

This slowdown reflects the fact that science only translates into a tangible impact on society with a long delay, as discoveries are translated into technologies and technologies diffuse around the world. Note that we will assume $g > 0$, so that some growth can occur even without science. In reality, the effects of a science pause would probably be distributed across many years, some nearer and others farther, but I assume this is not of first-order importance.

## 3.5 Population Growth

The evolution of $n(t)$, the number of people alive today reflects the joint forces of fertility and health. Health, in turn, is driven by both "natural" and "unnatural" risks. In this report, the primary way I think of the risks of science is via a "time of perils" framework. The time of perils refers to the idea that science and technology will eventually advance to a point where it becomes possible for small groups or even individuals to inflict unprecedented death and destruction on the rest of the human race. Conceptually, this is not too dissimilar to this paper's epistemic regime assumption; at some point in the future, historical trends, this time around health, are no longer applicable. However, unlike the epistemic regime assumption, which assumes that when we exit the current regime we can no longer forecast the impact of our policies, we will attempt to explicitly model the time of perils.

To start, if we set aside the time of perils, $n(t)$ increases each year as a function of the birth rate $b_t$, and decreases as a function of the (natural) death rate $m_t$ (for mortality). Thus, in each year, the number of people alive evolves according to:

$$n(t) = (1 + b_t)(1 - m_t)n(t - 1) \qquad (3)$$

Both the birth rate and the mortality rate are on a long-term decline. To keep the model tractable, it is convenient to assume that under the status quo, $b_t$ and $m_t$ evolve together such that we get a constant population growth rate $s$:

$$(1 + b_t)(1 - m_t) = (1 + s) \quad (4)$$

For small values of $b_t$ and $m_t$ we can ignore the second-order terms $b_t m_t$ and obtain $b_t - m_t = s$.

Basically, even though birth rates fall, mortality rates also fall in a way that the global number of DALYs increases (or decreases) by the same amount each year. For small rates this will occur so long as birth and death rates decline at roughly the same absolute level. Note, under this approximation, we do not require $b_t$ and $m_t$ to decline at a constant rate; just at the same (possibly time-varying) rate. For example, this model will still work if both $b_t$ and $m_t$ asymptotically approach 0 at the same rate. In section 6.0, we examine how the model's results are affected by a less simplistic model of the evolution of human health.

I will assume the time of perils commences after $t_1$ periods. In the status quo, once the time of perils commences, the annual population growth rate changes from $(1 + s)$ to $(1 + s)(1 - d)$, where $d$ is the annual probability of dying from unnatural causes in the time of perils.

Now we consider the impact of pausing science on the population growth rate. As with economic growth, I assume pausing science only has an impact after $T$ periods, because it takes time for technologies to be developed and deployed. I assume that pausing science for one year (eventually) slows the decline in mortality for one year, after which the mortality rate returns to its normal trajectory. If the birth rate $b_t$ remains on the same trajectory, and the mortality rate $m_t$ slows for one period, then because $b_t - m_t = s$ in the status quo, pausing science leads to a permanent decline in the net population growth rate $s$. This is because the decline in mortality is permanently "behind" where it would be if we had not paused science.[3] Thus, one of the primary effects of pausing science is that after $T$ periods, the population growth rate slows from $s$ to $\bar{s}$.

[3] I prefer to think in terms of the birth rate remaining on its status-quo trajectory, riding a wave of cultural and policy decisions, and science instead influencing the population solely via its impact on mortality. However, one could also presume a science pause has an impact on both the birth rate and the mortality rate, and the net impact on population growth delivers a slowdown in the overall growth rate $s$. On the one hand, science may reduce fertility, since rising incomes seem to be associated with lower birth rates; on the other hand, science may increase fertility, for example, by leading to better fertility treatment options. If the net effect on birth rates is zero, then again the primary channel of influence flows through the reduction in mortality.

## 3.6 Comparing Utility

Substituting in definitions of $u(t)$, $y_t$ and $n(t)$, into equation (1), we can obtain the formula for total utility in the status quo. For more details on this, see technical appendix A2. Once we have equations for the total utility in the status quo and the pause science scenarios, we can subtract the one from the other to obtain the increase in utility from a year of science today.

This results in the following equation, which I have broken down into a few parts in Table 3 that I think makes it a bit easier to understand. In the table, I rely on a few new variables and equations defined in appendix A2: $\Delta(t_1, t_2)$, $U_{T+1}$, $N_T$, and $L(s,d)$.

These four terms capture some useful intuitions. The term $\Delta(t_1, t_2)$ captures the total expected utility elapsing between periods $t_1$ and $T$, for each person alive today. The term $U_{T+1}$ captures the annual utility flow in period $T + 1$, assuming we make it there. The term $N_T$ can be interpreted as the number of people expected to survive to period $T$ in the current epistemic regime. It's driven by three competing dynamics. We remain in that regime each period with probability $p$, but in each period we remain, the population growth by $s$ percent and, once the time of perils begins, shrinks by $d$ per year.

The term $L(s,d)$ can be interpreted as the total number of future life years in the current epistemic regime that will be obtained for every person alive in year $T + 1$. It's value depends on the variables $s$ and $d$, which capture the annual growth (or decline) of life years via mortality and population growth, as well as $p$, which captures the annual probability of exiting the current epistemic regime.[4]

With these definitions in hand, the returns to science are captured by the following equation.

[4] I omit $p$ from the function because I assume it is fixed in this report.

| | | |
|---|---|---|
| $V_{SQ} - V_{PS} =$ | | The difference between total utility in the status quo, where we engage in science today, and the pause science scenario, where we do not |
| | $- d \times n(0) \times \Delta(t_1, t_2)$ | Pure peril effect: The expected share of utility that will be lost between when the time of perils commences and the positive effects of science commence, due to an extra year of science. |
| | $N_T \times \frac{1}{1-d} \times L(\bar{s}, d) \times (G - g) +$ | Pure income effect: For all future life years, if we keep science on, everyone alive earns a bit more income. |
| | $N_T \times U_{T+1} \times \left[L(s, d) - \frac{1}{1-d} \times L(\bar{s}, d)\right] +$ | Pure health effect: The value of utility with income at year $T+1$, multiplied by the change in expected life years; in the second term, we spend one less year in the time of perils (+), but thereafter have a lower annual growth rate $s$ (-) |
| | $N_T \times G \times [L(s, d)^2 - \frac{1}{1-d} \times L(\bar{s}, d)^2]$ | Health-income interaction: Utility in the more distant future is higher due to higher incomes, so this function puts more weight on more distant life years (quadratic in expected life years in this regime) |

Table 3. How Science Affects Utility

We can also make a few observations about this result.

The first component, the pure peril effect, is strictly negative. This component captures the notion that the risks from science may precede the benefits. As we discuss in more detail in section 4.4.1 and 4.4.2, the logic here is that if perils arise from the deployment of synthetic biology by bad actors, the risks may arise as soon as the technologies are invented in countries operating on the frontier. Just as the covid-19 pandemic was a global problem, genetically engineered pandemics developed in frontier countries would quickly cross borders and become an international problem. Meanwhile, beneficial new technologies enabled by science tend to diffuse slowly to the world at large.

The second component, the pure income effect, is strictly positive. This captures the increase in income that is enabled by a year of science.

The pure health and health-income effects, are ambiguous in sign. This reflects the fact that science reduces natural risks and increases unnatural risks in our model.

Another observation is that the larger is $d$ (the annual peril from being in the time of peril), the lower are the returns to science. This occurs through two distinct channels:

- When $d$ is larger, the cost of spending an extra year in the time of peril is larger, which increases the pure peril effect and reduces both the pure health and health-income interaction effects.
- When $d$ is larger, we have fewer expected life years in both scenarios. This compresses income gains and also reduces the gap between $L(s, d)$ and $L(\bar{s}, d)$: the effect of science on natural risks no longer matters as much when people are more and more likely to die of unnatural risks anyway.

A third observation, as the difference between $s$ and $\bar{s}$ grows and $p$ becomes close to 1, the returns to science are more likely to be positive. This is because a small value of $\bar{s}$ relative to $s$ maximizes the gap in $L(s,d)$ across scenarios, since a small value of $\bar{s}$ means science is very important for growth in life years. A larger value of $p$ also increases the gap in life years, because it becomes less likely that we will leave the current epistemic regime before enjoying different life years. A larger value of $p$ also implies we enter the time of perils with a greater probability, so that the policy can have more impact (though this does not affect whether the impact of policy is positive or negative).

To sum up, there are several competing factors, and so we cannot rule out positive or negative returns to science today. Instead, we must do our best to select parameter values. We turn to this task next.

# 4.0 Selecting Parameter Values

## 4.1 *d* and *t1*: peril parameters

I begin with one of the most important parameters, and the one that takes the most work to select: the annual excess mortality risk *d* associated with the onset of the time of perils. Selection of this is closely associated with the timing of the onset of perils, $t_1$ and so I discuss this parameter as well in this section.

In this report, I take three approaches to estimate *d*. First, to establish a baseline, I set $d = 0$ to illustrate what the returns to science would be in the absence of a time of perils. Second, I identify break-even values of *d*, for which the returns to science are positive below this value and negative above it.

My third approach relies on forecasts in the existential risk persuasion tournament (XPT). For these questions, I rely on the forecasts of up to 89 superforecasters and 14 biosecurity domain experts. Note that each group was able to present arguments to the other, and I use the median group forecast after exposure to the estimates of the other group. In other words, these are superforecaster estimates after they have been in extensive online dialogue with domain experts, and vice-versa.

As noted in section 2.3, while forecasting the future is very challenging and we do not even know if it can be reliably done at all over the long run, I think we have strong reasons to treat the results of this forecast as our best guess. I am not a domain expert in the relevant fields, but I think we have strong reasons to see these estimates as defaults, and to impose the burden of proof on critics who would seek to suggest alternatives. In this section, I briefly describe how I use forecasts from the Existential Risk Persuasion Tournament (XPT) to derive a variety of estimates. There's a lot of work in here, and I defer a lot of the details to technical appendix A3.

As discussed in section 2.3, forecasts in this tournament are consistent with the time of perils framework (though confidence intervals are large enough that the evidence in favor is weak). In this section, I go further and impose a lot of structure on the forecasts made in the XPT so as to derive the most credible estimate of *d* and $t_1$ for my model that is consistent with forecasts. Specifically, I perform the following adjustments (see the appendix for full details).

First, I use superforecaster and domain expert responses to questions about human extinction and global GDP growth in excess of 15% per annum to back out their implied annual forecasts that we exit the current epistemic regime. For example, if they think there is a low probability of a genetically engineered pandemic by 2100, but a high probability of global GDP growth above 15%, I take this as a belief that the *reason* we don't expect to see a pandemic by 2100 is because we will obtain transformative AI well before then. Since my model is interested in the risk of pandemic *conditional* on us remaining in the current epistemic regime, I need to adjust for that or I will understate respondents beliefs in the risks of pandemic.

Second, I turn to forecasts about the probability of major genetically engineered pandemics by 2030, 2050, and 2100 to infer respondents' beliefs about the increase in risk arising during the time of perils. I first use the implied beliefs about exiting the current epistemic regime to inflate these forecasts and arrive at conditional forecasts.

Third, I assume respondents actually believe the future conditional probabilities of genetically engineered pandemics follows the scheme diagrammed in Figure 1.

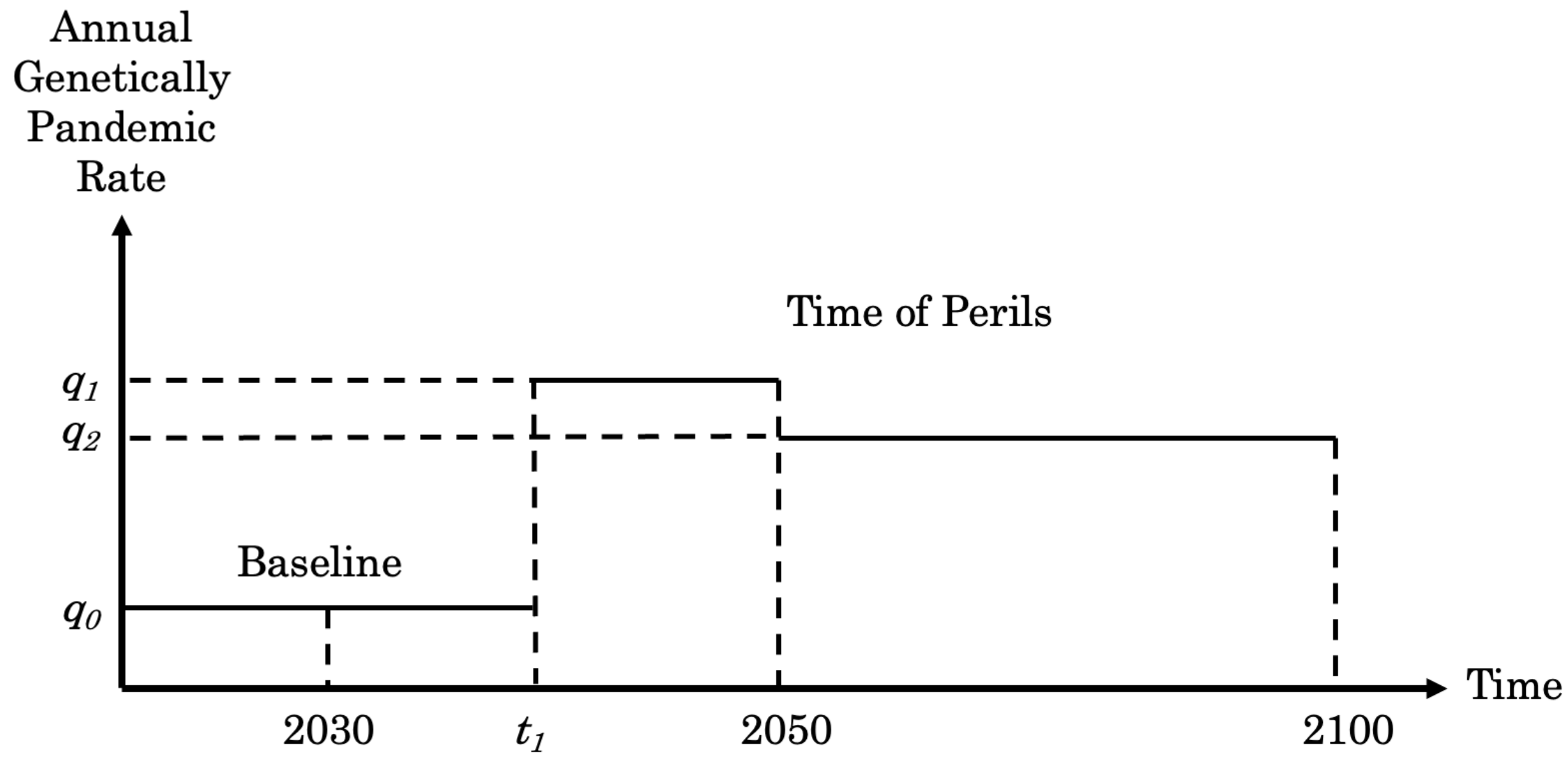

Figure 1. Annual genetically engineered pandemic probabilities framework

Using respondent forecasts about the likelihood of genetically engineered pandemics in 2030, 2050, and 2100, I back out mathematically consistent combinations of all these parameters (including $t_1$). I then select values that balance conservatism (by which I mean selecting parameter values that imply the time of perils is more dangerous) with plausibility (which means I try to avoid assuming forecasters believe a very short and sharp time of perils).

All of this gives me plausible estimates of the annual risk of a genetically engineered pandemic. But to fit my model, I also need to know the excess mortality associated with pandemics. To calibrate that, I use responses to a post-XPT survey about the probability a genetically engineered pandemic kills 10% or 100% of the population. A few key assumptions I make:

- Pandemics killing 1-10% of the population kill 2% on average
- Pandemics killing 10-99.9% of the population kill 20% on average
- It is only possible to design a virus capable of causing human extinction once the time of perils begins. This is a conservative assumption that amplifies the benefits of delaying the time of perils.

This is a lot of work, but here are the key takeaways.

The following set of conditions, while not explicitly predicted by superforecasters, are broadly consistent with a range of their estimates about the probability of various outcomes.

- We currently live in a world where there is an annual risk of 0.002% that an individual will die from a genetically engineered pandemic event.
- Sometime in the future - perhaps around 2038 - scientific and technological capabilities will evolve such that this rises by 0.0021% per year, to a total of 0.0041%. This represents, roughly, a doubling of the annual risk of death via genetically engineered pathogen.
- At the same time, the annual risk of human extinction due to a genetically engineered pathogen will rise from ~0% to 0.0002% per year.
- These risks will persist for as long as there is no major change to our civilization.

Such a scenario is broadly consistent with superforecaster estimates of the probability of genetically engineered pandemics, catastrophes, extinction, as well as major changes to our civilization.

Domain experts are more pessimistic. The following set of conditions are broadly consistent with a range of domain expert estimates about the probability of various outcomes.

- We currently live in a world where there is an annual risk of 0.0075% that an individual will die from a genetically engineered pandemic event.
- Sometime in the future - perhaps around 2037 - scientific and technological capabilities will evolve such that this rises by 0.0385% per year, to a total of 0.0460%. This represents, roughly, a six-fold increase in the annual risk of death via genetically engineered pathogen.
- At the same time, the annual risk of human extinction due to a genetically engineered pathogen will rise from 0% to 0.0228% per year.
- These risks will persist for as long as there is no major change to our civilization.

Such a scenario is broadly consistent with domain expert estimates of the probability of genetically engineered pandemics, catastrophes, extinction, as well as major changes to our civilization.

For both groups, I assume the time of perils commences around 2037-2038, or 14-15 years after 2023 (when this report was first drafted). In my baseline, I assume $t_1 = 15$.

For our model, we are primarily interested in the notion that science might bring forward in time the time of perils. Accordingly, in my model, I assume the baseline rates are captured in normal health indicators ($s$), and focus on the annual 0.0021% or 0.0385% increases in the mortality rates that occur in the time of perils.

In other words, in my model, I explore setting $d$ = 0.0021% (if we use superforecaster forecasts) and $d$ = 0.0385% (if we use domain expert forecasts).

For some context, the Economist estimated that in 2021, excess global deaths due to covid-19 were about 10mn. Against a global population of 7.91bn, this works out to an increase in the probability of death due to covid-19 equal to 0.1%. Thus, superforecasters forecasts are implicitly equivalent to an *additional* covid-19 level pandemic every 48 years, due to genetic engineering. Domain experts are more pessimistic and see the equivalent of this kind of event happening every 2.5 years.

As an aside, it is notable that these forecasts differ by more than an order of magnitude. I do not have the domain expertise to adjudicate between these positions, but in section 11.1 I give some reasons why my own preference leans towards the views of the superforecasters. Nonetheless, through the report, I present both estimates to establish the robustness of results across a wide range of opinion.

Finally, as I note in technical appendix A3.5, one can think of the $d$ = 0 and the domain expert forecasts as roughly corresponding to 10th and 90th percentile superforecaster estimates. Similarly, the superforecaster estimate roughly corresponds to the 15-20th percentile forecast among domain experts. While I do not estimate a 90th percentile estimate for domain experts, as I explain in technical appendix A3.5, I think these very pessimistic forecasts are actually not well characterized by a time-of-perils framework and instead think the analysis of section 7.2 is a better fit for the views of this group.

## 4.2 $p$: annual probability of leaving the current epistemic regime

I tried to estimate the annual probability of exiting the current epistemic regime in two ways.

First, in Appendix A3.1, I look at questions in the existential risk persuasion tournament (XPT) that are plausibly related to exiting the current epistemic regime. For example, I can look at forecaster's beliefs that global GDP growth will exceed 15% for at least one year (the kind of thing that some argue might be a consequence of achieving transformative AI), that the human race will go extinct or suffer a major catastrophe (via causes other than a biocatastrophe). In the appendix, I argue these deliver values of $p$ ranging from 99.3-99.95%. In other words, responses in the XPT are consistent with a belief that the world will change dramatically with only a 0.05-0.7% annual probability.

This is quite low, and tends to mean the distant future counts for a lot. For example, if the annual probability we enter a new epistemic regime is 0.2% (the geometric mean of 0.05% and 0.7%), then it is a coin flip whether the world looks mostly like today in 350 years. However, this range is probably too conservative for two reasons. First, the XPT only asked for forecasts through 2100 (or 77 years from now), and it may be inappropriate to extrapolate their forecasts forward by hundreds of years. Second, it is premised on the assumption that catastrophe and rapid global GDP growth are the only two ways we may enter a new epistemic regime. But there may be other ways this could happen.

For comparison, according to a 2023 Rethink Priorities report, GiveWell uses a discount rate of 1.4% to account for temporal uncertainty, including the possibility that forecast gains are not realized due to major changes in economic structure, catastrophe, or political instability. This is equivalent to assuming $p$ = 0.986. While close in spirit to the sense of $p$ used in this report, there are two issues with this estimate. First, it only covers the arrival of events that cause expected benefits not to be realized; I want to also include scenarios where the epistemic regime merely changes so much that we can no longer assess the impact of a policy, relative to the counterfactual. Second, it is probably referring to the probability of changes at the local level, where GiveWell interventions are occurring, whereas I am looking for the probability of global changes.

As a second approach, I used a mix of estimates that foresee much higher probabilities of change. To get an alternative perspective on the annual likelihood of this occurring, I turn to the Open Philanthropy AI Worldviews Contest, where Open Philanthropy assembled an internal panel and according to the program announcement, "[p]anelist credences on the probability of AGI by 2043 range from ~10% to ~45%." Given these credences, for transformative AI not to occur, we need to go through twenty periods without transformative AI. These are consistent with annual probabilities of transformative AI on the range of $p^{20} = (1 - 0.1)$ and $p^{20} = (1 - 0.45)$ which implies $p \in [0.971, 0.995]$.

Another place we can turn to is Ord (2020). Ord estimates a ⅙ probability of human extinction by 2100, which includes a 1/10 probability of extinction via unaligned AI. This suggests Ord sees a $1/6 - 1/10 = 1/15$ probability of extinction via other causes by 2100. This implies an annual probability of extinction due to non-AI factors of 0.08%.

Lastly, turning away from disasters, Thomas Phillipon's paper on additive growth argues there was a large trend break in the historical evolution of total factor productivity in the 1930s, the only such break since the industrial revolution, which I date to 1700. This implies one major economic break every 300 years or so, or an annual probability of 0.3%.

Taken together, I will assume an annual probability of transformative AGI of 1.6% (which implies a 27.5% probability of transformative AGI by 2043), plus an annual probability of extinction of 0.08% and annual probability of major economic break of 0.3%. Summing up, I assume an annual probability of 2% that we leave the current epistemic regime, so that historical trends are no longer applicable. This would imply $p = 0.98$.

In the following report, I make the conservative assumption that $p = 0.98$. Compared to the alternative assumptions implied by the XPT, this will put more weight on near-term events and less weight on what happens in the distant future. As we will see, this will tend to make the case for pausing science stronger.

## 4.3 *G*: *Status Quo per capita growth*

I interpret *G* as a measure of frontier growth, for which we can use the USA as a proxy. US growth has averaged around 2.0% per year since the 1950s, and inflation-adjusted long-term bonds imply similar growth in the years ahead (see the end of this post).[5] However, Chad Jones points out that about half the annual per capita growth since the 1950s has been driven by temporary boons: growth in education per worker, declines in misallocation, and the employment population ratio. If these are topped out, then TFP alone can only give us 1.0% annual growth; if that relies entirely on population growth, then only 0.3% growth can be counted on in the long-run! Robert Gordon, in contrast, forecasts productivity growth of just 1.2% per year from 2015-2040.

[5] Interest rates actually imply long-run growth of 2.6%, but assuming 0.5% population growth prevails (as it has recently), this implies growth in per capita income of 2.1% per year (since 1.026 / 1.005 = 1.021).

All told, I estimate in the status quo, growth derived from scientific and technological progress is 1% per year, or $G = 0.01$. An additional 1% likely will come from other factors, but since this growth is baked in, we can usually ignore it when comparing counterfactual policies.

## 4.4 *T: Lags between science and impact*

I break $T$ (the gap between when science takes place and when it has an impact on growth and health) into two components: the gap between science and invention, and the gap between invention and broad diffusion.

### 4.4.1: The gap between science and invention

As discussed in more detail in this post, we have two lines of evidence on the gap between science and technology. The first line of evidence tries to sniff out statistical correlations between changes in R&D funding and changes in total factor productivity in related industries in subsequent years. The second line of evidence comes from the citations patents make to the academic literature. Both lines of evidence converge on twenty years as being a reasonable rule of thumb for the gap between basic discoveries and their subsequent technological impact in frontier economies like the USA.

### 4.4.2: The gap between invention and broad diffusion

While 20 years is a reasonable measure of the gap between science and productivity or health impacts in the USA, our model is premised on global benefits and the USA contains only 4% of the Earth's population. While the costs of genetically engineered pandemics would plausibly spread in a matter of weeks and months around the world, historically the diffusion of technology around the world has proceeded at a comparatively glacial pace. Since this model is based on a representative average person, I estimate here the lag between invention and diffusion to an average person.

There are few ways to estimate that. Comin and Mestieri (2014) (p19) look at the rate of diffusion for a variety of different technologies and find the average adoption lags across countries and technologies is 44 years. While this is an informative first pass, it's not clear how representative their basket of technologies are. Since I am ultimately interested in the effects of science on income and health, I rely on an alternative approach that tries to estimate lags by looking at income and health directly. Since we are assessing the rate of diffusion going forward, I will also be interested in evidence on whether the rate of diffusion is speeding up or slowing down.

First, I compare global GDP per capita to GDP per capita at the scientific frontier. If technology is the ultimate driver of living standards, we can estimate the time until a technology is available to the typical global citizen by estimating how long it will take for global GDP per capita to converge to the current scientific frontier.

To do that, I pulled the Penn World Tables, and computed global GDP in 1970, 1980, 1990, 2000, 2010, and 2019, using expenditure-side real GDP at chained PPPs (in mil. 2017US$). Over this period, global GDP per capita grew from $5,760 (measured in US 2017 dollars) to $16,552. Average growth over this period was (16,552/5,760)^(1/49) = 1.022. I therefore use 2.2% as the average real GDP growth rate of global GDP per capita.

I then pulled GDP per capita for three frontier science countries: the USA, Germany, and the UK. I can then ask: how long would global GDP per capita have to grow at 2.2% per annum to reach the GDP per capita levels of these scientific frontier countries. Results are displayed in the table below.

| | Global per capita income | USA per capita income (lag at 2.2% annual growth) | German per capita income (lag at 2.2% annual growth) | UK per capita income (lag at 2.2% annual growth) |
|---|---|---|---|---|
| 1970 | 5,760 | 25,354 (68) | 16,183 (47) | 16,302 (48) |
| 1980 | 7,175 | 30,788 (67) | 22,180 (52) | 21,380 (50) |
| 1990 | 8,335 | 39,059 (71) | 27,886 (55) | 25,199 (51) |
| 2000 | 10,267 | 50,089 (73) | 37,226 (59) | 35,660 (57) |
| 2010 | 14,181 | 53,666 (61) | 43,696 (52) | 40,132 (48) |
| 2019 | 16,552 | 63,393 (62) | 51,593 (52) | 46,187 (47) |

Table 4. GDP per Capita (2017$, chained PPP), from Penn World Tables

In the USA, lags average around 67 years. For Germany, it's 53 years and for the UK its 50 years. This approach suggests lags longer than the 44 years implied by Comin and Mestieri's estimate. As an approximation, I think it's fair to let the UK/German average of 51.5 years stand in for the gap to the EU and the US gap of 67 stand in for the USA. The US funds about 23% of global R&D, compared to the EU's 17%, so a weighted average of USA and EU lags gives an average economic lag of around 60 years.

A second approach is based on health. Using Our World in Data, we can look at global life expectancy for a particular year, and then estimate the year life expectancy was at a similar level in our frontier countries. The gap between the two is another measure of how long it takes technology from the scientific frontier to impact global life expectancy.

| | Global Life Expectancy | US Life expectancy equivalent year (lag) | German Life expectancy equivalent year (lag) | UK Life expectancy equivalent year (lag) |
|---|---|---|---|---|
| 1970 | 56.1 | 1920 (50) | | 1920 (50) |
| 1980 | 60.6 | 1931 (49) | | 1933 (47) |
| 1990 | 64.0 | 1942 (48) | | 1943 (47) |
| 2000 | 66.5 | 1947 (53) | 1950 (50) | 1947 (53) |
| 2010 | 70.1 | 1964 (46) | 1963 (47) | 1954 (56) |
| 2019 | 72.8 | 1976 (43) | 1979 (40) | 1975 (44) |

Table 5. Life Expectancy Lags, from Our World in Data

Walking through this table, in 1970 global life expectancy was 56.1 years. That life expectancy was observed in both the UK and USA in 1920, or 50 years earlier. Over the whole period, the average lag has been 48 years in the USA, 46 years in Germany, and 50 years in the UK.

There is some evidence technological diffusion is happening a bit faster now than in the past: the lags to catch up with American income have shrunk from the upper 60-year range to the lower 60-year range. The lag in life expectancy has also shrunk from the upper 40-year range to the lower 40-year range over time. On the other hand, there is little or no reduction in the time to convergence to German and UK income levels. Moreover, in their study of the diffusion of a basket of major technologies, Comin and Mestieri (2014) find that while the time between when a technology is invented and when it *first* arrives in another country has shrunk over time, the lag between when a technology is invented and when it diffuses widely in a given country's population may actually have increased over time.

In the interests of being conservative, I won't assume a trend towards acceleration over this time period. Instead, I'll assume a lag on the order of 48 years for health and a lag of 60 years for economic growth. I'll split the difference and assume global lags of 54 years for the gains from science are a reasonable estimate.

### 4.4.3: The total gap

If we assume it takes 20 years to translate science into technology at the frontier, and a further 54 years for technology to diffuse across the globe and begin to impact average health and income, then the total lag between scientific research and average global impact on health and income is 74 years. Accordingly, I set $T = 74$.

## 4.5 *g: Pause Science per capita growth*

To get a handle on the share of technological progress that derives from science, we can turn to a few different lines of evidence. As discussed in this post, in a 1994 survey, R&D managers at major corporations estimated about 20% of research projects depended on public research. As discussed in the same post, about 26% of patents directly cited a scientific article in 2018.

There are reasons to believe science's contribution to technology are higher or lower than 20-26% though.

First, a variety of evidence consistently finds that patents that rely on science are more valuable and creative than those that do not. For example, Krieger, Schniter, and Watzinger (2022) find patents that directly rely on science are worth 26% more than otherwise similar inventions that do not. So even if roughly 20% of inventions directly rely on science, it is likely that these inventions account for a larger share of the total value of new technologies.

Second, as discussed in this post, even if a technology does not directly rely on science, it may indirectly benefit from science if it builds on technologies that do rely on science. Approximately 61% of patents over 1975-2015 *indirectly* cite scientific articles (for example, by citing a patent that cites a scientific article). This is, itself, probably on the low end because the dataset used lacks citation data for patents granted before 1975, so any patents that cite patents granted before 1975 are not counted as being indirectly related to science. Evidence on the predictability of technological advances that follow "upstream" technology provides additional evidence that these indirect citations are not spurious.

Other evidence comes from statistical correlations between public R&D proxies and subsequent total factor productivity growth. For a sample of manufacturing industries, Adams (1990) estimates the contributions of publicly funded research as follows:

| Period | Industry TFP Growth | Contribution of science (range) | Share of Growth Attributable to Science |
|---|---|---|---|
| 1953-1966 | 0.011 | 0.0063 - 0.015 | 57-136% |
| 1966-1973 | 0.006 | 0.0055-0.0056 | 92-93% |

Table 6. Adams (1990) Estimates of Science Contribution to Manufacturing TFP Growth

However, there are also reasons to revise the 20-30% range down. On the patent evidence, we do not know whether citation of a scientific article means that the invention would not have been possible, absent the underlying science. In some cases, citations may be issued for reasons besides direct intellectual debt (such as pointing to related, but not antecedent, work). In other cases, science may have accelerated development of a technology, but the technology would have been developed regardless. This case would seem to be stronger for patents that do not directly cite science at all, or only indirectly cite it.

Moreover, the data from Adams (1990) is quite old in two important respects: first, it literally describes the relationship between science and technology several decades ago, and second, it uses out-dated econometric methods. Unfortunately, I am unaware of directly relevant updates. But I think that means we shouldn't put too much emphasis on its extremely high findings.

Lastly, as a theoretical argument, we can conduct a thought experiment about what would happen to technology development if we paused science for a year. Let us suppose x% of new technologies have their roots in science. If we pause science for one year, we do not stop all further development work on these x% of technologies. Instead, it seems likely to me that we push forward those platforms using the existing science, which will not be as useful as new discoveries. Progress on those x% would be slower than it might otherwise be, but it would not fall to zero.

As a first pass estimate, in this report I assume 24% of patents directly cite science and these patents are worth 25% more than the average patent, so that they account for 0.24 x 1.25 = 30% of the value of all technological progress. I'll assume a bit arbitrarily, that stopping science for one year would reduce the value of this group by 50%. Second, I'll assume another 50% of inventions indirectly rely on science (that is equivalent to assuming 75% of patents have some indirect link to science, which is above the actually observed 61%, but I think that's an underestimate, especially looking forward) and that these inventions account for 50% of the value of technological progress. I'll assume pausing science for a year slows progress on this group by 20%. The remaining 20% of technological progress has no dependence on science. Taken together, I assume pausing science reduces technological progress by 30% x 50% + 50% x 20% = 25%. In other words, $g = 0.0075$, which is 25% less than $G = 0.01$.[6]

## 4.6 $n(0)$: *Initial population*

I take the initial population $n(0)$ to be 8.05bn, which is the estimate given by Our World in Data for 2023 under their medium fertility projection.

## 4.7 $s$: *Status quo healthy life year growth*

This model assumes constant population growth, at least while we remain in the current epistemic regime. Unfortunately, this is at odds with forecasts. According to Our World in Data, the UN high fertility scenario forecasts anticipate global population growth slowing from 1.1% per annum in 2023 to 0.6% by 2100, at which time global population will cross 14bn. The medium fertility scenario forecasts global population growth slowing by an even greater amount, to below 0% per annum by 2100, reaching a peak population slightly before then of 10.43bn.

In compromising between model realism and tractability, I set $s = 0.6\%$. This is significantly below the current population growth rate of around 1.1%, significantly above the long-run forecast growth rate in the UN's medium fertility scenario (which is slightly negative), and about equal to the growth rate forecast for 2100 in the UN's high fertility scenario. A growth rate of 0.6% implies the population reaches 10.9bn by 2073, or in about 50 years (about the length of a typical epistemic regime, if we set $p = 2\%$),, which is about 5% higher than the UN's medium fertility scenario. By 2100, a constant 0.6% growth rate implies the global population reaches 12.8bn, which is in between the medium and high-fertility scenarios.

While a lower value of $s$ would more closely match the UN's medium term long-run population projections, it would do this by underestimating the near-term population, which is projected to grow at faster than 0.6% per annum until around 2045. Since my model places more weight the near-term future, the cost of underestimating near-term population is higher than overestimating the long-run population.

[6] In a sense, this is closer to the value of science on the margin, rather than actual average value of science. In reality, I'm a bit of a science fundamentalist and believe that if we completely and permanently shut down science growth would slow very dramatically in the long-run. With the pause-science-for-one-year thought experiment I am implicitly assuming there are a lot of worthwhile scientific ideas that are never explored because we move forward; if we paused science, we would begin to turn to them.

I explore alternative growth assumptions in later sections of this report.

## 4.8 $\bar{s}$: Pause science healthy years growth

Recall $(1 + b_t)(1 - m_t) = s_t$, where $b_t$ is the birth rate and $m_t$ the mortality rate. For small values of $b_t$ and $m_t$ we can ignore the second-order terms $b_t m_t$ and obtain $b_t - m_t = s$. If we take the decline in the global mortality rate from 1985 to 2019 to be a good linear forecast for the future (see below), then the death rate falls about(10-7.5)/34 = 0.075 deaths per thousand per year or 0.000075 per person per year. We will assume for convenience that the birth rate will fall by the same linear amount so that we retain a constant value of *s* in the status quo.

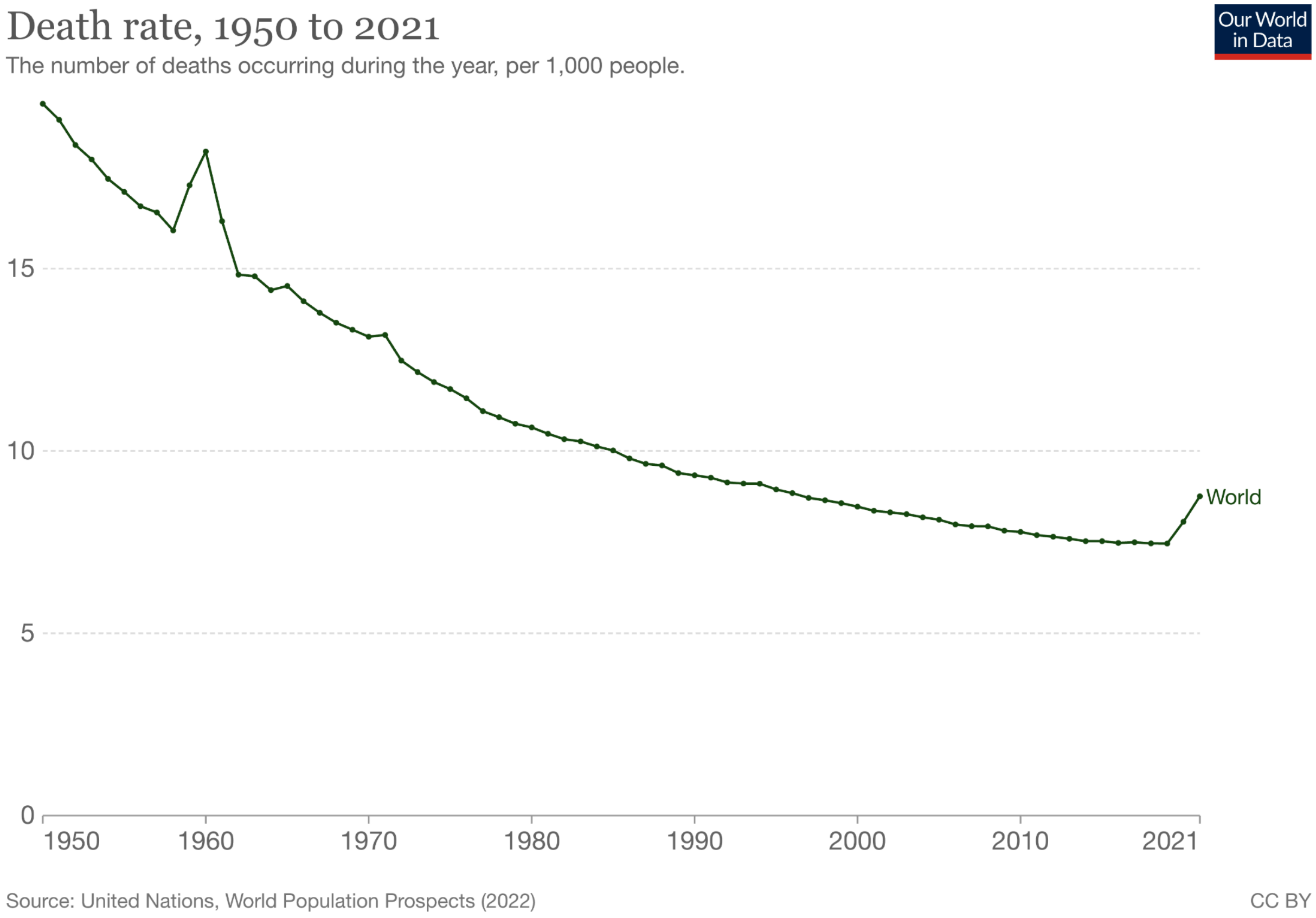

Figure 2. Annual mortality rate from Our World in Data

If we pause science for a year, I assume the impact on the growth rate of DALYs occurs through its effect on the mortality rate $m_t$. I assume $m_t$ declines as a function of scientific advance and income growth. Pausing science for a year slows the annual 0.000075 per person per year death rate.

Not all of the gains to survival are traceable back to science though. Many gains come instead from increases in income, and their impact on nutrition, access to care, and the provision of health public goods such as sanitation. Jamison et al. (2014, 2016) try to parcel out what fraction of gains to under-five mortality can be attributed to income, female education, more doctors, and a residual category that is labeled "technical progress." Between 68-80% of gains are attributed to technical progress in these reports (though this is actually just the label applied to the parts of declines in mortality that don't fall under other headings). However, even if all of these gains really do stem from technical progress, not all of that progress will be based on science (though I think quite a large share will be).

As a starting assumption, I assume 50% of the gains in life expectancy can be attributed purely to science, and the rest to economic growth (both in individual countries, and indirectly through growth's impact on global development aid). For the 50% attributable to economic growth, some portion will also ultimately be attributable to science, because I assume science causes economic growth. In high income countries, following the discussion of $G$ and $g$, I take annual frontier growth since 1965 to be around 2.0% per annum, of which 0.25% is directly attributable to science. Thus, about 12.5% of economic growth in high income countries is attributable to science, which implies 50% + 50% x 12.5% = 56.25% of life expectancy gains in high income countries can be attributed to science.

If we assume 56% of health gains are attributable to science, then if we pause science for a year, after $T$ periods we experience a decline of 0.00003342 for one year, instead of 0.000075. If we revert to the same linear decline in mortality thereafter, we are perpetually "behind" where we were in the status quo. This leads to a permanent decrease in the growth rate of DALYs $s$.

Status quo: $b_T - (m_{T-1} - 0.000075) = s$

Pause science: $b_T - (m_{T-1} - 0.000042) = \bar{s}$

Where $s = 0.006$ and $\bar{s} = 0.005967$.

# 5.0 Baseline Results: More (average) Science

By combining all of the above parameter value assumptions with the equation in Table 3 of section 3.6, I can compute changes in total utility attributable to one year of science. However, "total utility change" is not in units that are easy to interpret. To derive results that are at least somewhat interpretable, I convert my units using an approach inspired by Open Philanthropy. Full details of this conversion are in appendix A4, but in brief, my approach is:

1. Express utility in a standard unit equal to the change in utility experienced by someone earning $50,000/yr when they receive $1.
2. Divide this change in utility by global spending on science in one year.

The result is a measure of social impact that can be interpreted as how the change in utility brought about by the average dollar of spending on science compares to the change in utility brought about by simply giving that dollar to someone making $50,000/yr. For instance, a value of 100 can be interpreted as saying "if you can spend a dollar on science, or simply give it to someone making $50,000/yr, spending it on science will ultimately result in 100x as much utility." This value will be derived from a miniscule annual impact on a very large number of people spread over many decades. If we obtain positive values, that means our model is saying the benefits of science via income and health outweigh the cost of bringing forward the time of perils. If we obtain negative values, that means our model is saying the costs of bringing forward the time of perils by doing science today outweigh its benefits.

I summarize the results of this model, using the parameter choices described in section 4.0 in the table 7 below. The pure peril, pure income, pure health, and health-income interaction effect columns express the three channels through which science affects aggregate utility, expressed in terms of social impact as described above. The total impact column is the sum of these three columns and captures the overall return on R&D.

| Scenario | $d$ | Pure Peril Effect | Pure Income Effect | Pure Health Effect | Health-Income Interaction Effect | Total Impact |
|---|---|---|---|---|---|---|
| No time of perils | 0.0000% | 0 | 68 | 175 | 88 | 331 |
| | | | | | | |
| Superforecasters | 0.0021% | -2 | 68 | 173 | 88 | 326 |
| Domain Experts | 0.0385% | -33 | 65 | 135 | 73 | 239 |
| | | | | | | |
| Break-even | 0.1545% | -127 | 56 | 34 | 37 | 0 |

Table 7. Impact of Science Today (pen and paper model)

Each row corresponds to welfare effects under different values of $d$. These rows are grouped into three sets corresponding to different sets of assumptions around $d$, discussed in section 4.1. I briefly discuss these results below.

## 5.1 No Time of Perils

In the first row, we set $d$ equal to zero, to correspond to a world where there is no time of perils at all. This is mostly useful as establishing a benchmark against which to compare subsequent estimates.

In this scenario, the returns to science are very high. The pure peril rate is zero in this case, by definition. The pure income effect of science today is 68x, and the pure health effect is about 2.5 times as high, at 175x. The health-income effect lies in between these two at 88x. Collectively, science today has a 331x return.

It is worth pausing here to explain the intuition for why the pure health and health-income effects are higher than the (already reasonably high) pure income health effect. There are a few things going on here:

- First, following Open Philanthropy, I have placed a lot of weight on a healthy year of life. Today, saving a single year of life is worth increasing 200 people's income by 1% for a year. This will tend to mean the health effects of science are large, relative to income gains.
- While the long-run rate at which income grows ($G$) and mortality declines (0.000075 deaths per person per year) are unaffected by pausing science, because we assume birth rates are also declining, a pause in science leads to a temporary slowdown in mortality declines, which leads to a permanent decline in the growth rate of the population.
- The utility flow in year *t* can be thought of as $n(0)p^t(1 + s)^t(2 + ln(y_t/y_0))$. This is exponentially increasing in *s*, but only increasing in the log of income (or linearly in the growth rate).
- Even though we discount the future, the discount rate is relatively small given these parameter assumptions ($p(1 + s) = 0.986$). Meanwhile, economic growth means, all else equal, utility flows in the distant future are larger than today. A combination of low discount rates and compounding economic growth means small differences in the number of people alive in the distant future count for a lot. Pausing science has a large effect on this in the long run, via the compounding effect of growth rate $s$ or $\bar{s}$.

The importance of health in driving overall impact magnitudes is a reason I experiment with a more realistic model of the health impacts of science in section 6.0.

## 5.2 Forecasts

The next row, *d* = 0.0021%, is derived from median superforecaster estimates in the existential risk persuasion tournament, and the row after that, *d* = 0.0385%, is derived from the forecasts of domain experts in the existential risk persuasion tournament. The bottom line is this approach for estimating the dangers of biocatastrophes during the time of peril suggests the risks are too small to offset the (large) returns to science. However, in the most pessimistic forecast the returns fall by about 28% which is a substantial share of the benefits of science.

## 5.3 Breakeven Value

The final set of rows compute values of *d* that balance positive and negative effects of science. In "Break-even (welfare)" I compute the value of *d* where the returns to science are equal to zero. This occurs when *d* = 0.16%. Note, even in this setting, the income, health, and health-income effect remain positive, but they are exactly counter-balanced by a negative pure peril effect. Note *d* = 0.16% is about 77x the superforecasters estimate and about 4x the domain experts forecast.

For context, these break-even numbers are equivalent to the time of perils leading to one of the following:

- Biocatastrophes that destroy half of humanity once every 320 years.
- Bioterrorism incidents that kill 1% of the population every 6.5 years.
- Biocatastophes that kill 150% as many as the 2021 death toll of covid-19 in every year.

# 6.0 More Realistic Health

The model presented in section 3.0 results in a closed form solution, given by table 3, which allows us to (relatively) transparently understand what drives the model's results. But this model is also dissatisfying for a few reasons, which I begin to explore in the remaining sections of this report.

We may be particularly concerned about two assumptions made to keep the model tractable, that might be driving the results:

1. I assume a constant growth rate $s$, despite the fact that the population growth rate is on a long-run decline.
2. I assume pausing science leads to a small but permanent decline in the population growth rate.

These simplifying assumption, combined with a model where compounding income growth makes the welfare of future generations relatively important, means the returns to science are sensitive to what happens in relatively distant generations.

Another weakness of this baseline is related to moral philosophy. This model equates welfare with the total sum of utility flows, and is indifferent to the causes of different utility flows. In particular, it equates losses in welfare due to people prematurely dying with losses in welfare due to people never being born. Part of the large returns we are obtaining in this model are driven by the swings in long-run population. If less science leads to lower mortality, and that results in more people having children, then this model counts the utility of those newly created people the same as the utility that would be lost from premature death. It is not clear that we should equally weigh utility lost because someone is never born with utility lost because a living person dies.

This section develops a variant of the model that addresses these issues. Specifically, it employs a more realistic model of health gains from science, and adopts a population growth framework where in each year, the same number of people are born across scenarios, so that all losses in income stem from premature death. This model also implies population converges to a long-run steady state instead of undergoing constant growth.

The tradeoff is this model is too complex to be solved by hand. Instead, I write a program in python to estimate it.

## 6.1 Aggregate Utility, Population Growth, and Income

Here I briefly describe the key changes to this model. Readers who want more detail should consult technical appendix A5.

In this model, I retain all the same assumptions as in the baseline model, except for those related to health and the number of people alive.

First, rather than assuming a constant population growth *rate*, I now assume a constant number of annual *births*. This ensures that changes in science policy only impact utility via their impact on mortality, not birth rates. Notably, this assumption isn't too far off the truth for the period 1970-2080.

Second, I now assume each generation dies over time at a rate that varies with technological progress. To estimate the annual share of people, of a given age, who are alive in a particular year, I start with the US social security administration's Actuarial Study #120, by Felicitie C. Bell and Michael L. Miller. Published in 2005, this provides data on the share of US men and women born in each decadal year from 1900-2100, who are expected to be alive at every age. For example, for US men born in 1910, it has the share who were still alive at age 1, 2, 3, 4, and so on, all the way up to 120. For future birth cohorts, the social security administration forecasts mortality trends forward.

I fit a logistic regression to this data that allows me to estimate the share of people, born in any year after 1800, who live to reach any age up to 120. I fit my logistic regression so that the pace of medical progress slows over time, which matches the social security administration's actual forecasts. For example, my regression produces the following estimates of US life expectancy:

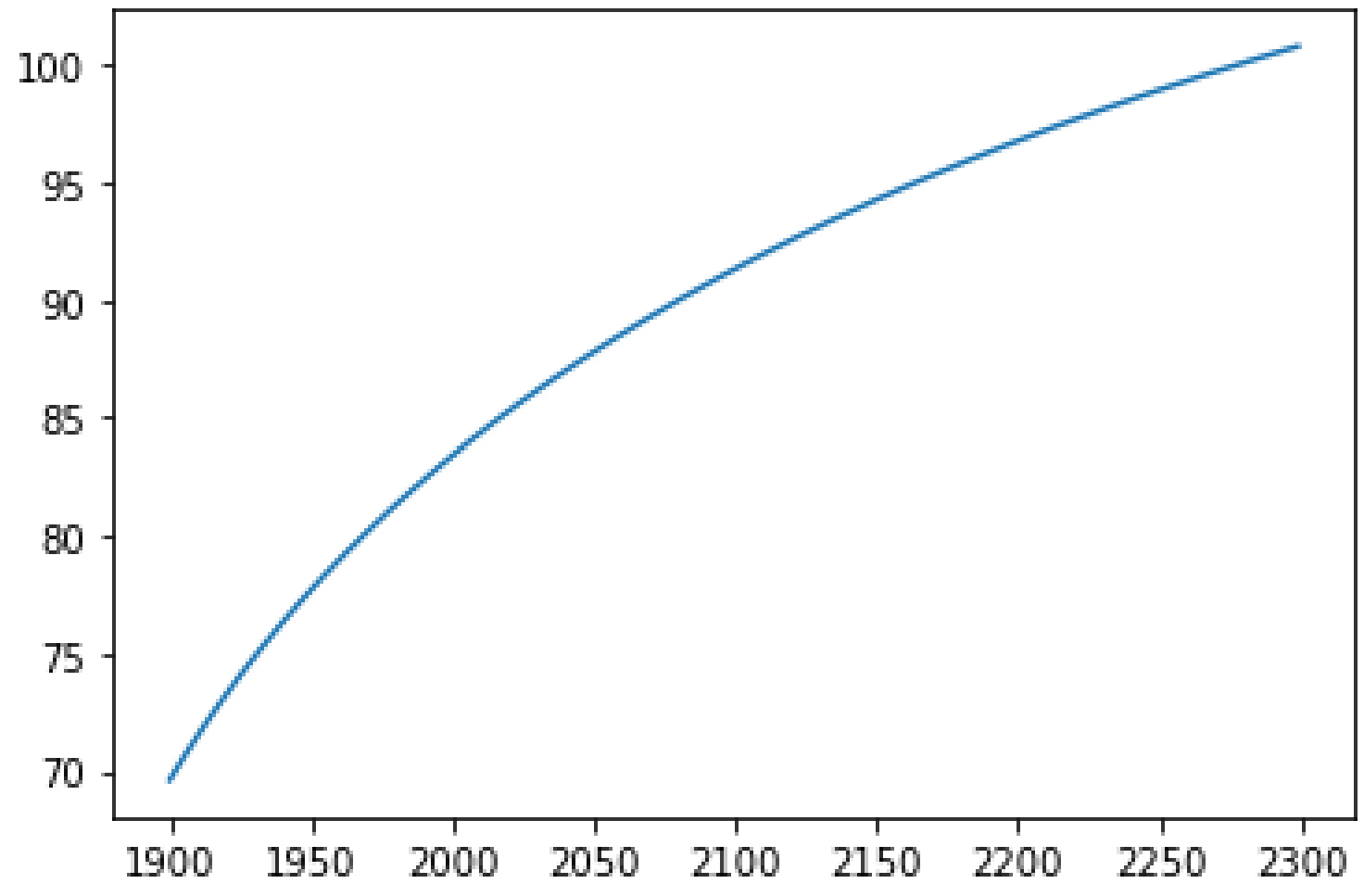

Figure 3. Forecast life expectancy under more realistic health model

Note that I forecast diminishing marginal returns to health benefits, with life expectancy gradually curving downwards.

Third, I have to modify my model a bit to account for the fact that people born in different eras will have different exposures to the time of perils. Most importantly, those born *before* the time of perils will benefit from its delay, while those born after do not.

Fourth and finally, when we pause science for a year, instead of assuming a permanent decline in the population growth rate, I now assume we fall back part of the way to the previous year's health trend. For example, in the pause science world, people in the year 2100 will have survival rates somewhere in between the survival rates of people in the status quo in 2099 and 2100. In practice, I assume pausing science for 12 months sets back health progress by about 7 months. Once again, I assume this change takes place only after $T$ years.

The net result of this is my model now envisions a world where population growth rises (due to declining mortality rates), but steadily converges to a long-run steady state around 13bn people as eventually, over several centuries, we squeeze all possible health gains out of science and everyone lives to an age of 120.

## 6.2 Results

I coded up a program in python to compute change in utility. I used the health parameters discussed in section 6.1, assuming once again that 56.25% of life expectancy gains per year in high income countries can be attributed to science. That is, I assume that if we pause science for one year, the probability of survival for some year $t$ is a weighted average of 56.25% the survival rate from the year $t$ - 1 (in the status quo where we do not pause science) and 43.75% the survival rate from the current year $t$. All other parameter choices are as in section 4.0. The following table is again expressed in multiples of OP's bar. See appendix A5.3 for definitions of the columns in this modeling framework.

| Scenario | d | Pure Peril Effect | Pure Income Effect | Pure Health Effect | Health Income Interaction Effect | Total Impact |
|---|---|---|---|---|---|---|
| No time of perils | 0.0000% | 0 | 39 | 25 | 5 | 69 |
| | | | | | | |
| Superforecasters | 0.0021% | -1 | 39 | 25 | 5 | 68 |
| Domain Experts | 0.0385% | -18 | 39 | 23 | 4 | 48 |
| | | | | | | |
| Break-even | 0.1315% | -61 | 37 | 20 | 4 | 0 |

Table 8. Impact of Science Today (python model)

This table is similar in spirit to Table 7. Once again, each row corresponds to a different value of *d*: a benchmark with no time of perils; variants based on the results of the existential risk persuasion tournament; and a break-even value. To facilitate comparison with Table 7, I numerically compute pure peril, pure income, pure health, health-income interaction, and the total impact of a scientific dollar today, again expressed in multiples of OP's preferred benchmark.

In comparing Tables 7 and 8, there are a few high-level takeaways.

First, while the direction of total impact is the same across each table, the magnitude is significantly smaller with more realistic models of health. This is mostly due to large changes in the pure-health and health-income effects. Recall, in this model of health, I assume survival curves improve with the log of years elapsed since the year 1800. This means health gains from science this year shrink over time, as compared to the baseline when they were constant. This shrinking of health gains over time means that changes in utility in the distant future, when income growth makes the flow of income much larger, no longer matter much.

This has the effect of also shrinking the returns to science in the pure-income and pure-health channels as well. The pure health effect is cut down by over 80%, reflecting the weaker long-run effects of science today on health. Meanwhile, the pure income effect is also reduced by about 40%, this time reflecting a more pessimistic outlook on the number of people alive in the future to enjoy income gains.

Notably, the break-even value using this model is more robust, dropping by just over 10%. Welfare from science is zero when the time of perils raises annual mortality risks by about 0.1315% per annum, rather than 0.1545% in the simpler model. One reason for this difference is that the more realistic version of the model involves more transitory benefits of pausing science. In the model from section 5.0, the benefits of pausing science persist in some form forever; by delaying the time of perils, in expectation, the population grows from a slightly higher level, forever. In the realistic health version of the model, the benefits of delay are transitory: those alive during the delay benefit, but after the last of their cohort dies out, future generations derive no benefit from the delay in the time of perils. Whether it begins in year $t_1$ or $t_1+1$, their whole lives have been subject to an annual mortality risk of $d$. With the benefits of pausing capped and the costs of pausing persistent, the break-even value balancing costs and benefits rises.

It is also notable that this version of the model delivers results that are reasonably consistent with other efforts to value the returns to science from health, as compared to the gains from income. For example, Nordhaus (2005) concludes that "value of improvements in life expectancy improvements is about as large as the value of all other consumption goods and services put together." I find the pure health effect is about two-thirds the pure-income effect, whereas in the baseline model from section 5.0, health benefits were 250% of income benefits.

While the results of section 6.0 do use a more "realistic" model of health, it's worth pointing out that they will *understate* the returns of science if we do in fact care about how science affects the total population size, since this model assumes no impact of science policy on the number of people born. Since the baseline model likely overstates the impact of science by assuming a constant and permanent impact on the population growth rate, the truth probably lies somewhere between these two estimates.

# 7.0 When Does the Time of Perils Begin?

Throughout this report, I have assumed the time of perils follows an analytically convenient form, discretely flipping on after $t_1$ periods. But there is a lot of uncertainty about when exactly the time of perils might commence, and about the shape of danger. Perhaps we'll see a gradual rise, rather than a discrete switch? Does this matter?

This section is meant to be a bounding exercise to demonstrate that the results are not too sensitive to these questions. Rather than considering a wide range of possibilities, here I consider what I believe to be an implausible "worst case scenario" where the time of perils will begin immediately, unless we pause science. More likely, the reality probably lies somewhere between this assumption and my baseline model. In any event, this section demonstrates that this assumption does not much matter. I also show what happens when we assume the time of perils has already begun and we cannot avert it.

## 7.1 Immediate Onset

Our first exercise is meant to provide a lower bound to the returns to science. Instead of assuming the time of perils begins in $t_1$ periods, we now assume in the status quo it begins next year. Moreover, we assume that, as unlikely as I believe this to be the case, that pausing science today can nonetheless delay the onset of the time of perils by one period. Essentially, we are now assuming a scientific discovery can be instantly translated into the capabilities to genetically engineer bioweapons by many actors. To formally implement this, we use the model from section 5.0 and set $t_1 = 1$.

In the following table, I estimate the returns to science in the scenario where the time of perils will commence immediately, unless science is paused, using the same parameter values as described in section 4.0.

| | | Impact: Time of Perils... | |
|---|---|---|---|
| Scenario | *d* | ...in 15yrs | ... Immediately |
| No time of perils | 0.00% | 331 | 331 |
| | | | |
| Superforecasters | 0.0021% | 326 | 325 |
| Domain Experts | 0.0385% | 239 | 227 |
| | | | |
| Immediate Peril Break-even | 0.132% | 43 | 0 |
| 15 yrs Peril Break-even | 0.1545% | 0 | -44 |

Table 9. The Return to Science Under Different Assumptions About Onset of Time of Peril

Each row of this table corresponds to a different peril rate. In the last two columns, I report the return on a year of science, in my preferred units (that is, relative to transferring cash to an individual earning $50,000/yr) under two assumptions. In the second-to-last column, I report the impact if we assume the time of perils begins in 15 years. In the last column, if it begins next year.

We can see first that if there is no time of perils (the first row), then (of course) there is no difference in the two estimates and investing in science has a return of 331x.

In the next two rows, we use the forecasts derived from the existential risk persuasion tournament. Using the peril rate derived from superforecaster forecasts, we see there is essentially no difference whether the time of perils begins immediately or in 15 years. It simply isn't very important either way for the results. We see a bigger effect using the peril rate implied by domain experts, but even here the effect isn't too serious. If the time of perils starts immediately, the return on science is about 5% lower than if its onset is delayed by 15 years.

We start to see bigger effects as we examine much larger peril rates. This makes sense, because the larger is the peril rate, the less likely it is that we survive to experience the gains of science and the more likely we are to lose the welfare of the first $T$ periods. In the last two lines, I compare two different break-even peril rates. In the second-to-last row, I estimate the break-even peril rate in the scenario where the time of perils commences next year. In the last row, I use the break-even peril rate when it commences in 15 years, as I assumed in section 5.3. A 15 year delay in the time of perils allows us to accept a peril rate about 17% higher than if it begins next year.

That said, my main take-away is that the main conclusions of my model are relatively robust to variation in the timing of the onset of the time of perils.

## 7.2 We're Too Late

As a second exercise, I consider the possibility that we are too late: we are living in the time of perils now, and pausing science does nothing to delay its onset. Formally, this is equivalent to assuming $d(t) = d$.

In the following table I estimate the returns to science under the assumption that we are too late to stop the time of perils. In appendix A6 I show how to modify the baseline model to perform this calculation.

| | | Impact | |
|---|---|---|---|
| Scenario | *d* | Just in time | Too late |
| No time of perils | 0.00% | 331 | 331 |
| | | | |
| Superforecasters | 0.0021% | 325 | 330 |
| Domain Experts | 0.0385% | 227 | 305 |
| | | | |
| Immediate Peril Break-even | 0.132% | 0 | 251 |
| 20 yrs Peril Break-even | 0.1545% | -44 | 239 |

Table 10. The Return on Science If We're Too Late

Here we reproduce the table from section 7.1, except the last two columns are now changed. The second-to-last column is now the same as the last column in the table of section 6.2. It corresponds to the returns to a year of science today, if the time of perils will begin now, unless we pause science. The last column now corresponds to the returns if we are already too late.

If we are too late, the returns to science are always positive: it improves health and income, and has no downsides. But as the time of perils gets more severe, the returns to science decline, basically because the value of everything declines when there is a higher probability we will die. Now there is no break-even where it is not worth doing science, except for $d = 100\%$, in which case utility falls to zero no matter what we do.

Note that while the above calculations correspond to the situation where we are already facing heightened annual technological perils, we would obtain similar results if we believed the science enabling the time of perils has already been discovered and disseminated, but not yet operationalized as a technology. In this situation, the time of perils is coming and pausing science today can no longer push back its arrival. This would describe our situation, for example, if all the knowledge necessary to genetically engineer a pandemic is already "out there" in the published academic literature, progress in AI will eventually make it feasible for bad actors to make use of that knowledge, and further progress in AI is not constrained by basic science (as arguably seems to be the case at the present moment).

# 8.0 Modeling Extinction Risks

So far, these models have implicitly adopted the view that biocatastrophes reduce utility during the years in which they occur, but do not otherwise impact humanity's long-term trajectory. Whether biocatastrophes happen or not, we have maintained the assumption that we exit the current epistemic regime with probability 1 - $p$ in every period.

This assumption is inappropriate for the worst kinds of biological peril. To take a simple, if extreme, example, if an engineered pandemic kills every human on the planet, then progress towards transformative AI will stop. This is also an outcome of pandemics that do not kill literally everyone, but kill enough to lead to the complete collapse of society (as is commonly envisioned in fiction like The Stand, The Last of Us, or The Last Man on Earth).

To explicitly account for this, let us define a new variable $d_x < d$ corresponding to the annual probability of extinction-level events that prevent us from transitioning into a new epistemic regime. This section discusses how to revise our modeling approach to account for this, and how beliefs about both the probability of civilization-killing biocatastrophes and the value of entering a new epistemic regime, affect the expected returns on science.

## 8.1 How to Model Existential Risk

In appendix A7, I show how to modify the baseline model to incorporate the idea that exiting the current epistemic regime requires not going extinct due to genetically engineered pandemics. To cut to the chase, I show that if the time of perils also includes an annual probability $d_x$ of going extinct, then the difference between the status quo and pausing science for a year can now be written as:

$$V_0 - V_0' = V_{SQ} - V_{PS} - d_x \frac{(1-p)}{1-p(1-d_x)} E[V^*] \tag{5}$$

Where $E[V^*]$ is the total value of all future utility if we exit the current epistemic regime.

The important thing about this equation is that the first part, $V_{SQ} - V_{PS}$, is the same[7] as in the baseline model or the model with more realistic health. But there is now an additional term $d_x \frac{(1-p)}{1-p(1-d_x)} E[V^*]$

[7] Technically there are small differences, but they do not substantively matter. See appendix A7.

, which captures the fact that the utility outside the current epistemic regime no longer entirely falls out of the model. In the baseline model, we assumed that whether we paused science or not, the probability we exit the current epistemic regime is unchanged. We have now abandoned that assumption. In particular, we are now taking into account the fact that doing science today brings forward the time of peril, which increases the prospects of going extinct and never realizing the utility that we would get if we left the current epistemic regime.

Intuitively, when we incorporate extinction risk, we face an additional penalty from the scientific status quo. If we die in the year of extra peril, we foreclose the chance of ever getting the value $E[V^*]$, which occurs with probability $(1 - p)$ in each period over $1/(1 - p(1 - d_x))$ expected periods.

How much this affects the returns on science now depends crucially on the value of $E[V^*]$.

## 8.2 Establishing A Utility Benchmark

By definition, it's very difficult to estimate the value of being in a new epistemic regime, since that refers to a time period during which recent historical trends are no longer applicable. But it is useful to get some kind of sense of what kinds of orders of magnitude may matter. To create a benchmark, we need some kind of unit of measure.

To my mind, the easiest unit to interpret is the total utility of everyone alive today, over the space of one year. Given roughly 8bn people and assuming utility of a healthy year of life today is worth 2 today, then the annual global utility is 16bn. Let $W$ = 16bn denote a year of total world utility today. For a bit of context, unless we assume the time of perils is quite severe, under the modeling and parameter choices in section 4.0, $V_{SQ} \approx 1.5 \times 10^{12}$. This is roughly $94W$, so that the total future utility we expect to have before we enter a new epistemic regime is equivalent to about 94 years of present global utility flows.

Let us define $E[V^*] = \lambda W$ so that the value of being in the next epistemic regime is equivalent to some multiple of years of current utility for everyone on Earth. Clearly we have a lot of guesswork going on here, but some might find these magnitudes useful.

## 8.3 Breakevens

To get a bit of intuition for the importance of extinction risk, I start by taking some of the values of $d_x$ from appendix A3.3.1, and compute break-even values of λ for which a year of science has net zero utility. For superforecasters, I estimate the implied extinction risk to be around 0.00016% per annum (1 in 625,000), once the time of perils begins. For domain experts, the figure is about 150x, at 0.02286% (1 in 4,375).

To compute λ, substitute $E[V^*] = \lambda W$ into equation (6), set this equation equal to zero, and rearrange to obtain:

$$\frac{V_{SQ}-V_{PS}}{W}\frac{1-p(1-d_x)}{d_x(1-p)} = \lambda \qquad (6)$$

Using this equation, I report values for different values of $d$ below, and different models.

| Forecast | Model | Breakeven Lambda |
| --- | --- | --- |
| Superforecaster (dx = 0.00016%) | Simplified health | 90,964 |
| | Realistic health | 18,967 |
| | | |
| Domain Expert (dx = 0.02286%) | Simplified health | 472 |
| | Realistic health | 95 |

Table 11. Break-even value of next epistemic regime, in current-population years utility

Using the Superforecaster estimates, we need the value of all future utility outside the current epistemic regime to be equivalent to tens of thousands of years at current consumption and population levels, approximately 19,000-90,000 population years. With domain experts we obtain much lower estimates. Given implied extinction risks, we would prefer to pause science if future utility is roughly equivalent to 100-500 years of current population-years utility.

## 8.4 Alternative Value Metrics

Is 500 years of current population-years utility a lot or a little? What about 20,000 or more? One approach a reader can take to this is to imagine what they think the most likely futures are for humanity. For example, the economist Tyler Cowen once stated that he thinks the human population, with something on the order of 10 billion people living on Earth, will last for centuries, but not 50,000 years. Under this view, the domain experts forecast would suggest extinction risks should be taken seriously; but would not necessarily be a primary concern under superforecaster views.

Unless he also believed we would be much richer over that period. We can also think about how this measure changes with income. Given our utility function, if Cowen believed the world would, on average, be 10x as rich in the long run as it is today, then in a new epistemic regime flow utility rises from 2 to $2 + ln(10) \approx 4.3$. This is equivalent to 2.2x increase in the flow rate of utility. So each year where 8bn humans are 10 times as rich as today is worth $2.2W$. If we thought the population would be 10x as rich, we would want to roughly halve the number of years we consider. If Cowen was confident this richer human race would last more than 10,000 years, he might think faster science is no longer worth the risk, even if the superforecasters are right.

This approach has a zero discount on the future: you simply value people's utility the same in every period. This is an attractive moral stance, but it has important ramifications. If one believes there is a one in a trillion chance that humans will eventually colonize the galaxy then the potential value of that expansion could be a billion-fold expansion of the population, lasting for billions of years. A one in a trillion chance of billions of billions of population-years is easily worth at least a million population-years in expected value terms. In that case, we would be much more sensitive to much more remote extinction risks than implied by superforecaster estimates of risk.

But another approach to valuing the future is to suppose we would continue to use a discount rate to value utility in the next epistemic regime. For example, if we thought there was a persistent, constant probability the human race of the future might go extinct, it would be appropriate to discount future utility by this probability. Alternatively, we might justify a discount rate by again allowing it to stand in for epistemic limits. We might want to believe we simply shouldn't weigh things in the very distant future as much because we simply know so little about what the moral value of the distant future will be.

What kinds of discount rates are consistent with the above views? To approximate them, let's assume income growth remains 1% per year, and that the population is constant, as happens in the long-run of our realistic health model. But now we discount the future by some $\rho$ annually, to capture the idea that we may view future utility as possibly less valuable. In other words, we are assuming the value of the future can be written as:

$$E[V^*] = \sum_{t=0}^{\infty} n(0)\rho^t(2 + tG)$$

In appendix A7.1 I show that, for this equation to equal 18,967$W$, and using $G = 0.01$, we would require $\rho = 0.9994596$. In other words, we would need to discount utility flows like our own at 0.05% per year, to value such a future at 18,967 population years. This is substantially lower than, for example, the 0.2% used to discount the long-run benefits of R&D in Davidson 2022 or the 1.1% rate used in the longest-run recommendations of the 2023 Circular A-4 guidance on how the US government should conduct cost-benefit analysis. It suggests we would need to use extremely low discount rates to prefer pausing science, if extinction imperiled continued existence like our own, at rates implied by superforecaster estimates.

However, for this equation to equal to 95$W$, we would require merely that $\rho = 0.99526$. In other words, we would need to discount utility flows like our own at 1.4% per year, to value such a future at 95 population years. This is higher than both Davidson (2022), and the lowest rate recommended in Circular A-4. It suggests reasonable valuations of the distant future would prefer pausing science, if extinction imperiled our existence at rates implied by domain expert estimates.

# 9.0 Can Better Science Reduce Risks?

Throughout the report so far, we have mostly thought of the innovation policy program as leading to *more* science, and further assumed that this science will be of average quality. But it may also be that the innovation policy program will additionally lead to *better* science. The reason this may matter is that whereas *more* science will tend to accelerate the onset of the time of peril, *better* science may reduce the risks of science. In formal terms, we may bring forward the time of perils, while also reducing the annual peril rate $d$.

I wrote a large defense of this argument, but am relegating the entire discussion to appendix A8 because (to my surprise) it turns out that it probably does not matter quantitatively. But very briefly, we might expect a philanthropic program to improve science to reduce $d$ for a few reasons:

- If more effective scientific institutions generically speed up scientific progress this will shift the ratio of defensive to offensive technological capabilities towards defense (appendix A8.1)
- If more effective scientific institutions increase scientific state capacity, this may enable us to respond more effectively to new biocatastrophe challenges. (appendix A8.2)

## 9.1 Modeling Better Science

To model the potential gains from more effective science, I modify my baseline model to allow for differences in the peril rate $d$ across the more science and less science scenarios. Whereas, in the baseline model, the more science scenario corresponded to the scientific status quo and the less science corresponded to a pause science scenario, now I think of the more science scenario as corresponding to a world where we have made science sufficiently efficient that we get the equivalent of an extra year of science over some indeterminate time period. This now has four effects:

1. We get the equivalent of an extra year of economic growth due to science (i.e., income in every period is increased by $G - g$
2. We get the equivalent of an increased in the growth rate of healthy life years from $\bar{s}$ to $s$.
3. We bring forward the time of perils by one year
4. We reduce the annual peril rate $d$ to $\bar{d}$

I furthermore assume this increase in the quality of science is temporary, so that we do not get a permanently higher economic growth rate, nor ongoing declines to mortality. I show how to modify my baseline model to perform these calculations in appendix A8.3.

To estimate the returns to more and better science, we can use the same parameter values as discussed in section 4.0, but we additionally need a value for $\bar{d}$, the reduced peril rate due to better science.

As a starting point, I assume that a years' worth of better science is as effective at reducing the annual excess mortality from unnatural risks as it is at reducing natural risks in a given year. This is arguably a lower bound, since science appears to respond disproportionately to crises (see section 10.0). On the other hand, I suspect the returns to one year of science on any particularly novel challenge may be quite limited.

To ballpark this, note that in 2019 global life expectancy was 72.8, and had been increasing by a fairly consistent 0.338 years per year for several decades. Let us continue to assume science contributes 56% of this decrease, or 0.19 years. Let us model life expectancy as emerging from a constant mortality risk $x = 1/72.8$, which is decreased to $\bar{x} = 1/(72.8 + 0.19)$ by science. We then assume:

$$\bar{d} = \frac{\bar{x}}{x}d = \frac{72.8}{72.8+0.19}d = 0.9974d \qquad (7)$$

In sections 5.0 and 6.0, we computed the return to science in terms of social impact per dollar spent on science. That's not possible in this case, because we don't have a good sense of how much it would cost to accelerate science by the equivalent of a year via better metascience (with gains presumably spread over multiple years). So we are missing the denominator that would be needed to compute an ROI of accelerating science.

Instead, I compute:

$$[Utility\ scaling] = \frac{V_{BS} - V_{LS}}{V_{SQ} - V_{PS}} \tag{8}$$

This is the ratio of the returns to a year of better science, when science also reduces *d* by 0.26%, compared to the returns to a year of science when it does not. Below, I display these results, taking the values of *d* derived in section 4.1.

| Scenario | *d* | Utility Multiple |
|---|---:|---:|
| No time of perils | 0.00% | 1.00 |
| | | |
| Superforecasters | 0.0021% | 1.00 |
| Domain Experts | 0.0385% | 1.06 |
| | | |
| Break-even | 0.1835% | 0.00 |

Table 12. Relative Utility Gains from Better Science

The final column in the above table calculates how much more utility we obtain from a year of science, if that year comes from efficiency gains that allow us to also reduce *d*. We can see, across most values under consideration, the effects are pretty small. For example, if we use the domain experts estimate, the ability to reduce *d* with more efficient science only increases utility by 6% relative to the model where we assume *d* is not affected. In general, these small effect sizes are due to the fact that *d* is already sufficiently small that reductions even to zero do not dramatically change the results.

The break-even value for overall utility increases slightly more from 0.1545% to 0.1835%. In other words, if we assume a metascience program pulls forward the time of peril but also (slightly) reduces its severity, we would take this trade so long as the peril rate was less than 0.1835%, but not if it was higher.

## 9.2 Modeling Reductions in Existential Risk

The preceding focuses only on utility effects in the current epistemic regime. But as discussed in section 8.0, we may also want to consider the effect better science has on reducing extinction risks.

Let us again suppose that a metascience program has the effect of accelerating scientific progress by the equivalent of a year while reducing the annual peril rate $d$ to $\bar{d}$. Recall in section 8.0 we introduced $d_x$, the annual risk of extinction. I'll now assume better science results in an equal proportional decrease in $d_x$ to $\bar{d}_x$, such that $d/\bar{d} = d_x/\bar{d}_x$.

As discussed in appendix A8.4, it can be shown that under these assumptions, the expected change in utility due to more and better science can be written as:

$$V_0 - V_0' = V_{SQ}(\bar{d}) - V_{PS}(d) + (1 - p)E[V^*]\left\{\frac{1}{1-p(1-\bar{d}_x)} - \frac{1+pd_x}{1-p(1-d_x)}\right\} \qquad (9)$$

Note that if the final term is positive, then the impact of metascience on extinction risk is actually positive! This occurs when the additional year spent in the time of perils is outweighed by permanently lower extinction risk. If we define the proportional decrease in existential risk as $\lambda_x = \bar{d}_x/d_x$ I show in appendix A8.4 that this condition holds if:

$$\frac{p}{1-pd_x} > \lambda_x \qquad (10)$$

Given $d_x = 0.00016\%$ for superforecasters, and $d_x = 0.00385\%$ for domain experts, this implies that a metascience program that accelerates science by the equivalent of a year also needs to reduce annual extinction risk by about 2% for it to be net positive, in terms of its impact on extinction risk. For comparison, I estimated above a year of science typically reduces conventional mortality risks by something like 0.26% per year of science.

As a final exercise, we can repeat the break-even exercise of section 8.3, but using equation (10) and again assuming $\bar{d}_x = 0.9974d_x$. The results are displayed below:

| Forecast | Model | Breakeven Lambda |
|---|---|---|
| Superforecaster (dx = 0.00016%) | Simplified health | 106,681 |
| | Realistic health | 22,246 |
| | | |
| Domain Expert (dx = 0.02286%) | Simplified health | 553 |
| | Realistic health | 111 |

Table 13. Break-even value of next epistemic regime, in current-population years utility, if we science can reduce existential risk

Compared to the values in section 8.3, assuming a better metascience program (that accelerates overall scientific progress by a year) would also reduce existential risk by about 0.26%, results in a small increase in breakeven lambda of about 17%.

# 10.0 Racing to Safety

In this section I consider more seriously an argument related to Aschenbrenner (2020), which argued that in some models of technological peril, it was desirable to accelerate economic growth to "rush past" the time of perils. Throughout this report, I have so far adopted the stance that the time of perils is only ended by a new epistemic regime, such as transformative AI. But this is not the only path out of the time of perils. It could also be that normal scientific and technological progress eventually develops workable solutions to new technological dangers.

For example, consider the following figure, from Sampat, Buterbaugh, and Perl (2013), which shows 2008 research funding at the NIH across 107 different disease categories, relative to the US mortality burden associated with these diseases.

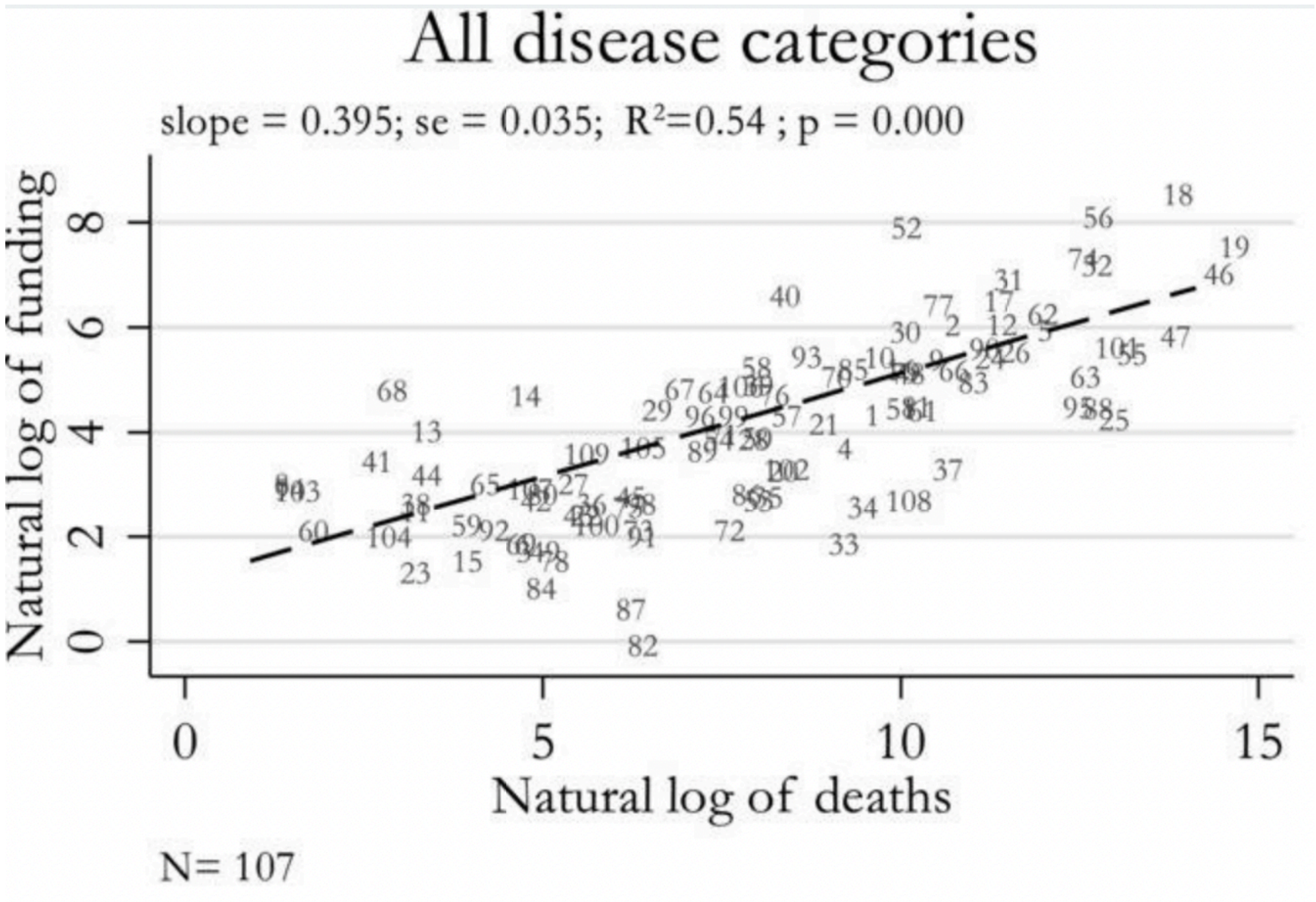

Figure 4. NIH funding by disease, relative to US mortality burden, from Sampat, Buterbaugh, and Perl (2013)

There is no official policy to match funding levels to the mortality burden of deaths, but this outcome nonetheless roughly attains as an outcome of the decentralized process of legislators responding to lobbyists for different diseases setting different funding levels for NIH institutes, and grant applications and grant managers seeking to identify the most important grants. In other words, science responds reasonably well to health risks. If genetically engineered pandemics become a new health risk, we can anticipate the scientific ecosystem to eventually structure itself around trying to understand and mitigate them.

In fact, if there is a "warning shot" where there is a genetically engineered pandemic that we survive, this process is likely to be quite rapid. Consider the following figure from [Sampat (2012)](), which plots funding levels at different NIH centers in 1980 vs 2010.

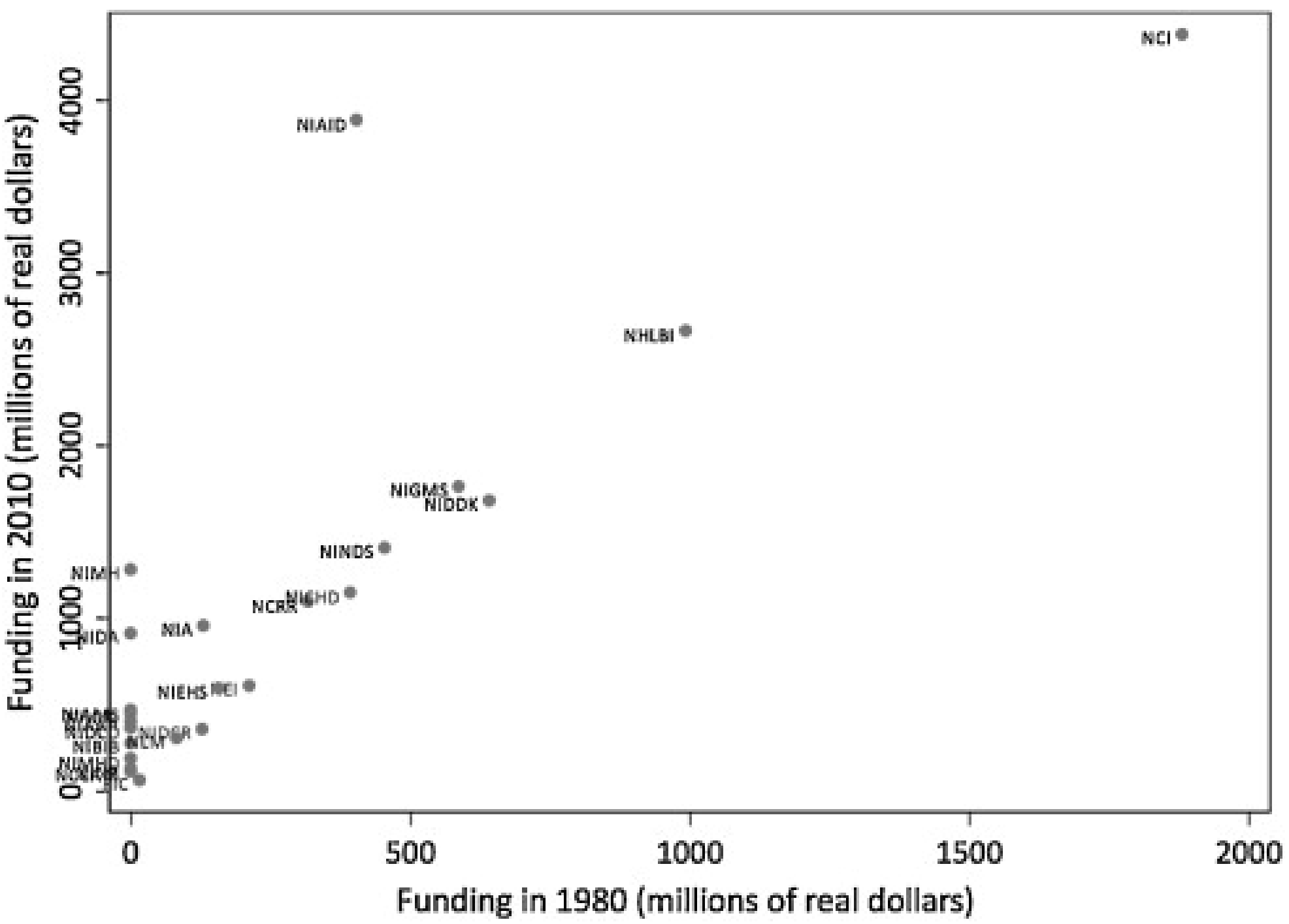

Figure 5. NIH Funding by Center across Time

In general, funding has a lot of inertia, but we do see one major outlier on this chart: this is the NIAID, which saw large funding increases in the 1980s in response to the AIDS epidemic, and again in the 2000s, in anticipation of bioterrorism dangers. These represent two of the three most salient biological crises of the last half-century (the other is Covid-19 - more on that shortly). My read of these two figures is that science will gravitate towards health issues that are important, and that the movement towards these issues can be very fast in crisis situations.

The previous paragraph alluded to two out of three of the most salient biological crises of the last century. The third very salient biological crisis of the last half-century is probably the covid-19 pandemic. This again illustrates a *more than proportionate* increase in scientific attention. Agarwal and Gaule (2022) examines how the global R&D system responded to this threat. They find covid-19 induced a more-than-proportional response: as large as the potential mortality for covid-19 was (they estimate 70% of the global population gets covid-19, and it has an infection mortality rate of 0.5%), R&D effort, as measured by new clinical trials, was even larger by historical standards.

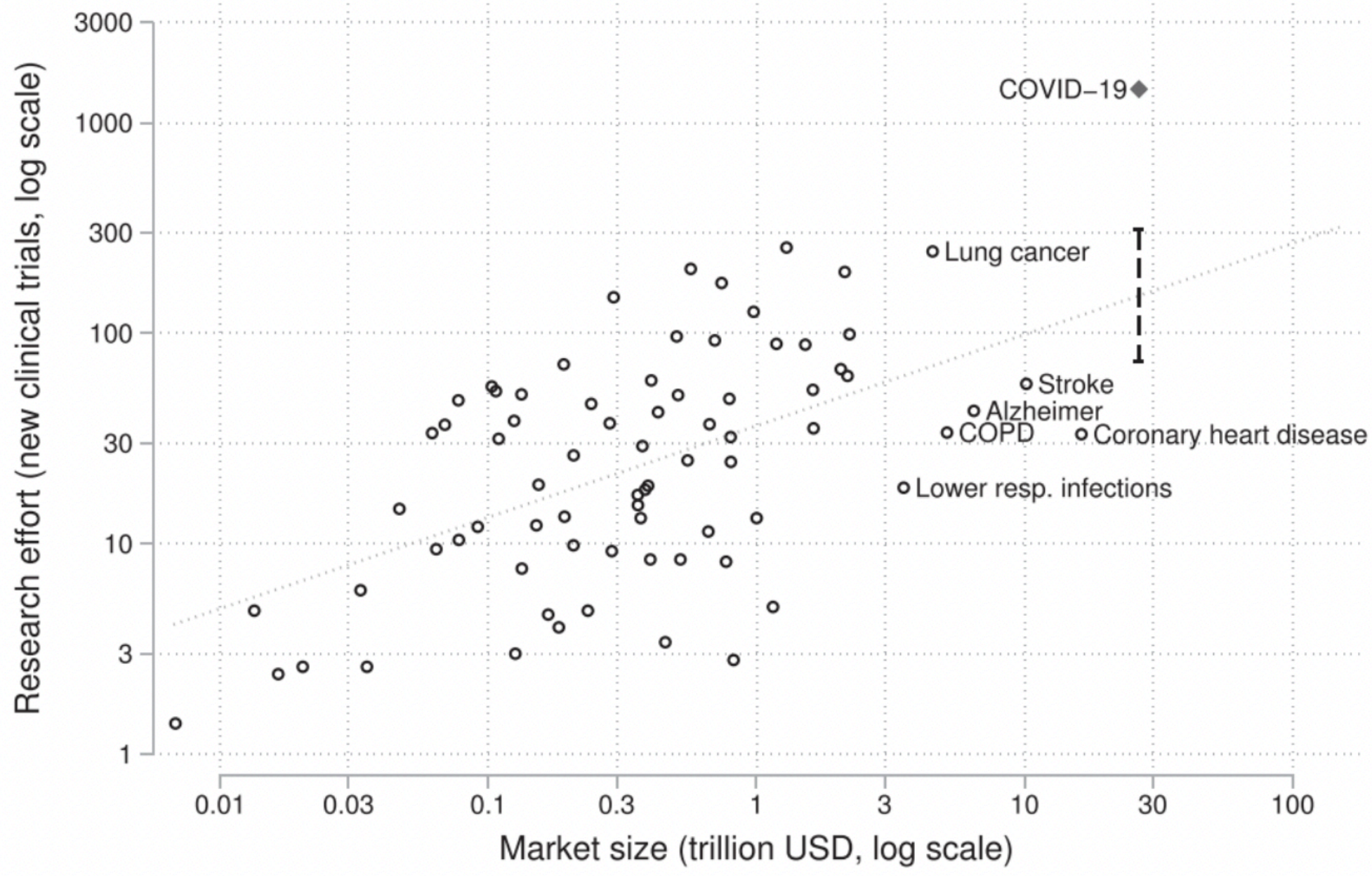

**Fig. 5.** COVID-19 R&D Effort Compared with Other Diseases.
*Notes*: The figure displays the relationship between potential market size and research effort in the cross-section of 75 diseases in our sample, plus COVID-19. Research effort is defined as the average yearly number of new trials per disease from 2015 to 2019; for COVID-19 we use the actual number of 2020 COVID-19 trials instead. Potential market size is the 2017 disease-level mortality at the national level weighted by national GDP per capita and a value of statistical life of USD 1 million for the mean global citizen, expressed in logs. For COVID-19, our estimate of potential market size is based on the hypothetical case in which in the absence of non-pharmaceutical control measures 70% of the world population eventually would get infected, and COVID-19 has an infection fatality rate of 0.5%. The dashed vertical line represents a 95% prediction interval for the number of COVID-19 trials. See main text for a detailed discussion of the COVID-19 potential market size and Table A.3 for sensitivity to different elasticity estimates and COVID-19 market size assumptions.

Figure 6. R&D response by disease, including covid-19, from Agarwal and Gaule (2023)

Elsewhere, Hill et al. (2021) document a similarly large pivot among basic science, with roughly 5% of new academic papers published after May 2020 related to covid-19. I suspect this larger than proportional response in response to crises stems from dynamics related to the critical mass necessary to launch new research fields: highly salient events can create public knowledge and consensus about which areas will be promising future research topics (see Building a New Research Field for more discussion).

All this suggests that if genetically engineered pandemics become a highly salient risk to human health (as would especially be the case if it had the potential to lead to extinction), then the global R&D system would respond rapidly. Note table A3.8 in appendix A3.4 does indicate participants in the existential risk persuasion tournament believe it is much more likely we encounter a genetically engineered pandemic that we survive to one that we do not. This would have the potential to reduce the long-run risk from these hazards.

Indeed, more generally, forecasts in the existential risk persuasion tournament indicate some kind of long-run risk reduction is the expected outcome. As discussed in detail in section 4.1 and appendix A3, it is possible to use a variety of questions in the existential risk persuasion tournament to derive forecasts about the risks of a genetically engineered pandemic, conditional on the human race not going extinct via other causal pathways or experiencing transformative AI. Below, I pull out my estimates of forecasts that there will be a genetically engineered pandemic, conditional on us remaining in the current epistemic regime, from appendix A3.2.

| Conditional Forecast | Group | Year | | |
|---|---|---|---|---|
| | | 2030 | 2050 | 2100 |
| Genetically-engineered pathogen killing >1% population | Superforecasters | 0.25% | 1.52% | 4.16% |
| | Domain experts | 1.26% | 9.06% | 14.62% |

Table 14. From Appendix A3.2

Using the method discussed in section 2.3 and appendix A1, we can infer the implied annual probabilities of genetically engineered pandemics using these conditional probabilities. Doing so yields the following:

| Forecast | Group | Year | | |
|---|---|---|---|---|
| | | 2023-2030 | 2030-2050 | 2100 |
| Genetically-engineered pathogen killing >1% population | Superforecasters | 0.04% | 0.06% | 0.05% |
| | Domain experts | 0.18% | 0.41% | 0.13% |

Table 15. Annual Conditional Probabilities of Genetically Engineered Pandemic

Both superforecasters and domain experts anticipate that the annual rate of genetically engineered pandemics will decline in the long-run, after rising in the medium term. For superforecasters, the decline is pretty small, from 0.06% to 0.05% (a roughly 17% decline). But for domain experts, this decline is quite large, from 0.41% to 0.13% (a nearly 70% decline). And note, these are the declines forecast after I have attempted to strip out probabilities associated with transformative AI occuring (which I define as either AI-led extinction or global growth above 15% per annum), or other non-biological extinction risks.

To sum up, historical evidence suggests that global science is likely to respond rapidly to novel threats. Historically, science is also pretty good at mitigating health issues, on average. Moreover, forecasters also anticipate a reduction in the risks of genetically engineered pandemics in the long run. We don't know how much of the decline forecasters attribute to scientific and technological solutions to pandemics, but this is directionally consistent with the story that science may eventually reduce risks.

If this model is true, then a philanthropically supported program that accelerates scientific progress is net positive for much higher values of $d$, given a sufficiently low discount rate. While it brings forward the time of perils, it also brings forward their diminution. Assuming both advance at roughly similar rates, the total time spent in the time of perils is compressed.

# 11.0 Discussion

To begin, I will briefly summarize this report's conclusions.

- As scientific understanding and technological capabilities advance, it is sensible for science policy to pay increasing attention to issues related to downside risks of technology. In this report, I focus on evaluating the risk that accelerating scientific progress could be net negative via its impact on bioterrorism and biowarfare.
- Biocatastrophes enabled by advanced technology could reduce social welfare both through their impact on mortality and on the continuation of advanced civilization.
- If we focus only on mortality effects and ignore the potential for biocatastophes to upend civilization, then:
    - In this report's model the social impact of science ranges from +48-331x the social impact of cash transfers to people earning $50,000/yr.
    - Most of the variation in this range stems from different assumptions about the long-run impacts of science on health and population.
    - The benefits of science outweigh the costs if we expect the annual increase in mortality due to new bioweapons to be below 0.13-15%.
    - Since the forecasts of both superforecasters and biosecurity domain experts imply excess mortality well below these break-even rates, under these assumptions, the expected value of science is strongly positive.
- If we take into account forecasts of human extinction caused by biocatastrophes, then science is welfare *reducing* if:
    - The annual probability of extinction during the time of perils is 0.02286% (a number I estimate is implied by domain expert forecasts) and the total expected value of future

utility outside our current epistemic regime is more than roughly 100-500 years of today's global utility (depending on how health gains from science are modeled)
    - Or if the annual probability of extinction during the time of perils is 0.00016% (a number I estimate is implied by superforecasters) and the total expected value of future utility outside our current epistemic regime is more than roughly 20,000-100,000 years of today's global utility (again, depending on how health gains from science are modeled).
- The report explores some sets of alternative assumptions. Some of these do not much matter for the main results about the returns to science. In particular, results are not much changed if we assume:
    - The time of perils begins sooner than assumed in the baseline models.
    - A philanthropic program that accelerates science would reduce the annual mortality risk during the time of perils by an amount comparable to the annual reduction in "normal" health risks, even as it still accelerates the arrival of the time of perils.
- Other alternative assumptions do matter. In particular, the returns to science remain positive over a significantly wider range of possible risks during the time of perils if:
    - The arrival of the time of perils does not depend on making new scientific discoveries (i.e., we are already "too late.")
    - Faster science accelerates *both* the onset and cessation of the time of perils, resulting in a net contraction in the time spent in the time of perils.

In the rest of this discussion, I give my personal views on:
1. Why I prefer the superforecaster forecasts to the domain experts.
2. How metascience policy should think about extinction risk.
3. My preferred value for the social impact of science.

## 11.1 Superforecasters or Domain Experts?

Superforecasters and domain experts differ pretty dramatically in their forecasts. My estimate for the annual mortality risk during the time of perils, implied by the domain expert forecasts, is 18x the corresponding estimate from superforecasters. My estimate for annual extinction risks during the time of perils, implied by the domain expert forecasts, is 143x the corresponding estimate from superforecasters.

While the variation in annual mortality risks does not actually matter much (both see it as too low to substantially affect the return on science), the variation in extinction risks plausibly matters for evaluating the desirability of faster science. In particular, many readers would probably agree that human civilization quite likely has more than 100 years of prosperity ahead of it, if we don't go extinct in the near term. Setting aside some complications I'll discuss in the next section, a reader who has this view would probably not agree to a trade to get more income and health benefits via more effective science, if that meant imperiling this future. So it does matter who is right on this question.

While I do not have domain expertise to independently evaluate the arguments advanced by participants in the XPT, I prefer superforecaster estimates to the domain experts for three reasons.

### 11.1.1. Intersubjective accuracy

Because long-run forecasts cannot be scored for a long time, as a surrogate endpoint, the XPT asked forecasters to also predict each other's forecasts. For example, a forecaster would be asked to give their own forecast about some event, but also what they think other superforecasters will say, and what they think domain experts will say. The hope was that this would require an understanding of the position held by others. On average, forecasters with the most accurate beliefs about others had lower forecasts of existential risk. And these effects were not small, as can be seen in the figure below. For example, among superforecasters in the top quintile for "intersubjective accuracy" estimated all-cause extinction risk by 2100 at 0.4%, compared to 7% for those in the bottom quintile. Among experts, the ranges were even wider, from a low of 1.1% in the top quintile to 32% in the bottom.

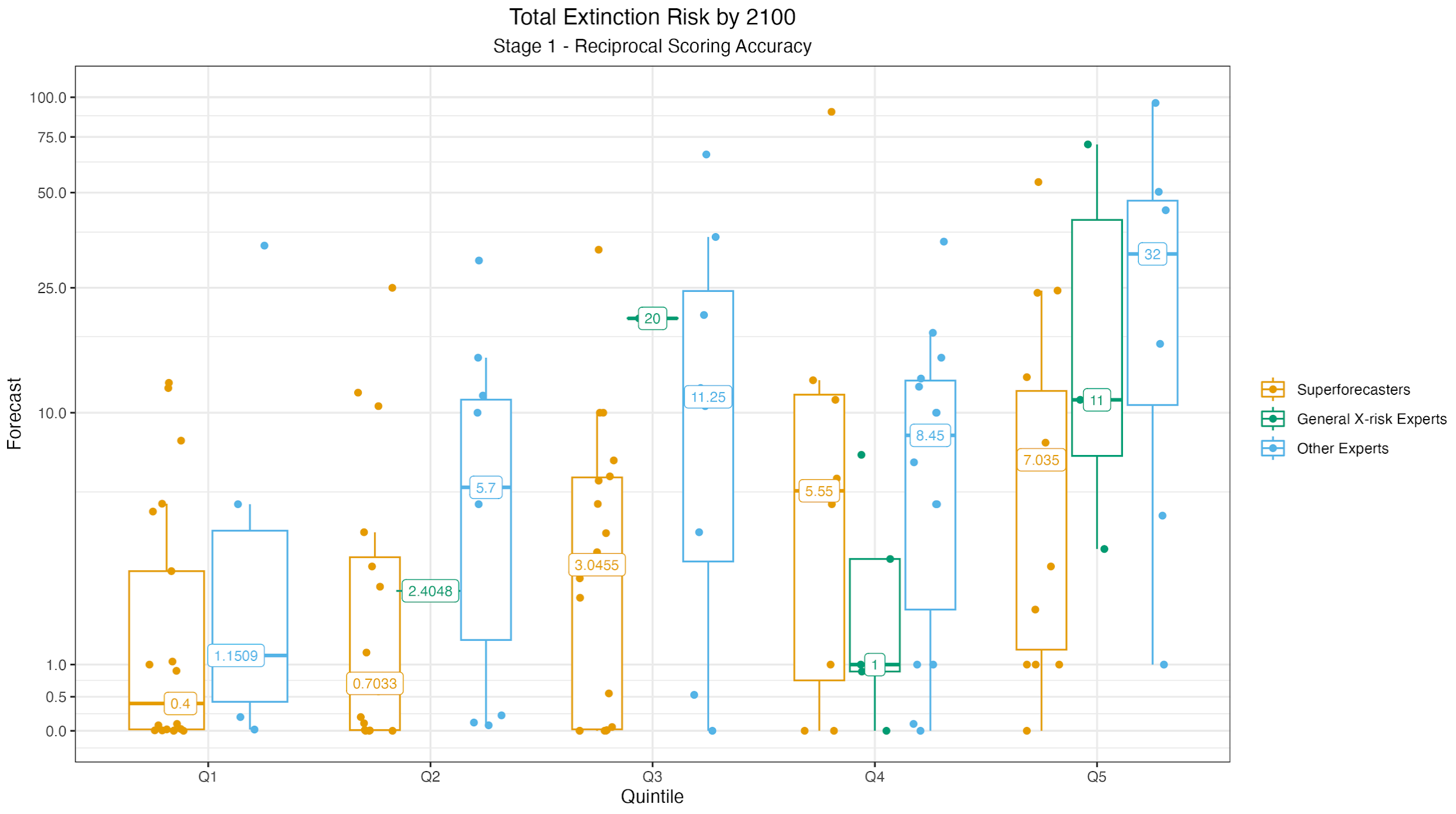

Figure 7. Intersubjective accuracy and total extinction risk forecasts, from XPT report

In short, the groups that had the best handle on what others would say about existential risks, and who therefore plausibly best understand their opponents' arguments, opt for lower existential risks.

### 11.1.2. Correlated pessimism

A second reason I prefer the superforecaster estimates is that there is extensive cross-domain correlation related to general pessimism and optimism among groups in ways that imply biases. Those who foresee a higher risk of catastrophe due to biocatastrophes, are also more likely to foresee higher risks of catastrophe from artificial intelligence or nuclear weapons. This is not that surprising, because as noted in section 2.2 of this report, we have reason to believe the advance of AI will probably be correlated with dangers from bioweapons.

But we also see a high degree of correlation in beliefs about catastrophic risk even for categories that seem likely to be uncorrelated. For example, consider how respondents answered this question:

*What is the probability that non-anthropogenic causes (e.g., asteroid or comet impacts, solar flares, a supervolcanic eruption, or a stellar explosion) will be the cause of death, within a 5-year period, for more than 10% of humans alive at the beginning of that period...*
*...by the end of 2030?*
*...by the end of 2050?*
*...by the end of 2100?*

This doesn't seem likely to be a question that is implicitly about AI risk perceptions. Yet the third of respondents most concerned about AI risk (which is correlated with a higher probability of genetically engineered pandemic) foresaw a 0.14% chance of such a catastrophe by 2100. The third of respondent least concerned foresaw a 0.01% chance, more than an order of magnitude less.

This implies either that domain experts are systematically too pessimistic, or that superforecasters are systematically too optimistic. Of these two possibilities, I think it is more likely domain experts are too pessimistic, rather than that superforecasters are too optimistic. I find it plausible that an individual who is "too" pessimistic about catastrophic risks in general would be disproportionately likely to seek to learn more about catastrophic risk, and hence become a domain expert. But I cannot see any reason why individuals who are "too" optimistic about catastrophic risks should tend to enroll in and succeed in forecasting tournaments. Similarly, superforecasters who are overly optimistic in general (beyond catastrophic risks) will receive corrective feedback on this trait when their overly optimistic forecasts go astray. Domain experts do not necessarily receive corrective feedback about their pessimism, in the course of their work.

It is plausible that domain experts may have an initial advantage, by being aware of ways that catastrophic risks can manifest that are unknown to superforecasters. But the forecasts we use are after superforecasters and domain experts have exchanged views and arguments.

### 11.1.3. Domain Expert Track Records

We don't have much evidence, at present, about the forecasting acumen of generalists (which broadly applies to superforecasters), relative to specialists (which broadly applies to domain experts), in this kind of long-run forecasting context. But the scant evidence that does exist is consistent with preferring the estimates of superforecasters to domain experts.

Rhys (2023) conducted an incentivized forecasting survey where he asked participants to forecast the impact of various economic development interventions after 2-9 years. In his study, there was no statistically significant difference in the accuracy of single-point best guesses of the impact of these interventions between academics who studied these issues and superforecaster generalists. But superforecasters outperformed academics in estimating *ranges* into which the impacts were likely to fall.

Forecasting the impacts of development interventions several years out is still a far cry from forecasting biocatastrophes over the next century. Perhaps more applicable is a [2014 report] by Gary Ackerman, which describes a research project (chapter 6) to get domain experts (futurists and biosecurity experts) to forecast the probability of biological weapons attacks by various groups over 2013-2022. One simple way to check their accuracy is to add up all their attack probabilities for different groups and compare to the actual number of attacks that happened over this time period, which was [apparently] 1.

Adding up the lower bound estimates, we would expect to see 1.23 attacks, which is pretty close to what was in fact observed (see calculations on the Section 11 tab of this [spreadsheet], pulled from 6.9a on page 205-206 of Ackerman's [report]). Adding up the midpoint probabilities estimates, we would expect to see 2.75 incidents over 2013-2022. And adding up the higher range, we would anticipate seeing 5.4 attacks. This isn't a lot to go on, but it's a closer fit to the lower-bound estimates. And as noted in section 4.1, the median superforecaster estimate is roughly comparable to the 15-20th percentile estimates of the domain experts.

For these three reasons - greater intersubjective accuracy among those with lower forecasts of catastrophic risk, indications of a pessimism bias among those with high probabilities of catastrophic risk, and some scant evidence on the track record of experts versus superforecasters - I think the superforecaster estimates are more likely to be true than the domain experts.

## 11.2 How much attention should we pay to extinction risk from science?

This report argues that the question of extinction (or some other kind of collapse of advanced civilization) is the main crux for determining whether it is desirable to accelerate scientific progress in the future. So how should we think about extinction risk?

I have three primary reasons why I do not view extinction risk via faster science as the primary axis on which we should think about metascience.

1. As noted in section 11.1, I think the superforecaster estimates are more likely to be correct. If this is so, the risks of extinction are so remote that I do not think they should drive our decision-making.
2. In section 7.2, I discuss the possibility that the scientific discoveries that would enable new biocatastrophes have already been made and merely await their activation from generic technological advance (most likely via the continued advance of AI). If so, the returns to science are positive for all levels of risk, because it is "too late" for science to make a difference on questions of extinction risk but science can still help in more traditional ways.
3. In section 10.0, I provide evidence that if technological peril becomes salient, then it is likely the scientific ecosystem will leap to developing solutions to it. In such a world, the ability to efficiently use our scientific resources to maximize the rate of discovery would speed us to the end of the time of perils. In such a world, so long as the probability of exiting the current epistemic regime is not too high, then faster science *reduces* the probability of extinction.

My personal view is that these three issues are largely separate from each other. I would estimate something like a 3 in 4 chance that the superforecasters are correct, rather than the domain experts, especially about extinction risks. If the domain experts are correct about the scale of danger of the time of perils, I further think there is probably a 50:50 chance that further scientific discoveries in fundamental biology are not necessary to realize the time of perils. Put together, I think we are in a world where extinction risks will be high and where the pace of science can affect this risk with about 13% probability. But on top of that I think there is probably a 50:50 chance that faster science can reduce existential risk. All told, that gives us about a 6% chance[8] that we live in a world where faster science is not worth the extinction risks (recognizing that this level of precision is a bit of a fiction).

In my view, that also implies it's very likely (more than 90% probable) that we live in a world where faster science is good for us. Of course, that doesn't settle the question, because if the losses in that 6% of worlds where science is bad are sufficiently large relative to the gains in the 94% of worlds where science is good, then it could dominate the decision-making. In principle, one could use the above probabilities, in conjunction with some views about the expected value of not going extinct, to derive an all-things-considered view on the value of science.

I am hesitant to do that for a few reasons though. To begin, I have a general aversion to model-based extrapolation a long way away from historical experience. These aren't models of physical systems for which we have very strong evidence of their predictive power. They are more like thought experiments, to help us think through whether our quantitative judgments are internally consistent with each other. My preferred interpretation of this report's quantitative work is that historical trends around the benefits of science very likely overpower sensible forecasts of the risk. At the same time, this report highlights a non-negligible probability that this story is wrong and that faster science could actually be bad for us, if new technologies are so dangerous they can lead to extinction with sufficiently high probability.

Given the conclusions that (1) extinction risk could be important and (2) the evidence is weak that faster science leads to meaningfully more extinction risk, I think we should view accelerating scientific progress and reducing new technological risks as separate and independent objectives. While there will be exceptions to this rule, in general we should not view one objective as coming at the expense of another. Very probably faster science makes the world a better place, though there is some probability it does not.

In this setting, I think the best policy outlook is to treat extinction risk seriously, but separately. When evaluating a potential intervention that can accelerate science, in most cases the impact of faster science on catastrophic biological risk can be set aside.[9] But at the same time, potential interventions that can reduce risks are also valuable and should be implemented (since I think these risks are non-negligible and very costly if we're wrong). The optimal metascience strategy will pursue both ends in parallel (though I think with more weight put on the goal of accelerating science), but without much attention to their interaction.

---

[8] 25% probability that the domain experts are right x 50% chance that it's not too late for science to affect the onset of the time of perils x 50% chance that science cannot accelerate us to safety = 6.25%

[9] Not always; sometimes you'll be dealing with a scientific area where the risks are unusually salient, such as gain-of-function research. But I do not think this is the norm.

There is some precedent for this kind of position. This is essentially how Jones (2016), discussed briefly in section 2.1, frames the problem of danger from science. In his model, faster economic growth is not dangerous per se, but as a society grows richer it will rationally reallocate more of its R&D budget towards mitigating hazards rather than accelerating growth. This has the effect of slowing growth, but only because some R&D effort that could have been allocated to faster growth instead goes to mitigating risk. In his model, we would take faster growth if we had the resources to invest in it! Similarly, I am suggesting the optimal metascience policy should allocate a portion of its effort towards reducing risks from new technologies. That will mean slower scientific progress than if it could allocate 100% of its efforts to faster science, but only because our efforts are finite and tradeoffs exist.

Another precedent comes from climate change, another context where there is a potential tradeoff between the damages of technology and faster economic growth. A mainstream policy stance in the climate change debate argues we should pursue *both* faster economic growth *and* a pivot towards renewable energy, using different instruments in parallel. For example, one might argue that we should lower taxes to boost growth and simultaneously subsidize investments in renewable energy to combat climate change. Another policy stance in this debate, associated with the degrowth movement, rejects the separability of these issues and argues slowing growth is necessary to combat climate change. In this report, I recommend a metascience policy closer in spirit to the mainstream policy stance rather than the degrowth one.

Lastly, as a general economic principle, it is better to address separate market failures with separate instruments, rather than to use one instrument to try and attain multiple (possibly conflicting) goals. Of course, this depends on there actually *being* instruments that can address each policy separately. While this has not been a focus of this report (except a brief note in section 2.1 that Von Neumann was skeptical such a process was feasible in the context of picking technological sectors to focus on), I do believe there are a variety of options for tackling each problem. More research on these policy options would be a valuable contribution though.

## 11.3 The Social Impact of Science

Finally, if we are to approach accelerating science and reducing risk as two separable goals, we can then ask: what's the value of science? Sections 5.0 and 6.0 are most directly relevant to this question, as they estimate the social impact of science when we set aside extinction risks. In these sections, whether we use domain expert or superforecaster implied values of $d$, or if we ignore $d$ all together, we obtain values on the order of 300x in section 5.0 and 60x in section 6.0. These differ by pretty significant amounts. Which is it?

I prefer a value of around 60x for three main reasons.

As discussed in section 6.2, the key reason these models differ is because they have different implications for the long-run impact of science on the size of the healthy population. The forecasts implied by the model in section 5.0 are analytically convenient, but do not well match our best guess of where global population is going and that matters a lot for the final value of science. In line with actual global population forecasts, the model of section 6.0 predicts a global population whose growth slows and eventually stops.

Another virtue of the section 6.0 model is that it is less sensitive to impacts in the very long-run. This model tries to account for our declining confidence about the very long run with the epistemic discount rate $p$, an annual probability that the world of tomorrow will be different enough from the world of today that we can no longer predict policy consequences. But one simplifying assumption is that I assume $p$ is constant in every year. This is, itself, something I do not have confidence about! A model where the effects of science policy wither away, as eking out further health gains becomes harder and harder, is a model where effects in the far distant future decay more rapidly than in the alternative. This better matches my own confidence about predicting the distant future.

Finally, whereas both models obtain broadly similar results for the income benefits of science, as discussed in section 6.2, the model of section 6.0 more closely matches other academic attempts to measure the value of health, as compared to income. On a subjective level, it also feels roughly right to see health and income benefits of science as being comparable, and not substantially weighted towards health (though neither do I think that position is unreasonable).

This is not to say the baseline model is not useful in other ways. The section 5.0 model is more useful as an analytical tool to see how parameter choices influence the value of science. But when it comes to estimating the value of science, rather than understanding the forces that shape it, I prefer the model from section 6.0, aptly named the "more realistic health" model.

To sum up. In this report, I argue that spending a dollar on science increases the wellbeing of society by roughly 60x as much as simply giving a dollar to someone making $50,000/yr. That effect mostly

comes from the fact that science leads to a better understanding of the world, and that understanding in turn leads to the development of new technologies that can improve the lives of billions of people who live on the planet now, and in the decades to come. A bit more than half the value comes from technologies that make these people just a tiny bit richer, in every year, for a long time to come. The other part comes from technologies that make people just a tiny bit healthier, again in every year, for a long time to come. Aggregating up tiny improvements over billions of people and many decades, the dollar spent on science eventually tends to do far more good than it would do for one materially comfortable individual.

That said, spending a dollar on science isn't all good. Sometimes we discover things that bad actors might be able to use to create new and powerful technologies, most likely biological, that can unleash mass suffering either intentionally or by accident. In some versions of the model, these costs are so large that they erase almost a third of the value of science! Even in this pessimistic scenario, these aren't large enough to offset the benefits of science, but they are large enough to take seriously and try to reduce.

# Technical Appendices

## A1. Annual Pandemic Risks Implied by Forecasters

As discussed in section 2.3, the existential risk persuasion tournament asked two questions directly relevant for assessing the plausibility of a time of perils framework.

*What is the probability that a genetically-engineered pathogen will be the cause of death, within a 5-year period, for more than 1% of humans alive at the beginning of the period...*
*... by the end of 2030?*
*... by the end of 2050?*
*... by the end of 2100?*

And:

*What is the probability that a non-genetically-engineered pathogen will be the cause of death, within a 5-year period, for more than 1% of humans alive at the beginning of the period...*
*... by the end of 2030?*
*... by the end of 2050?*
*... by the end of 2100?*

In the following table, I present the median answers submitted in the final round of the tournament, i.e., after people have had an opportunity to discuss and debate.

| Forecast | Group | Year | | |
|---|---|---|---|---|
| | | 2030 | 2050 | 2100 |
| Genetically-engineered pathogen killing >1% population | Superforecasters | 0.25% | 1.5% | 4% |
| | Domain experts | 1.22% | 8% | 10.25% |
| Non-genetically-engineered pathogen killing >1% population | Superforecasters | 0.5% | 1.69% | 3.62% |
| | Domain experts | 1% | 5% | 8.14% |

Table A1. Forecasts of pandemics. Excerpted from Table 23 in the XPT Report

To annualize the forecasts, let $X \in (t_1, t_2)$ denote a pandemic event occurring in between years $t_1$ and $t_2$, and $X \notin (t_1, t_2)$ denote a pandemic event *not* occurring in the same time interval. Using this notation write:

$$Pr(X \in ('23, '30)) = 1 - Pr(X \notin ('23, '30))$$

Where '23 denotes 2023. Even though the XPT took place in 2022, it concluded in November of that year, so that it is closer to the forecasting submission date to assume forecasts begin in 2023 than 2022. If we assume the probability of a genetically engineered pandemic is constant in every year between 2023 and 2030, we can write:

$$Pr(X \notin ('23, '30)) = Pr(X \notin (t, t + 1) | t \in ('23, '30))^{2030-2023}$$

That is, the probability no pandemic occurs is simply the probability no pandemic occurred in each year, over seven years. Noting that $Pr(X \notin (t, t + 1)) = 1 - Pr(X \in (t, t + 1))$, we obtain a formula for computing the implied annual probability of a pandemic over the period 2023-2030.

$$Pr(X \in (t, t + 1) | t \in ('23, '30)) = 1 - (1 - Pr(X \in ('23, '30)))^{1/(2030-2023)}$$

For later years, we need to adjust for the fact that the annual rate of pandemics might change after 2030 (indeed, we are interested in precisely this case).

Note by the same argument, we can also write the annual probability of an genetically engineered pandemic in 2030-2050 as:

$$Pr(X \in (t, t + 1) | t \in ('30, '50)) = 1 - (1 - Pr(X \in ('30, '50)))^{1/(2050-2030)}$$

Of course, we do not directly observe forecasters beliefs about $Pr(X \in ('30, '50))$, but it is implied by the following:

$$Pr(X \notin ('23, '50)) = Pr(X \notin ('23, '30)) \times Pr(X \notin ('30, '50)).$$

That is, the probability of no pandemic occurring between 2023 and 2050 is the joint probability of no pandemic occurring between 2023 and 2030, and then between 2030 and 2050. Noting $Pr(X \notin ('30, '50)) = 1 - Pr(X \in ('30, '50))$, we can write:

$$Pr(X \in (t, t+1) | t \in ('30, '50)) = 1 - \left(\frac{1 - Pr(X \in ('23, '50))}{1 - Pr(X \in ('23, '30))}\right)^{1/(2050-2030)}$$

And by the same argument, we can also write:

$$Pr(X \in (t, t+1) | t \in ('50, 2100)) = 1 - \left(\frac{1 - Pr(X \in ('23, 2100))}{1 - Pr(X \in ('23, '50))}\right)^{1/(2100-2050)}$$

Using these formulas, the implied average rate of pandemic risk is:

| Forecast | Group | Year | | |
|---|---|---|---|---|
| | | 2023-2030 | 2030-2050 | 2100 |
| Genetically-engineered pathogen killing >1% population | Superforecasters | 0.04% | 0.06% | 0.05% |
| | Domain experts | 0.18% | 0.35% | 0.05% |
| Non-genetically-engineered pathogen killing >1% population | Superforecasters | 0.07% | 0.06% | 0.04% |
| | Domain experts | 0.14% | 0.21% | 0.07% |

Table A2. Implied Annual Risks of Pandemic Events

These forecasts are discussed in section 2.3.

## A2. Baseline Model Total Utility Calculations

Substituting in definitions of $u(t)$, $y_t$ and $n(t)$, given in sections 3.1-3.5, into equation (1), we obtain the formula for total utility in the status quo, given by equation (A2.1). If we pause science, the formula for total utility is instead given by equation (A2.2). I present both equations below, to make it easier to compare between the two.

$$V_{SQ} = n(0)\left\{\sum_{t=0}^{t_1} (p(1+s))^t (2+tG) + \sum_{t=t_1+1}^{\infty} (p(1+s))^t (1-d)^{t-t_1} (2+tG)\right\} + \sum_{t=0}^{\infty} (1-p^t)\, \hat{n}(t)\, \hat{u}(t)$$

$$V_{PS} = n(0)\left\{\sum_{t=0}^{t_1} (p(1+s))^t(2+tG) + \sum_{t=t_1+1}^{T} (p(1+s))^t(1-d)^{t-(t_1+1)}(2+tG)\right\}$$

$$+ n(0)\left\{(1+s)^T \sum_{t=T+1}^{\infty} p^t(1+\bar{s})^{t-T}(1-d)^{t-(t_1+1)}(2+(t-1)G+g)\right\}$$

$$+ \sum_{t=0}^{\infty}(1-p^t)\,\hat{n}(t)\,\hat{u}(t)$$

(A2.2)

Let's start with equation (A2.1). The term in brackets corresponds to annual utility flow (approximated as 2 + *tG*), multiplied by population growth and the probability we remain in the current epistemic regime. Note we split the summation into two components, with an additional 1 - *d* term entering after period $t_1$, corresponding to the dangers posed by the time of perils. The last term refers to the (unknown) utility obtained upon exiting the current epistemic regime.

Equation (A2.2) has the same summations for periods 0-$t_1$, during which time the impact of pausing science has had no effect, and for the summation corresponding to exiting the current epistemic regime. However, for period $t_1$+1 on, in the current epistemic regime, the exponent on the (1 - *d*) term is lower by 1 year, since the time of peril is delayed by one year. At year *T* additional changes take place. At this point, the pause in science begins to affect the rate of population growth, which drops from *s* to $\bar{s}$. Moreover, utility flows are now lower, to account for persistently lower growth levels.

To compute the impact of science today, we compute the difference in utility under these scenarios. If we take the difference of equations (A2.1) and (A2.2) and simplify, we obtain:

$$V_{SQ} - V_{PS} = n(0)\left\{-\frac{d}{1-d}\sum_{t=t_1+1}^{T} (p(1+s))^t(1-d)^{t-t_1}(2+tG)\right\}$$

$$+ n(0)\sum_{t=T+1}^{\infty} (p(1+s))^t(1-d)^{t-t_1}(2+tG)$$

$$- n(0)(1 + s)^T \sum_{t=T+1}^{\infty} p^t (1 + \bar{s})^{t-T} (1 - d)^{t-(t_1+1)} (2 + (t - 1)G + g)$$

(A2.3)

Note, all the terms associated with period 0 - $t_1$ have dropped out of this difference, because the policy has no impact on them. Similarly, the terms associated with utility from exiting the current epistemic regime have also disappeared, as they are the same in either policy.

Let us define $t_2$ as the gap between the onset of the time of perils and the positive impact of science, i.e., $t_2 = T - t_1$. We can rewrite these terms as summations beginning in period 0, which lets us use some summation rules to simplify them:

$$V_{SQ} - V_{PS} = n(0)\left\{- d(p(1 + s))^{t_1+1} \sum_{t=0}^{t_2-1} (p(1 + s)(1 - d))^t (2 + (t_1 + 1)G + tG)\right\}$$

$$+ n(0)(p(1 + s))^{t_1+t_2+1} (1 - d)^{t_2+1} \sum_{t=0}^{\infty} (p(1 + s)(1 - d))^t (2 + (t_1 + t_2 + 1)G + tG)$$

$$- n(0)p^{t_1+t_2+1} (1 + s)^{t_1+t_2} (1 + \bar{s})(1 - d)^{t_2} \sum_{t=0}^{\infty} (p(1 + s)(1 - d))^t (2 + (t_1 + t_2)G + tG + g)$$

(A2.4)

These are three arithmetico-geometric sums. They can be simplified as:

$$\begin{aligned} V_{SQ} - V_{PS} = & - dn(0)\Delta(t_1, t_2) \\ & + N_T\left\{L(s, d)(U_{T+1} + GL(s, d)) - \tfrac{1}{1-d} L(\bar{s}, d)(U_{T+1} - (G - g) + L(\bar{s}, d)G)\right\} \end{aligned}$$

(A2.5)

Where:

$$\Delta(t_1, t_2) = (p(1 + s))^{t_1+1} \sum_{t=0}^{t_2-1} (p(1 + s)(1 - d))^t (2 + (t_1 + 1)G + tG)$$

$$U_{T+1} = 2 + (T + 1)G$$

$$N_{T+1} = n(0)(p(1 + s))^{T}(1 - d)^{t_2}$$

$$L(s, d) \equiv \frac{p(1+s)(1-d)}{1-p(1+s)(1-d)}$$

The term $\Delta(t_1, t_2)$ captures the total utility flows that happen between periods $t_1$ and $t_2$: after the onset of the time of periods, but before any benefits from technology accumulate.

The term $U_{T+1}$ can be interpreted as the annual utility flow individuals in period $T + 1$, when the individual is wealthier.

The term $N_T$ can be interpreted as the number of people expected to survive to period $T$ in the current epistemic regime. We start with $n(0)$ people, and remain in that regime each period with probability $p$, but in each period we remain, the population growth by $s$ percent. At some point we enter the time of perils and annual mortality increases by $d$ in each year.

The term $L(s,d)$ can be interpreted as the total number of future life years in the current epistemic regime that will be obtained for every person alive, beginning in the "next" year. In that period we will have $p(1 + s)(1 - d)$ more (or fewer) life years to spawn new life years; this term is captured by the numerator. The terms $s$ and $d$ capture the annual growth (or decline) of life years via mortality and population growth, while $p$ (which I omit from the functions arguments because I assume it is fixed in this report) captures the annual probability of exiting the current epistemic regime.

## A3. Quantifying the Time of Perils

The key parameter in this model is the annual peril rate $d$, the increased mortality risk that arises during the time of perils. In this section, I describe how I use forecasts from the Existential Risk Persuasion Tournament to derive a variety of estimates.

My basic approach is to use forecasts about the probability of major pandemics by 2030, 2050, and 2100 to estimate the implied increase in risk arising during the time of perils. I then use forecasts of that a biocatastrophe kills 1%, 10%, or 100% of the human population by 2100 to infer plausible distributions of mortality risks during pandemic events. For example, if forecasters assign a 5% probability of a pandemic by 2100, and a 5% probability of human extinction due to a pandemic by 2100, that implies a belief that pandemics will always be extinction events. In practice, we'll see people think extinction level pandemics are rare, but this illustrates the idea.

## A3.1 Backing out beliefs about epistemic regimes

Before we can proceed to this estimation though, we need to additionally adjust for the possibility of leaving the current epistemic regime. It will be recalled that in the models of section 1, the time of perils only occurs so long as we are in the current epistemic regime. It is important to adjust for this. For example, suppose a forecaster believes we will experience a pandemic event every year with 100% probability so long as we stay in the current epistemic regime and a 0% chance of a pandemic once we leave it, but also that we will leave this regime with an annual probability of 90%. In that case, this forecaster would list just a 10% annual chance of a pandemic. To recover their 100% belief in a pandemic during the current epistemic regime, we would need to take into account their forecasts about leaving the current epistemic regime.

In my model, a change in the epistemic regime primarily occurs though extinction from the collapse of civilization due to factors besides biocatastrophes, or due to a dramatic transformation of the economy, most likely due to transformative AI. A few questions from the XPT allow us to roughly estimate forecaster beliefs that these outcomes will occur by 2100.

First, we have summary forecasts about the prospects of human extinction, inclusive of all causes.[10]

[10] A shortcoming of the above analysis is that domain experts and AI domain experts are not the same population as the biosecurity domain experts. But the XPT report finds strong correlation among the views of domain experts, across domains, on catastrophic risks, which should reduce the bias implied.

| | Median (95% confidence interval) | |
|---|---|---|
| Question | Superforecasters | Experts |
| Total Catastrophic Risk by 2100 | 9.04% (6.13%, 10.25%) | 20% (15.44%, 27.60%) |
| Total Extinction Risk by 2100 | 1% (0.55%, 1.23%)[11] | 6% (3.41%, 10.00%) |

Table A3.1. Total Risks, drawn from table 1 of the XPT report

Second, after the XPT, the tournament organizers also asked participants beliefs about the prospects of human extinction, specifically via genetically engineered pathogens.

| Forecast | | Group | Pre-XPT | Post-XPT |
|---|---|---|---|---|
| Catastrophic risk by 2100 | Engineered pathogens | Superforecasters | 1.3% | 0.85% |
| | | Domain experts | 3.9% | 4% |
| | Natural pathogens | Superforecasters | 1% | 1% |
| | | Domain experts | 1% | 1.5% |
| Extinction risk by 2100 | Engineered pathogens | Superforecasters | 0.1% | 0.01% |
| | | Domain experts | 1% | 1% |
| | Natural pathogens | Superforecasters | 0.01% | 0.0018% |
| | | Domain experts | 0.1% | 0.01% |

Table A3.2. Total Biorisk Probabilities. Drawn from Table 22 of the XPT report.

Noting $Pr(extinction) = 1 - Pr(no\ extinction)$, we can use this data to back out the implied belief about the probability of extinction arising from non-biological causes as follows:

$$Pr(no\ extinction) = Pr(no\ bio\ extinction) \times Pr(no\ NON\ bio\ extinction)$$

[11] In this report, when we present median forecasts, we present them with the same precision the median forecaster provided (e.g. 1% versus 1.00%), up to two decimal places or two significant digits, whichever is more precise (e.g. 0.0123% becomes 0.012%, 1.23% stays 1.23%). For confidence intervals, we add trailing zeroes if one end of the interval has more implied precision than the other.

Thus:

$$Pr(NON\ bio\ extinction) = 1 - \frac{1-Pr(extinction)}{1-Pr(bio\ extinction)}$$

This implies superforecasters believe the probability that the human race goes extinct before 2100 due to factors other than genetically engineered pandemics is 0.9%. For domain experts the number is 5.1%.

We can also exit the current epistemic regime via transformative AI. I think the best question to capture that belief in the possibility of transformative AI is the belief that annual real GDP growth will ever exceed 15% for at least one year by 2100. Superforecasters assign a 2.75% median probability to this occurring, while AI domain experts assign a 25% median probability to this.

I assume transformative AI is not possible if the human race is extinct, and that extinction is not possible if we obtain transformative AI. This implies the two events (extinction and transformative AI) are mutually exclusive and so we can simply add the probabilities of each occurring to infer the estimated probability of leaving the current epistemic regime. My best guess at each groups' forecast for the probability of leaving the current epistemic regime by 2100 is given by:

- Superforecasters: 0.9% (extinction) + 2.75% (transformative AI) = 3.65%
- Domain experts: 5.1% (extinction) + 25% (transformative AI) = 30.1%

Following the modeling assumption made in my baseline model, I assume there is a constant annual probability of leaving the current epistemic regime. This likely does not reflect the actual beliefs of superforecasts, but I think it's appropriate for the purposes at hand, which is to determine how much we should adjust each group's forecasts of pandemic risk by the probability that we go extinct or experience transformative AI. We can work out each group's implied annual belief as follows:

Superforecasters' annual probability of remaining in the current epistemic regime:

$$p = (100\% - 3.65\%)^{1/(2100-2023)} = 99.95\%$$

Domain experts' annual probability of remaining in the current epistemic regime:

$$p = (100\% - 30.1\%)^{1/(2100-2023)} = 99.54\%$$

We will use these figures to estimate the implied beliefs in annual pandemic risks, conditional on the world remaining in the current epistemic regime (i.e., humans haven't gone extinct or had society radically transformed by AI).

Before proceeding, it's worth pausing to discuss how reliable these estimates are. On the one hand, there are arguments to be made that they over-estimate the implied beliefs about staying in the current epistemic regime. Probably the biggest reason is that we could exit the current epistemic regime if a disaster does not lead to extinction, but does knock us "back to the stone age." One way we can bound this risk is by using the forecasts for catastrophes, rather than extinction, noted above. In the XPT, a catastrophe is an event killing 10% of the human population; I suspect this is not enough to permanently knock humanity off it's current trajectory.

Using this alternative measure, we obtain:[12]

- Annual superforecaster probability of remaining in epistemic regime: 99.85%
- Annual domain experts probability of remaining in epistemic regime: 99.3%

As noted above, since a catastrophe is probably not enough to knock human civilization off it's current course, the above probably underestimates the probability of staying in the same epistemic regime. At the same time, there are other reasons to believe the original estimates are too low. Most specifically, while global GDP growing at least 15% in a year is probably a necessary condition for transformative AI, it may not be sufficient. For example, this also covers cases where growth is 15% per year for just a few years before plateauing. Thus, the true estimate of a civilizational break due to transformative AI may be lower than implied by the forecast probability of 15% annual growth.

## A3.2 Implied Timelines and Severity of Time of Perils

Next, I want to back out the implied beliefs about the time of perils that are consistent with both the scenario envisioned in my baseline model, with XPT forecasts. We'll begin with the evidence briefly reviewed in section 2.3.

[12] The probability of a non-bio catastrophe by 2100 for superforecasters is 1 - (1 - 0.0904)/(1 - 0.0085) = 8.26%. Adding the probability of transformative AI, we obtain 11.01% of leaving the regime. Note (100% - 11.01%)^(1/77) = 99.54%.
The probability of a non-bio catastrophe by 2100 for domain experts is 1 - (1 - 0.20)/(1 - 0.04) = 16.67%. Adding the probability of transformative AI, we obtain 41.67% of leaving the regime. Note (100% - 41.67%)^(1/77) = 99.3%.

| Forecast | Group | Year | | |
| --- | --- | --- | --- | --- |
| | | 2030 | 2050 | 2100 |
| Genetically-engineered pathogen killing >1% population | Superforecasters | 0.25% | 1.5% | 4% |
| | Domain experts | 1.22% | 8% | 10.25% |

Table A3.3. Excerpted from Table 23 in the XPT report

These forecasts are unconditional forecasts. To make them applicable to our model, we need to adjust them to take into account the probability of leaving the current epistemic regime. That will give us the forecast probabilities of genetically engineered pandemic events in a world like our own - one not transformed by AI or where humans have gone extinct.

To make this adjustment, let $X$ denote a pandemic occurring. Then:

$$Pr(X) = Pr(X|current\ regime) \times Pr(current\ regime) + Pr(X|new\ regime) \times Pr(new\ regime)$$

We observe $Pr(X)$ only in the above table, but we can easily back out $Pr(X|current\ regime)$. To compute $Pr(current\ regime)$, we just use the annual probabilities of staying in the current epistemic regime discussed in section 5.1. I assume $Pr(X|new\ regime) = 0$; pandemics are impossible if we are extinct or experience transformative AI.

After this adjustment, we obtain each group's forecast probability of genetically engineered pandemics, conditional on remaining in the current epistemic regime.

| Conditional Forecast | Group | Year | | |
| --- | --- | --- | --- | --- |
| | | 2030 | 2050 | 2100 |
| Genetically-engineered pathogen killing >1% population | Superforecasters | 0.25% | 1.52% | 4.16% |
| | Domain experts | 1.26% | 9.06% | 14.62% |

Table A3.4. Implied Forecasts for Pandemics in Current Epistemic Regime

Note the effect of our adjustment is we now see higher probabilities of pandemic events occurring by 2100; it is conservative in the sense that it implies greater peril rates compared to the unadjusted numbers.

The next step is to estimate how the risk of a pandemic evolves over time. My approach is to find the set of genetically engineered pandemics probabilities consistent with the above conditional forecasts and the scenario diagrammed below.

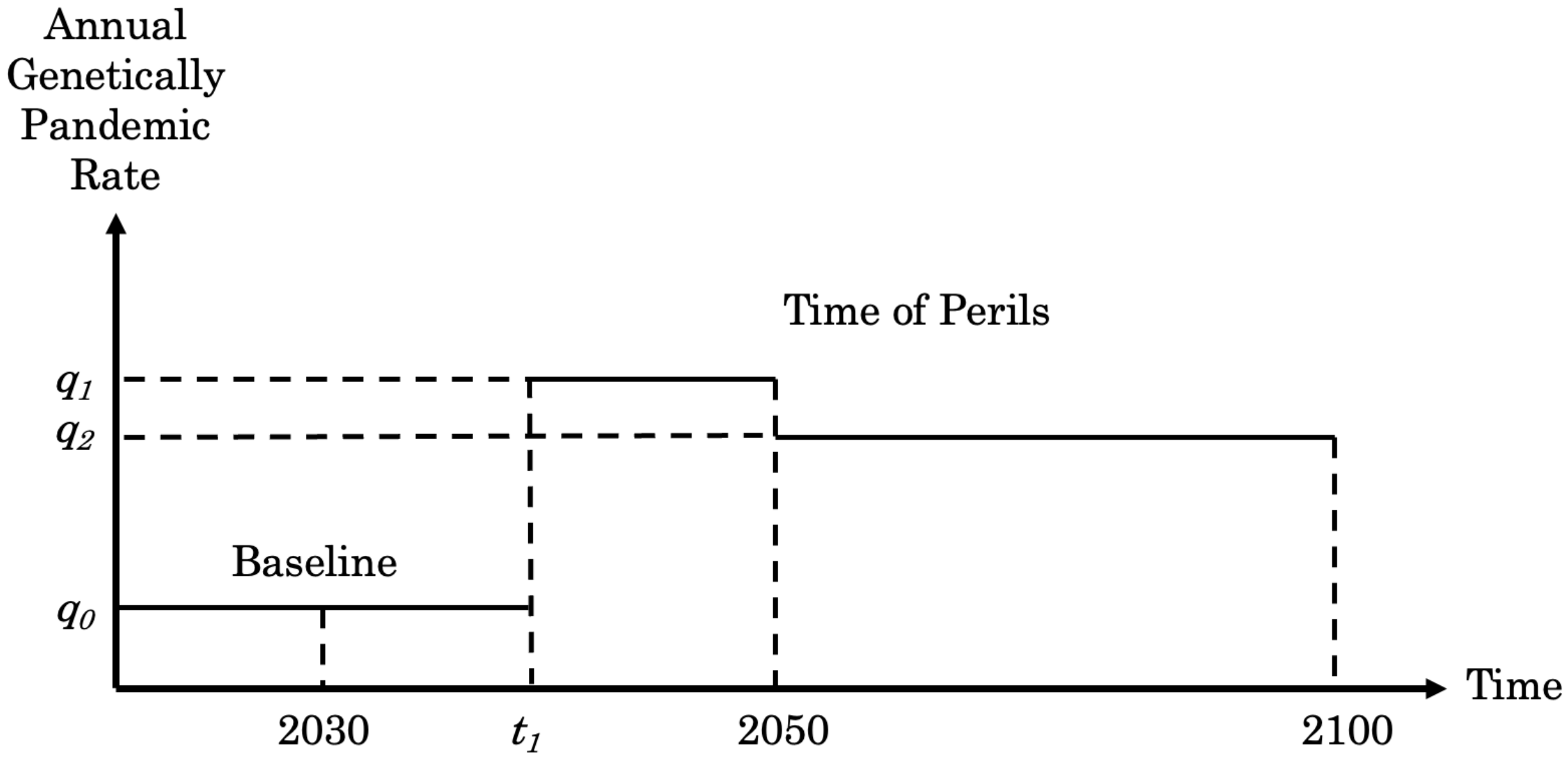

Figure A3.1. Annual genetically engineered pandemic probabilities

In the diagram above, we have a baseline rate of genetically engineered pandemics that occur at rate $q_0$, then in year $t_1$ (which is after 2030), we enter the time of perils, at which point the pandemic rate jumps to $q_1$. The difference between $q_0$ and $q_1$ is my main variable of interest, as it is my preferred measure of the increased risk caused by the onset of the time of perils. But it is also informative to compute $q_2$ as this gives us information about the implied long-run rate of genetically engineered pandemics in the current epistemic regime. My model assumes this is constant, so values that deviate strongly from this assumption may be less informative.

To compute value of $q_0$ and $q_1$ that align with conditional forecasts, we can use the same formulas given in section 2.3, but using the conditional forecasts given in Table A3.4.

To compute the value of $q_1$ that matches the conditional forecasts above, we need to know $t_1$. This is a free parameter, and below I consider a number of values from 2030-2050. For any given value of $t_1$, we can compute the value of $q_1$ that is consistent with the above conditional forecasts as follows:

$$Pr(no\ pandemic\ in\ 2050|current\ regime) = (1 - q_0)^{t_1 - 2023}(1 - q_1)^{2050 - t_1}$$

Alternatively:

$$q_1 = 1 - [\frac{1 - Pr(pandemic\ in\ 2050|current\ regime)}{(1-q_0)^{t_1 - 2023}}]^{1/(2050 - t_1)}$$

For superforecasters, the baseline rate $q_0$ implied by forecasts is 0.04%, and the long-run rate $q_2$ is 0.05%. For domain experts, the baseline rate $q_0$ is 0.18% and the long-run rate $q_2$ is 0.13%. The value of $q_1$ depends on the choice of $t_1$. In general, if we think the time of perils will begin at a later date, then that implies a higher peril rate $q_1$, since the average pandemic rate over 2030-2050 is higher than over 2023-2030. I plot these values below.

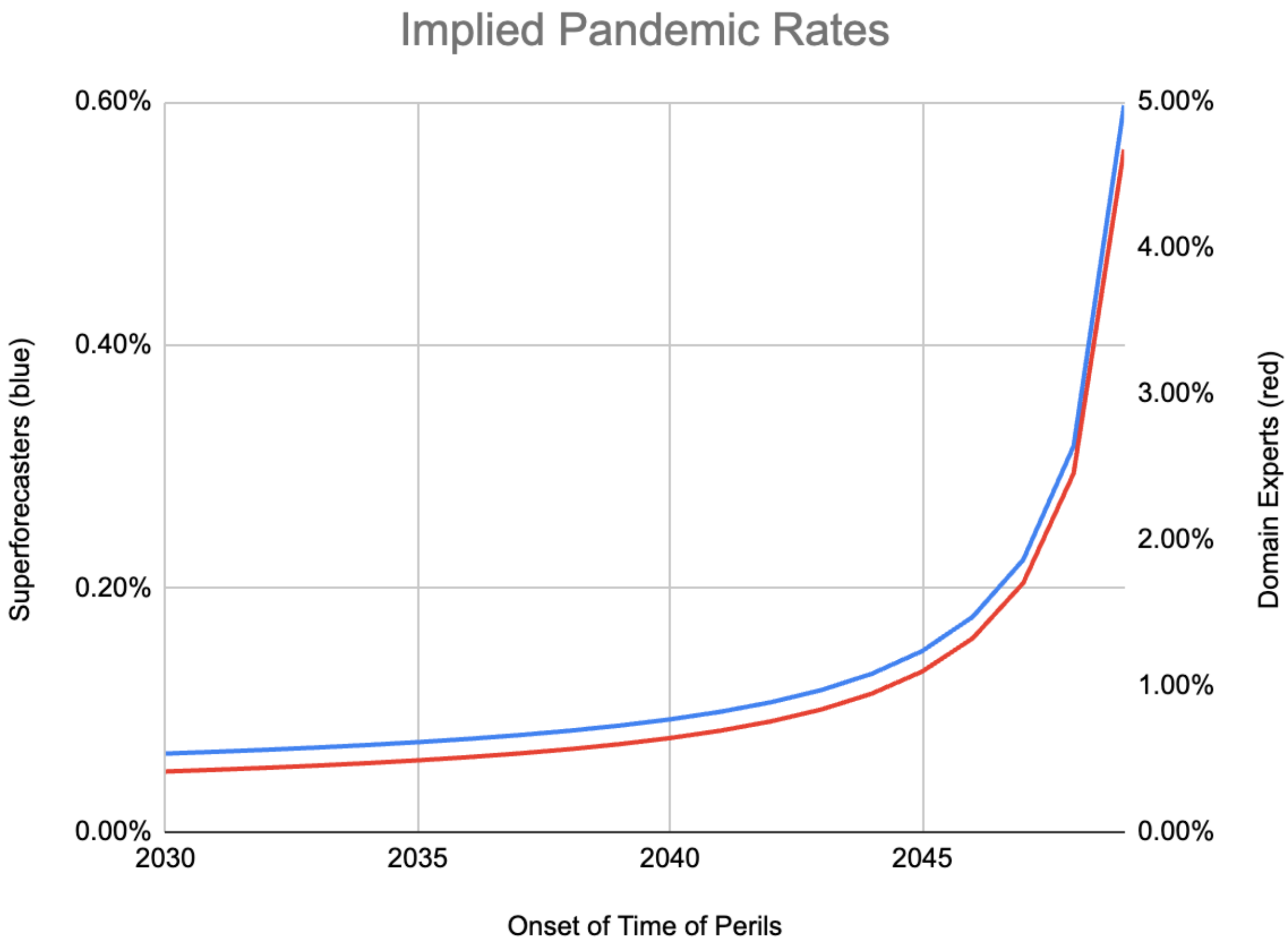

Figure A3.2. Implied peril rate, relative to onset time

In selecting my preferred value for $q_1$, there are two opposing factors to consider. On the one hand, a higher value of $t_1$ and therefore a higher value of $q_1$ will imply a greater difference between our baseline and the time of perils. That will tend to make science less desirable, and so a conservative estimate should be biased towards higher values of $t_1$ and $q_1$. On the other hand, large divergences between $q_1$ and $q_2$ seem implausible, as my model assumes a constant peril rate over time. My preference is to take the highest value of $q_1$ that is not super sensitive to the choice of $t_1$ and doesn't imply extreme volatility in the pandemic rate.

For superforecasters, my preference is $q_1 = 0.08\%$, which is consistent with $T$ = 2038, but we obtain similar values (within 25%) for $T$ in 2030-2041. For domain experts, my preference is $q_1 = 0.53\%$, which is consistent with $T$ = 2037 but we obtain similar values (within 25%) for $T$ in 2030-2040.

To sum up, my preferred estimates of the annual pandemic rate, which I think delivers results broadly consistent with forecasts in the XPT tournament are:

| | Superforecasters | Domain Experts |
|---|---:|---:|
| Baseline ($q_0$) | 0.04% | 0.18% |
| Time of Perils ($q_1$) | 0.08% | 0.58% |
| Time of Perils Onset | 2038 | 2037 |

Table A3.5. Preferred Genetically Engineered Pandemic Rates

Note I do not actually use $q_2$ in my modeling, only to help justify a choice for $q_1$ on the lower end of the possible range. Since in my baseline model I do not assume science can accelerate the cessation of the time of perils, in practice the long-run value of $q_2$ should largely drop out of my model when I difference across the status quo and pause science scenarios. To the extent it has an effect, the choice to maintain a consistent (higher) time of perils, rather than to let it drop in line with forecasts, will act to reduce the measured benefits of science. This is because a higher mortality rate for longer will reduce the benefits of higher income that accrue in the status quo. Alternatively, the decision not to use lower values of $q_2$ can be justified by a decision to ignore the 2050-2100 forecasts.

## A3.3 Converting Pandemic Risks to Mortality Risks

The preceding analysis gives us estimates of the annual rate of genetically engineered pandemics in the current epistemic regime. The final step is to convert annual pandemic risks into annual excess mortality risks. For that, we need to estimate the probability of dying, conditional on a pandemic happening.

A few questions from the XPT again allow us to calibrate this conditional risk to forecaster beliefs. Here, we rely on two questions from the XPT: the probability of a genetically engineered pathogen kills more than 10% of the global population over a five year period by 2100 and the probability it kills 100% of the global population by 2100.[13] The former is dubbed a catastrophe, and the latter extinction.

Below, I give the unconditional forecasts for these events, drawn from Table 23 of the XPT report, as well as conditional forecasts, which follow the procedure described in the beginning of section 5.2 to adjust these to account for the possibility of leaving the current epistemic regime.

| | | Unconditional Forecasts | Conditional Forecasts |
|---|---|---|---|
| Catastrophic Event by 2100 | Superforecasters | 0.85% | 0.88% |
| | Domain Experts | 4% | 5.70% |
| Extinction Event by 2100 | Superforecasters | 0.01% | 0.01% |
| | Domain Experts | 1% | 1.43% |

Table A3.6. Conditional Forecasts of Catastrophe and Extinction

We now work out some implied conditional probabilities to assess how the annual mortality risk changes during the time of perils. I'll begin with extinction risk, for which there is no ambiguity about the share of people killed.

### A3.3.1 Extinction Risk

I will conservatively assume that it is not currently possible to genetically engineer a pathogen to cause human extinction, and that such capabilities are only available once the time of perils begins. Given an assumption about when the time of perils begins, and assuming a constant peril rate, we can back out annual extinction risk as:

[13] Technically, extinction in the XPT is defined as an event reducing the human population to below 5,000.

$$1 - Pr(extinction \in (2023, 2050) | current\ regime) = (1 - Pr(annual\ ext. | Time\ of\ Perils))^{2100 - t_1}$$

This implies an annual conditional extinction risk of 0.00016% during the time of perils for superforecasters and 0.02286% for domain experts. Put another way, conditional on a genetically engineered pandemic occurring, the probability it leads to human extinction is 0.00016%/0.08% = 0.2% for superforecasters and 0.02286%/0.58% = 3.9% for domain experts. As noted, I assume the conditional and unconditional probability of extinction is zero during the baseline period.

This implies that during the time of perils, the annual probability of a pandemic that does not lead to the extinction of humanity is 0.08%-0.00016% = 0.07984% for superforecasters and 0.55714% for domain experts. We now turn to estimating the mortality risk for these non-extinction events.

### A3.3.2 Non-Extinction Risk

I will assume it is possible to genetically engineer a disease that can lead to a catastrophe during the baseline period (defined as a disease killing at least 10% of the global population). I will also make the simplifying assumption that, conditional on a pandemic event killing at least 1% of the population, the probability it kills 10% is the same in the baseline and during the time of perils.

Let $\hat{q}_1$ denote the probability of a pandemic occurring that is not an extinction event during the time of perils, and *x* the annual unconditional probability of extinction. Then we can write:

$$1 - Pr(catastrophe) = (1 - cq_0)^{t_1 - 2023}(1 - c\hat{q}_1 - x)^{2100 - t_1}$$

Essentially, this says that the probability there is no catastrophe over 2023-2100 is the joint probability that there is no catastrophe during 2023-$t_1$, and no catastrophe or extinction event during $t_1$ through 2100. This is a non-linear equation with no simple closed form solution. I numerically find values of *c* that solve the equation in the attached google spreadsheet. I find *c* = 0.16 fits best for superforecasters, and *c* = 0.12 is the best fit for domain experts.

## A3.4 Summary Risk

We now have estimates of the annual probabilities of different mortality events.

| | | Annual probability genetically engineered pandemic kills | | |
|---|---|---|---|---|
| | | >1% of population | >10% of population | 100% of population |
| Superforecasters | Baseline | 0.04% | 0.0064% | 0.0000% |
| | Time of Perils | 0.08% | 0.0129% | 0.0002% |
| Domain experts | Baseline | 0.18% | 0.0216% | 0.0000% |
| | Time of Perils | 0.58% | 0.0897% | 0.0228% |

Table A3.7. Summary of Greater-Than Risks

As our final step, we need to pick the average mortality risk for a pandemic event killing 0-1% of the population, 1-10% of the population and 10-100%. My preference is given in the following table:

| Pandemic Event | | 0-1% | 1-10% | 10-100% | 100% | Expected Annual Mortality |
|---|---|---|---|---|---|---|
| | Mortality Rate | 0 | 2% | 20% | 100% | |
| Superforecasters Annual Probabilities | Baseline | 99.96% | 0.03% | 0.0064% | 0.0000% | 0.0020% |
| | Time of Perils | 99.92% | 0.07% | 0.0128% | 0.0002% | 0.0041% |
| | Change | | | | | 0.0021% |
| Domain Experts Annual Probabilities | Baseline | 99.82% | 0.16% | 0.0216% | 0.0000% | 0.0075% |
| | Time of Perils | 99.42% | 0.49% | 0.0669% | 0.0228% | 0.0460% |
| | Change | | | | | 0.0385% |

Table A3.8. Computing Expected Annual Excess Mortality

This has been a lot of work, but here are the key takeaways (this text is also in the main report).

The following set of conditions is broadly consistent with a range of superforecaster estimates about the probability of various outcomes.

- We currently live in a world where there is an annual risk of 0.002% that an individual will die from a genetically engineered pandemic event.
- Sometime in the future - perhaps around 2038 - scientific and technological capabilities will evolve such that this rises by 0.0021% per year, to a total of 0.0041%. This represents, roughly, a doubling of the annual risk of death via genetically engineered pathogen.

- At the same time, the annual risk of human extinction due to a genetically engineered pathogen will rise from ~0% to 0.0002% per year.
- These risks will persist for as long as there is no major change to our civilization.

Such a scenario is broadly consistent with superforecaster estimates of the probability of genetically engineered pandemics, catastrophes, extinction, as well as major changes to our civilization.

Domain experts are more pessimistic. The following set of conditions is broadly consistent with a range of domain expert estimates about the probability of various outcomes.

- We currently live in a world where there is an annual risk of 0.0075% that an individual will die from a genetically engineered pandemic event.
- Sometime in the future - perhaps around 2037 - scientific and technological capabilities will evolve such that this rises by 0.0385% per year, to a total of 0.0460%. This represents, roughly, a six-fold increase in the annual risk of death via genetically engineered pathogen.
- At the same time, the annual risk of human extinction due to a genetically engineered pathogen will rise from 0% to 0.0228% per year.
- These risks will persist for as long as there is no major change to our civilization.

Such a scenario is broadly consistent with domain expert estimates of the probability of genetically engineered pandemics, catastrophes, extinction, as well as major changes to our civilization.

For our model, we are primarily interested in the notion that science might bring forward in time the time of perils. Accordingly, in my model, I assume the baseline rates are captured in normal health indicators, and focus on the annual 0.0021% or 0.0385% increases in the mortality rates that occur in the time of perils. In other words, in my model, I explore setting $d = 0.0021\%$ and $d = 0.0385\%$.

## A3.5 Robustness

Throughout these calculations, I have relied on the median forecasts. But it is useful to inquire how representative these forecasts are. Below, I present the distribution of forecasts for the question, what is the probability a genetically engineered pandemic will kill >1% of the population within a five-year time period by the year 2100.

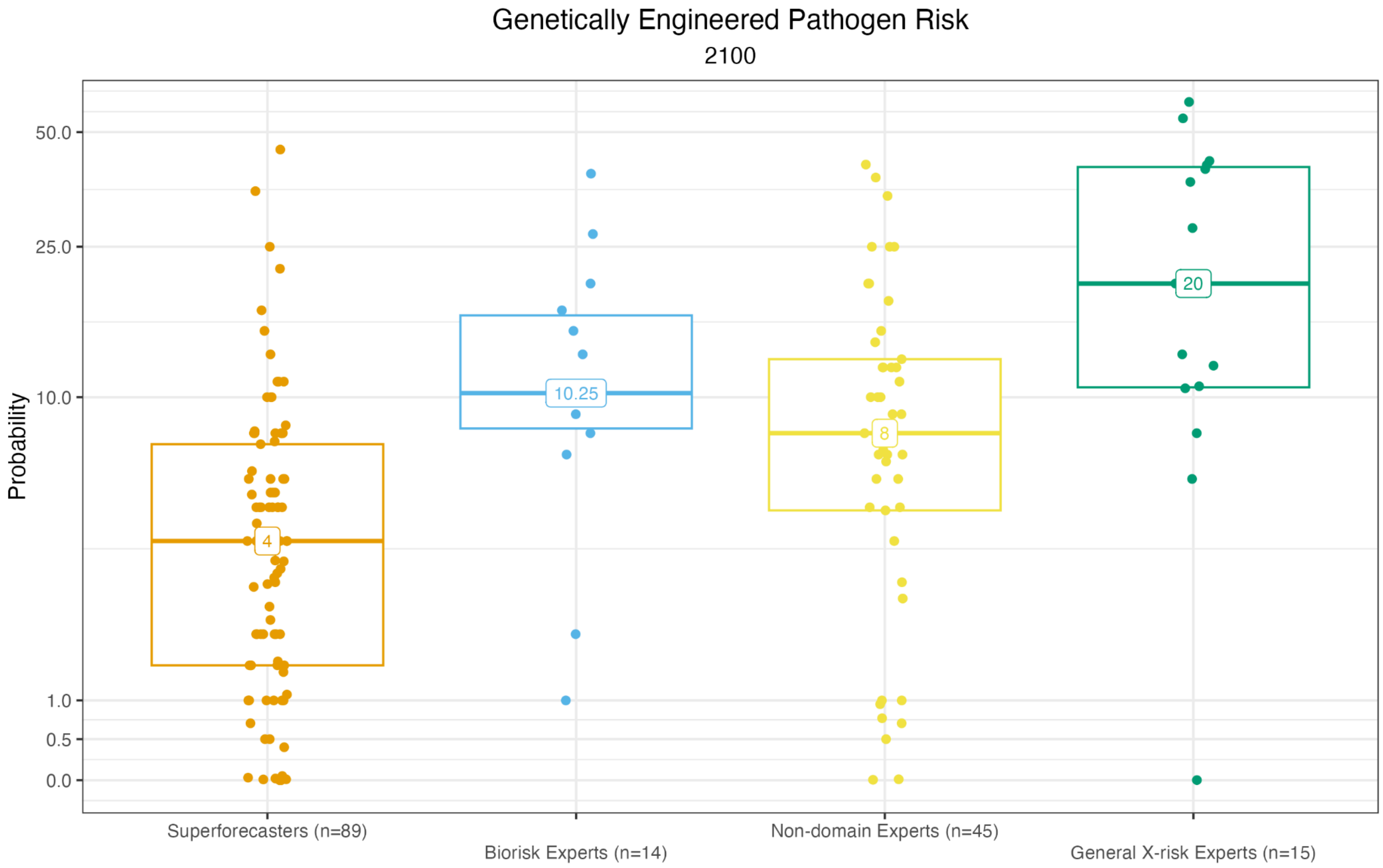

Figure A3.3. Distribution of Forecasts

The preceding draws on the forecasts of superforecasters and domain experts. Note that the median biorisk expert forecast of 10.25% is at approximately the 90th percentile for superforecasters, while 0.0% is approximately the 10th percentile range for superforecasters. Since I always report the outcomes when there is no time of perils, and for domain experts, these are approximately the same as presenting the 10th and 90th percentile superforecaster estimates.

Among domain experts, the median superforecaster estimate lies between the 14th and 21st percentiles. So we can think of the median superforecaster estimate as serving as a reasonable lower bound estimate for domain experts.

However, some domain experts see significantly high probabilities of a genetically engineered pandemic. For at least two domain experts, there is a greater than 25% chance of a genetically engineered pandemic event by 2100.

By visual inspection of similar charts for a pandemic by 2030, 2050, and 2100, I explored creating a pessimistic forecast that was close to the 90th percentile for domain experts. Specifically, I investigated a pandemic occurring by 2030 with 5% probability, by 2050 with 10% probability, and by 2100 with

25% probability.

However, it turns out these pessimistic forecasts are quite inconsistent with a time of perils framework. Even after inflating them using the methods described in sections A3.1 and A3.2, they are heavily front-loaded. They see the annual rate of a pandemic on the order of 0.75% over 2023-2030, which then *falls* to roughly 0.34% over 2030-2050, before *rising* to 0.64% per annum. I think these pessimistic forecasts are perhaps best captured by section xx, which explores the implications of assuming we are already inside the time of perils, so that pausing science is too late to have any useful impact.

# A4. Converting Utility to Impact ROI

To compute the ROI, we need to divide by spending on science, to compute utility per dollar.

To start, about $2.4trn was spent globally on R&D in 2019. However, most of this was spent on applied research and development, not spent on scientific research. To estimate spending on science, I use spending on basic research as a proxy. In 2019, we have data on the breakdown between basic and overall R&D spending for OECD countries whose total R&D spending is equal to $1.3trn. Collectively, this group spent about $230bn on basic research (17% of total R&D spending). In 2019, China spent $60bn on basic research, out of a total of $526bn (6% of total R&D spending). Adding together data on China and the OECD, we get 15.5% of R&D is spent on basic science which I apply to the $2.3trn spent on R&D in 2019 to obtain $357bn spent globally on science.

I therefore use $357bn as my denominator when computing "utility per dollar" of science today. To be more precise, I should adjust this figure by the knock-on effects of future growth, as in Jones and Summers (2020): if a constant share of GDP is spent on science than science today has a secondary cost in enabling growth, which, in turn raises the bill for science (since we will be taking a constant share of a bigger economy's resources). In practice though, given my parameter choices, this adjustment is negligible, increasing the cost of science by less than 6%, so I ignore this adjustment.

Once we have obtained a measure of utility per dollar, we need to put it in terms of OP's preferred measure of impact, by comparing the ROI of scient to the ROI of giving $1 to a person earning $50,000 per annum. This is:

$$ROI_{benchmark} = ln\, 50,001 \; - \; ln\, 50,000 \; \approx \frac{1}{50,000}$$

Thus, to compute the ROI of science today we use the following formula:

$$\frac{ROI_{science}}{ROI_{OP}} = \frac{50{,}000}{357{,}000{,}000{,}000}(V_{SQ} - V_{PS})$$

# A5. Modeling Health More Realistically

This appendix describes a variant of the baseline model that employs a more realistic model of health gains from science, and adopts a population growth framework where in each year, the same number of people are born across scenarios, so that all losses in income stem from premature death.

The tradeoff is this model is too complex to be solved by hand. Instead, I write a program in python to estimate it.

## A5.1 Aggregate Utility, Population Growth, and Income

To begin, we modify the welfare function given by equation (1) to be:

$$V_0 = \sum_{t=0}^{\infty} \left\{ b^t \sum_{a=0}^{120} n(t,a)u(t) + (1 - b^t) \sum_{a=0}^{120} \widehat{n}(t,a)\,\widehat{u}(t) \right\} \qquad \text{(A5.1)}$$

Let's take this equation from left-to-right:

- The sum $\sum_{t=0}^{\infty}$ signifies that we'll be summing utility over all future periods.
- The term $b^t$ corresponds to the probability we are still in the current epistemic range in period $t$. If so:
  - The population is given by $\sum_{a=0}^{120} n(t,a)$, where $n(t,a)$ is the number of people aged *a* alive in period *t*, and we assume the maximum age is 120.
  - $u(t)$, which is the utility obtained in period *t* in the current epistemic regime.
- The term $(1 - b^t) \sum_{a=0}^{120} \widehat{n}(t,a)\,\widehat{u}(t)$ corresponds to the utility we expect to obtain in the event we have exited the current epistemic regime. This happens with probability $1 - b^t$, in which case we use $\widehat{n}(t, a)$ and $\widehat{u}(t)$ to compute the number of people we expect to be alive and their lifetime utility.

I will continue to assume the utility function is given by equation (2) and that in the status quo, per capita income grows in each period by $1 + G$. If we pause science for one year, per capita income growth again slows to rate $g$ for one year, after $T$ periods.

Lastly, I will assume the same number of people are born in every period, equal to 130mn. This is roughly true, at a number that fluctuates between 120mn and 140mn, between 1970 and 2082 (UN medium fertility scenario). More importantly, this assumption means that any loss of utility, due to health, between the status quo and pause science scenarios will be due to deaths, rather than births that do not occur.

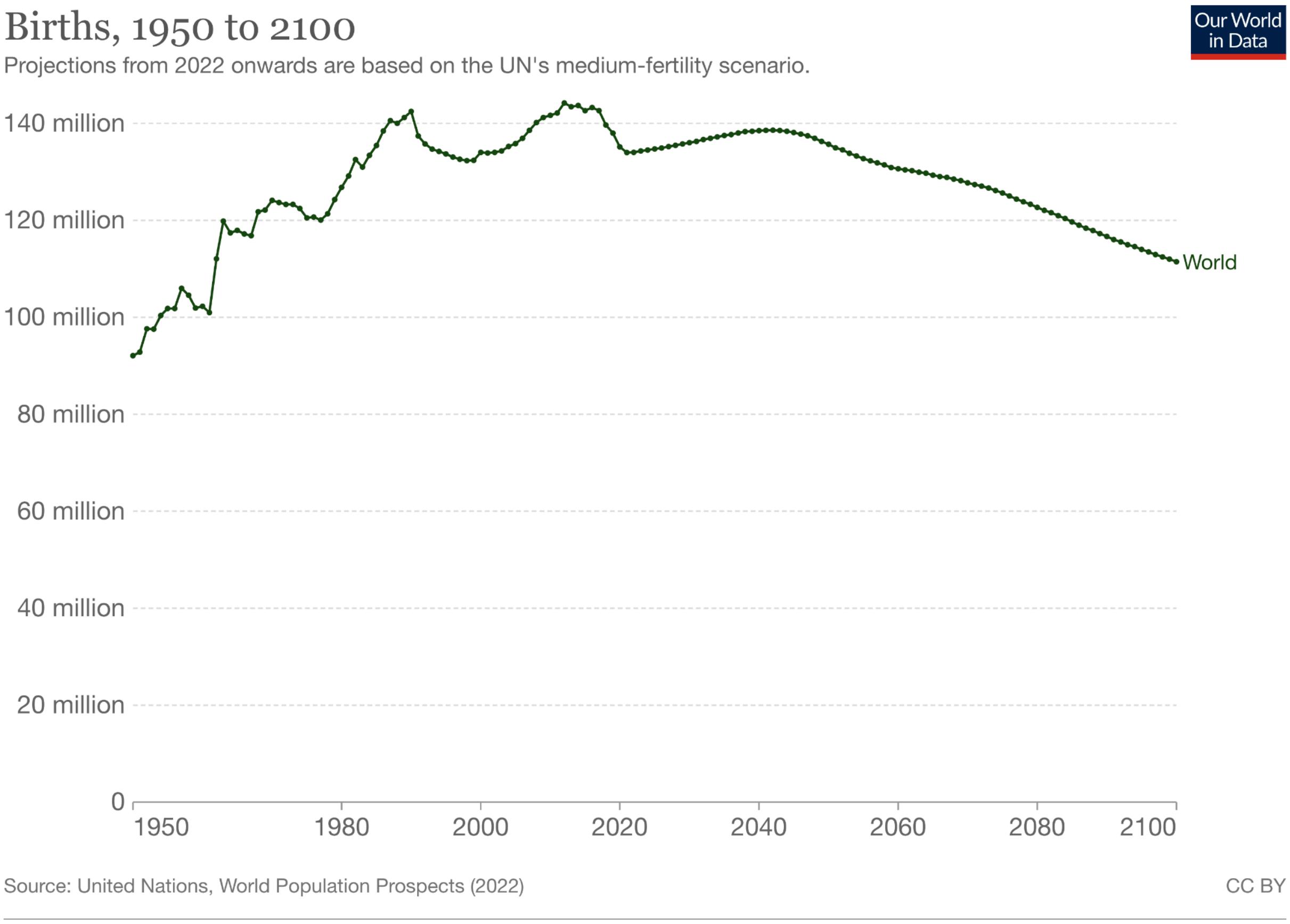

Figure A5.1. Annual Births Forecast from Our World in Data

## A5.2 Modeling Health

Let the number of people alive in period *t*, born in period *t'* be determined by three factors, so that $n(t, t') = N(t')S(t, a)D(t, a)$:

- $N(t')$ : the number of people born in period *t' = t - a*. As noted in section A5.1, I will assume $N(t') = N$, so that the same number of people are born in every period.
- $S(t, a)$ : The share of people aged *a* we would expect to be alive in period *t*, if historical trends in health were the only factor that mattered. This mixes the deleterious effects of age and the health-improving effects of advancing science, technology, and income. In general, it will be declining in *a* and increasing in *t*, if we hold the other term constant.
- $D(t, a)$ : the additional mortality risk stemming from new risks to health arising due to advances in science and technology during the time of peril.

To estimate $S(t, a)$, the annual share of people, of a given age, who are alive in a particular year, I start with the US social security administration's Actuarial Study #120, by Felicitie C. Bell and Michael L. Miller. Published in 2005, this provides data on the share of US men and women born in each decadal year from 1900-2100, who are expected to be alive at every age. For example, for US men born in 1910, it has the share who were still alive at age 1, 2, 3, 4, and so on, all the way up to 120. For future birth cohorts, the social security administration forecasts mortality trends forward.

Unfortunately, I cannot simply use this data for two reasons. First, I am interested in global health rather than US health. US life expectancy in 2019 was 79.1 versus 72.8 for the world. I assume US survival data from earlier years, when US life expectancy was 72.8, is an acceptable stand-in for global survival data. In practice, I will be assuming the survival share of global citizens aged 72.8 today (born in 1950) can be proxied by the survival rate of US citizens born in 1926.

A second issue is that to capture the survival share of the global cohort aged 99 and over, I need data predating the US 1900 birth cohort. Additionally, given the importance of long-run effects, I need estimates of the survival share persisting past the end of the social security administration's data.

Accordingly, I fit the following logistic model to the data from Actuarial Study #120:

$$survival.share = \frac{1}{1+exp(-(a+b\times age+c\times(log(birth.year-1800)))}$$

This generates a logistic survival curve in age, which is a standard simplified model of survival rates. Note that I also include a parameter for the log of the birth year, less 1800 (which is when historical life expectancy data seems to have a trend break). The assumption of a log structure embeds a strong assumption of diminishing marginal returns to life expectancy gains, which is a useful bounding assumption compared to the model in section 1.2, which assumes the population growth rate was invariant.

Moving from left to right, estimating this data on the social security administration's data and forecasts yields the following survival share estimates for the birth cohorts in 1900, 2000, 2100, 2200, and 2300 (note I make an assumption that no one survives past age 120):

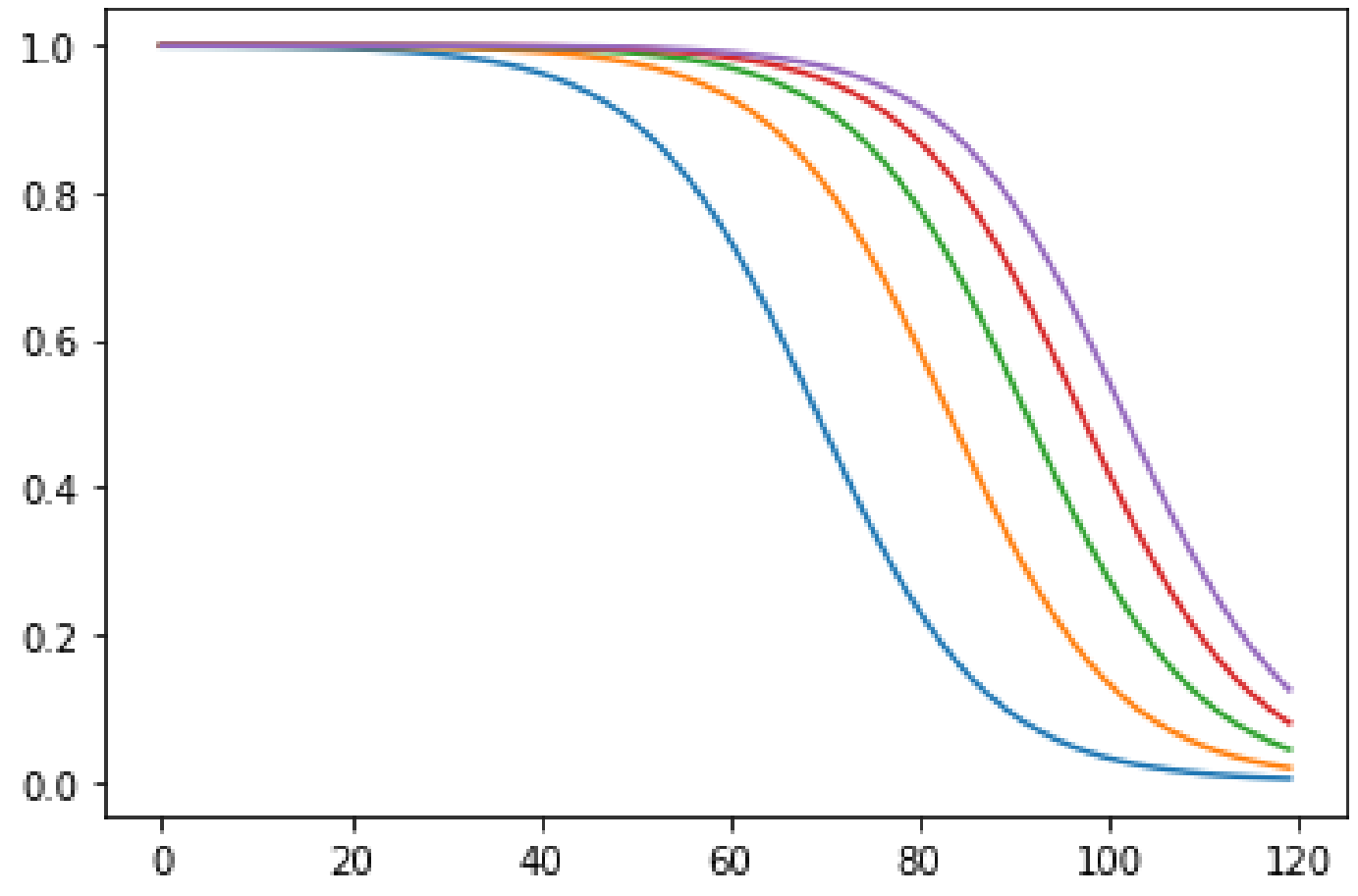

Figure A5.2 Logistic Survival Curves: 1900, 2000, 2100, 2200, 2300 (left to right)

This model shows how (1) the probability of survival increases over time, but at a diminishing rate (which matches the social security administration's assumptions) and (2) that the increases in survival after 2000 mostly accrue to those aged 60 and above.

Alternatively, these estimates produce the following estimates of US life expectancy:

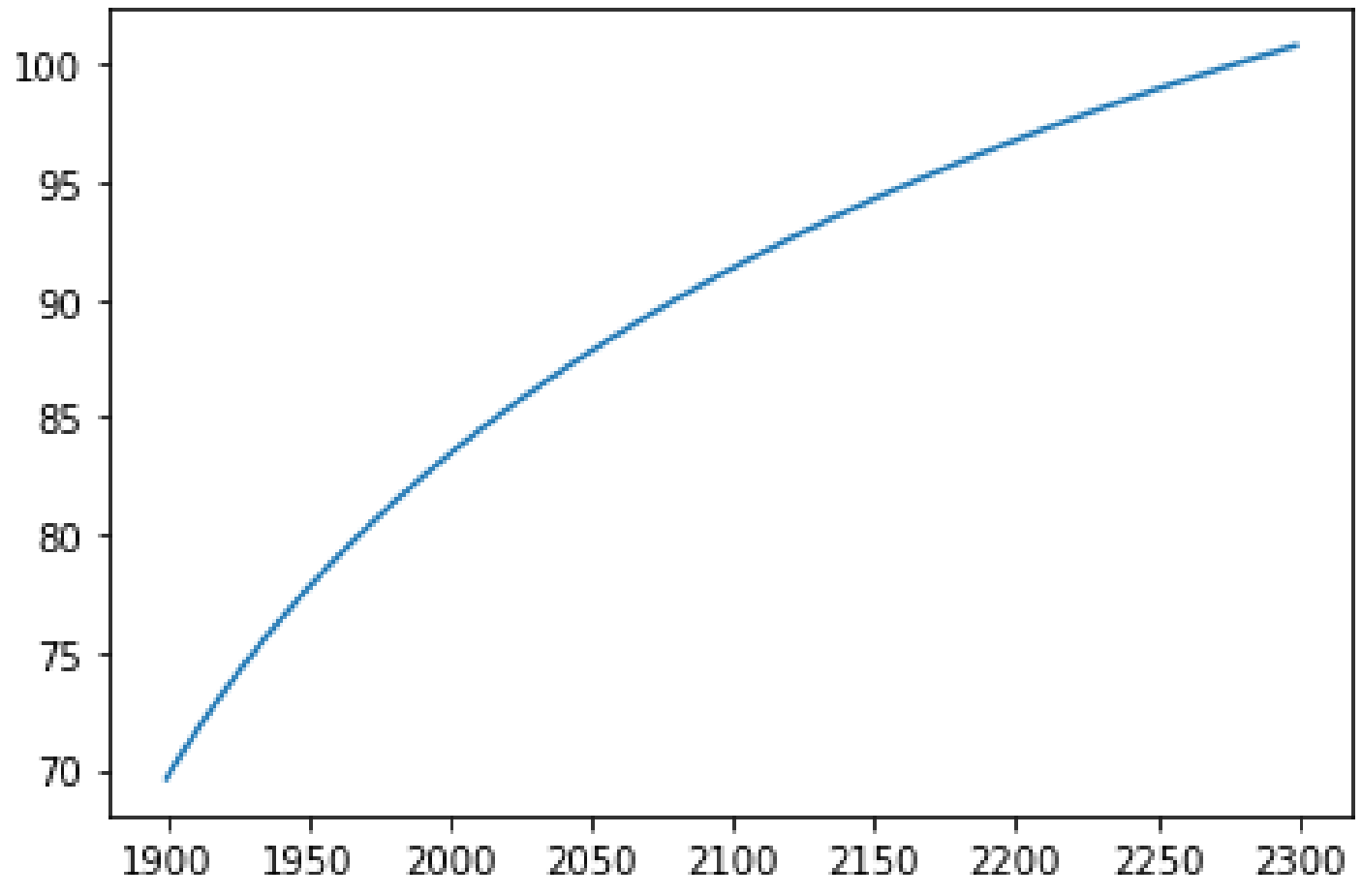

Figure A5.3. Forecast life expectancy

Again, note that we forecast diminishing marginal returns to health benefits, with life expectancy gradually curving downwards.

To this baseline model I now need to add a model of the health effects of the time of perils. Assuming that the time of perils results in heightened disease risk that affects people of all ages equally, I employ a very similar model as the baseline. Specifically, I again assume that we enter the time of perils in year $t_1$, and that each year in the time of perils, individuals have a constant probability of dying from one of these unnatural risks. Formally:

If $t \leq t_1$ then $d(t, a) = 1$.

If $t > t_1$ then $d(t, a) = (1 - d)^{min(t - t_1, a)}$

In other words, the probability of surviving the unnatural dangers of the time of perils to period *t* is *d* raised to the number of years in the time of period, which is $t - t_1$ for those born before $t_1$ and *a* for those born after (since they are only affected for years they are alive, which is their age).

Putting all these pieces together, expected status quo utility can be written as:

$$V_0^{SQ} = N \sum_{t=0}^{\infty} p^t \sum_{a=0}^{120} d(t, a) S(t, a)(2 + tln(1 + G) +$$
$$\sum_{t=0}^{\infty} (1 - p^t) \sum_{a=0}^{120} \widehat{n}(t, a)\, \widehat{u}(t)$$

(A5.2)

Where $ln(y_t / y_0) = ln(1 + G)^t = tln(1 + G)$ for small values of *G*. Here, we split the status quo utility flow into two components, prior to the time of perils (periods 0 to $t_1$) and after (periods $t_1$ + 1 to infinity).

For health, when we pause science for a year, I assume we fall back part of the way to the previous year's health trend. Define:

$\widehat{S}(t, a) = (1 - h)S(t, a) + hS(t - 1, a)$ where $0 < h \leq 1$.

Note $\hat{S}(t, a)$ is a mix of survival probability for someone the same age in year $t$ and year $t$ - 1, reflecting the fact that pausing science slows the advance of medical advances. If $h$ = 1, then pausing science shifts us back one year, in the sense that the probability of survival for people aged $a$ is what it was for people last year at that age. When $h < 1$, as we assume, then we don't lose a full year of advance. Note again, I assume the impact of slowing science takes effect only with a delay of $T$ years.

Let us define $\bar{S}(t, a)$ as the probability of surviving natural risk when we pause science today for one year. This value takes on different values for different cohorts.

If $t \le T$, then $\bar{S}(t, a) = S(t, a)$.
That is, before the impact of pausing science is felt, those living in the pause science scenario are not affected.

If $t > T$ and $t - a > T$ then $\bar{S}(t, a) = \hat{S}(t, a)$
That is, for those born after the impacts of pausing science are felt (that is, individuals where $t - a > T$), they live their whole lives as if medical science is held back part of a year.

If $t > T$ and $t - a \le T$ then $\bar{S}(t, a) = S(T, T - (t - a))\frac{\hat{S}(t,a)}{\hat{S}(T,T-(t-a))}$
This one takes a bit of unpacking. Individuals who were born prior to period $T$ but living after period $T$ experience the impact of delaying science on only a part of their life. These individuals were born in period $t$ - $a$. Thus, in period $T$ their age is $T$ - ($t$ - $a$). The first part of this equation, $S(T, T - (t - a))$, captures the notion that these individuals survive to period $T$ with the same probability as those living in a world where we do not pause science. However, going forward, their probability of survival declines.

In a world where we pause science, out of all the people who live to age $T$ - ($t$ - $a$), the share who subsequently live to age $a$ is given by $\frac{\hat{S}(t,a)}{\hat{S}(T,T-(t-a))}$. Out of all the people who lived to age $T$ - ($t$ - $a$) before the impact of pausing science took effect, we assume the same share survive to age $a$ once the effects of pausing science take place.

Pausing science also delays by one period the onset of the time of perils. To capture this, define $\bar{d}(t, a)$

If $t \le t_1 + 1$ then $\bar{d}(t, a) = 1$.

If $t > t_1 + 1$ then $\bar{d}(t, a) = (1 - d)^{min(t - t_1 - 1, a)}$

Putting all these pieces together, expected utility can be written as:

$$V_0^{PS} = N \sum_{t=0}^{T} p^t \sum_{a=0}^{120} \bar{d}(t, a)\, \bar{S}(t, a)(2 + tln(1 + G)) +$$
$$N \sum_{t=T+1}^{\infty} p^t \sum_{a=0}^{120} \bar{d}(t, a)\, \bar{S}(t, a)(2 + (t - 1)ln(1 + G) + ln(1 + g)) +$$
$$\sum_{t=0}^{\infty} \sum_{a=0}^{120} (1 - p^t)\, \hat{n}(t, a)\, \hat{u}(t)$$

(A5.3)

The key differences, compared to the status quo path begin with using $\bar{d}(t, a)$ rather than $d(t, a)$. After $T$ periods, we also begin using $\bar{S}(t, a)$ rather than $S(t, a)$ and utility reflects a slightly lower income.

## A5.3 Comparing scenarios

To estimate the returns to science, I numerically compute $V_0^{SQ} - V_0^{PS}$ over 3,000 periods. To compute analogues to the pure peril, pure health, pure income, and health-income interaction effects, it is convenient to define the following terms:

- $P_{SQ}^t = N \sum_{a=0}^{120} d(t, a)S(t, a)$: This is the population alive in period *t*, under the status quo.
- $P_{PS}^t = N \sum_{a=0}^{120} \bar{d}(t, a)\, \bar{S}(t, a)$: This is the population alive in period *t*, under the pause science regime.
- $U_{SQ}^t = 2 + tln(1 + G)$: This is utility for an individual alive in period *t*, under the status quo.
- $U_{PS}^t = 2 + tln(1 + G) + 1(t > T)(ln(1 + g) - ln(1 + G))$: This is the utility of an individual alive in period *t*, under the pause science scenario. $1(t > T)$ is an indicator function equal to 1 for period $t > T$ and otherwise zero.

With these terms, we can define $V_0^i = \sum_{t=0}^{\infty} p^t P_i^t U_i^t$: where *i* = *PS, SQ*.

Now write:

$$V_0^{SQ} - V_0^{PS} = \sum_{t=0}^{\infty} p^t P_{SQ}^t U_{SQ}^t - \sum_{t=0}^{\infty} p^t P_{PS}^t U_{PS}^t$$

Let us add and subtract $\sum_{t=0}^{\infty} p^t P_{PS}^t U_{SQ}^t$ to this equation. Then we can write:

$$V_0^{SQ} - V_0^{PS} = \sum_{t=0}^{\infty} p^t (P_{SQ}^t - P_{PS}^t) U_{SQ}^t + \sum_{t=0}^{\infty} p^t P_{PS}^t (U_{SQ}^t - U_{PS}^t)$$

Note that up through period *T*, $U_{SQ}^t = U_{PS}^t$. We can accordingly rewrite the above as:

$$V_0^{SQ} - V_0^{PS} = \sum_{t=0}^{T} p^t (P_{SQ}^t - P_{PS}^t) U_{SQ}^t + \sum_{t=T+1}^{\infty} p^t (P_{SQ}^t - P_{PS}^t) U_{SQ}^t + \sum_{t=T+1}^{\infty} p^t P_{PS}^t (U_{SQ}^t - U_{PS}^t)$$

Lastly, note that $U_{SQ}^t = U_{SQ}^{T+1} + (t - T - 1) ln(1 + G)$. Use this to rewrite the above as:

| $V_0^{SQ} - V_0^{PS}$ | | |
|---|---|---|
| | $\sum_{t=0}^{T} p^t (P_{SQ}^t - P_{PS}^t) U_{SQ}^t +$ | Pure peril effect. This is the change in utility due to an extra year of science due purely to the delay in the onset of the time of perils. |
| | $\sum_{t=T+1}^{\infty} p^t P_{PS}^t (U_{SQ}^t - U_{PS}^t) +$ | Pure income effect: This is the change in utility due entirely to the extra income (and hence utility) attributable to an extra year of science. |
| | $\sum_{t=T+1}^{\infty} p^t (P_{SQ}^t - P_{PS}^t) U_{SQ}^{T+1} +$ | Pure health effect: This is the change in utility, holding income fixed at $T + 1$ levels, attributable entirely to changes in health. |
| | $\sum_{t=T+1}^{\infty} p^t (P_{SQ}^t - P_{PS}^t)(t - T - 1) ln(1 + G)$ | Health-income interaction: This is the extra income that is jointly attributable to greater health and greater income in the more distant future. |

Table A5.1. Realistic Health Returns to Science Decomposition

# A6. Too Late Calculations

Let us now assume $d(t) = \overline{d}(t) = d$, so that we are too late to stop the time of perils. We can now modify equations A2.1 and A2.2 to compute the value of the status quo and pause science scenarios as follows:

$$V_{SQ} = n(0)\left\{\sum_{t=0}^{\infty} (p(1 + s)(1 - d))^t (2 + tG)\right\}$$

$$+ \sum_{t=0}^{\infty} (1 - p^t) \widehat{n}(t) \widehat{u}(t)$$

(A6.1)

$$V_{PS} = n(0)\left\{\sum_{t=0}^{T} (p(1 + s)(1 - d))^t (2 + tG)\right\}$$
$$+ n(0)\left\{(1 + s)^T \sum_{t=T+1}^{\infty} (p(1 - d))^t (1 + \bar{s})^{t-T} (2 + (t - 1)G + g)\right\}$$
$$+ \sum_{t=0}^{\infty} (1 - p^t)\, \hat{n}(t)\, \hat{u}(t)$$

(A6.2)

Now, the pause science scenario only differs from the status quo after period *T*, and only with respect to income and the population growth rate *s*. There is no impact on the peril rate *d* which now applies in every period. If we difference these two equations, we obtain:

$$V_{SQ} - V_{PS} = n(0)\left\{\sum_{t=T+1}^{\infty} (p(1 + s)(1 - d))^t (2 + tG)\right\} -$$
$$n(0)\left\{(1 + s)^T \sum_{t=T+1}^{\infty} (p(1 - d))^t (1 + \bar{s})^{t-T} (2 + (t - 1)G + g)\right\}$$

(A6.3)

This can be simplified to two infinite sums starting in period 0 as follows:

$$V_{SQ} - V_{PS} = N_T\left\{p(1 + s)(1 - d) \sum_{t=0}^{\infty} (p(1 + s)(1 - d))^t (U_{T+1} + tG)\right\} -$$

$$N_T\left\{p(1 + s)(1 + \bar{s}) \sum_{t=0}^{\infty} (p(1 - d)(1 + \bar{s}))^t (U_{T+1} + (g - G) + tG)\right\}$$

(A6.4)

Where

$$U_{T+1} = 2 + (T + 1)G$$

$$N_T = n(0)(p(1 + s)(1 - d))^T$$

This can, in turn, be simplified to:

$$V_{SQ} - V_{PS} = N_T \left\{ L(s,d)(U_{T+1} + L(s,d)G) - L(s,\bar{d})(U_{T+1} - (G - g) + L(s,\bar{d})G \right\}$$

(A6.5)

Where

$$L(s,d) \equiv \frac{p(1+s)(1-d)}{1-p(1+s)(1-d)}$$

This is the same as equation (A2.5), which is the core equation describing the returns to a year of science today, except for two changes. First, we no longer have a term corresponding to a pure peril loss. Whether we pause science or not, we suffer the same mortality effects from the time of peril. Second, we drop the the term $1/(1 - d)$ term from the second term in parentheses to capture the fact that pausing science no longer leads to one fewer periods in the time of perils.

# A7. Incorporating Extinction Risk

To incorporate existential risk, it is helpful to approach the problem slightly differently than in section 3.2. Let $V_t$ denote the present discounted total utility, in period *t*, if we begin the period in the current epistemic regime. Meanwhile, let $E[V^*] = \sum_{t=0}^{\infty} \hat{n}(t)\, \hat{u}(t)$ denote the expected present discounted total utility we obtain if we exit the current epistemic regime. For simplicity, I assume this value is constant over time; the utility we get in the next regime doesn't depend on when we make the transition, only that we do.

Lastly, we now will assume that $d_x$ corresponds to the risk of a civilization-killing event. If it occurs, the present discounted value of all future utility is zero (note, we don't need it to be literally zero, just very small relative to alternatives).

Let us assume we are entering the time of perils. Then we can write the following:

$$V_t = n(t)u(t) + p(1 - d_x)V_{t+1} + pd_x \cdot 0 + (1 - p)E[V^*] \quad \text{(A7.1)}$$

Equation (A7.1) states that the present discounted utility we obtain is equal to the flow of utility $u(t)$ that is enjoyed by $n(t)$ people, plus different future outcomes. With probability $p(1 - d_x)$ we stay in the current epistemic regime, and do not experience a civilization-killing event, so that next period we will obtain the present-discounted utility that accrues in period $t + 1$. With probability $pd_x$, we again stay in the current epistemic regime, but we also experience a civilization-killing event, and obtain 0 utility forever after. With probability $(1 - p)$ we exit the current epistemic regime (I assume this transition occurs prior to any potential civilization-killing event). We then obtain $E[V^*]$.

We can use equation (A7.1), to define $V_{t+1}$ and substitute it in to obtain:

$$V_t = n(t)u(t) + p(1 - d_x)\{n(t + 1)u(t + 1) + p(1 - d_x)V_{t+2} + (1 - p)E[V^*]\} + (1 - p)E[V^*] \quad \text{(A7.2)}$$

We could repeat this exercise by using equation (A7.1) to substitute in $V_{t+2}$ and then $V_{t+3}$ and so on. In the limit, setting $t = 0$, we would obtain the following:

$$V_0 = \sum_{t=0}^{\infty} (p(1 - d_x))^t n(t)u(t) + \sum_{t=0}^{\infty} (p(1 - d_x))^t (1 - p)E[V^*] \quad \text{(A7.3)}$$

Note, unlike in equation (1), $d_x$ is now part of the probability that we exit the current epistemic regime. That means policies that impact $d_x$ will now also impact the probability we obtain this utility. Meanwhile, the payoff in the current epistemic regime is unchanged from the general form (in the time of perils).

Let us consider the expected value of pausing science for one year. Suppose at time $t = 0$, we do not face the risk of a civilization killing, since we paused science. However, in the next period, we do:

$$V_0' = n'(0)u'(0) + pV'_1 + (1 - p)E[V^*]$$
$$V_t' = n'(t)u'(t) + p(1 - d_x)V'_{t+1} + (1 - p)E[V^*] \text{ for } t > 0 \quad \text{(A7.4)}$$

Where the apostrophe on *V, n,* and *u* denote the fact that we are in the pause science scenario. Performing substitutions, in the limit, $V_0'$ can be written as:

$$V_0' = n'(0)u'(0) + \sum_{t=1}^{\infty} p^t(1 - d_x)^{t-1}n'(t)u'(t) + (1 - p)E[V^*] + p\sum_{t=0}^{\infty} (p(1 - d_x))^t(1 - p)E[V^*]$$

(A7.5)

This is not transparent, but the important point is that (1) the payoffs in the current epistemic regime are the same[14] as in the pause science scenario of section 1 and (2) the necessity of surviving the time of perils in order to obtain $E[V^*]$ is pushed off one period.

We can now difference equations (A7.3) and (A7.5) to obtain the payoff to maintaining science (versus pausing for a year), but now including the potential losses due to extinction.

$$V_0 - V_0' = V_{SQ} - V_{PS} + (1 - p)\sum_{t=0}^{\infty} (p(1 - d_x))^t(1 - p)E[V^*] - (1 - p)E[V^*]$$

(A7.6)

Using a geometric sum and simplifying we obtain:

$$V_0 - V_0' = V_{SQ} - V_{PS} - d_x\frac{(1-p)}{1-p(1-d_x)}E[V^*] \quad \text{(A7.7)}$$

## A7.1 Discount Rates and the value of the future

We begin with the definition:

$$E[V^*] = \sum_{t=0}^{\infty} n(0)\rho^t(2 + tG)$$

[14] Actually, this is not quite true, since we should be using slightly lower values of d, to account for the fact that we the biocatastrophes that occur in the normal regime no longer result in extinctions. However, as Tables 5 and 6 make clear, small changes in d, relative to the superforecaster and domain expert estimates, will not lead to major revisions in expected payoffs.

First, note that $W = 2n(0)$. This implies, we can rewrite the above as:

$$E[V^*] = W \sum_{t=0}^{\infty} \rho^t (1 + tG/2)$$

We can next use the rules for a geometric and arithmetico-geometric sum to obtain:

$$E[V^*] = W\left(\frac{1}{1-\rho} + \frac{G}{2}\frac{\rho}{(1-\rho)^2}\right)$$

We can set this equation equal to some $\lambda W$ to back out the necessary value of ρ.

# A8. Can Better Science Reduce Risks?

In this section, I consider three arguments that metascience could reduce risks during the time of perils. I then illustrate how this would quantitatively change the expected value of science.

## A8.1 Accelerating Science for Safety

My preferred definition of "technology" is the orchestration of regularities in nature that enable new capabilities. Science teaches us about the regularities that occur in nature, and thus add to the menu of capabilities available to us. But importantly, there is a long process of iterative experimentation and development that goes into the development of new technologies, and this means new regularities in nature do not instantly result in new capabilities. As noted in section 4.4, I think we have pretty good evidence that, historically, the norm was a twenty-year gap between scientific discoveries and new technologies.

Secondly, the typical pattern is that new technological capabilities get cheaper over time, as technologies are improved and refined. This has an important corollary: on average, younger technologies, which are closer to science and have newer capabilities, will be disproportionately available to those with more resources - either wealthy individuals, or large groups of people jointly contributing resources. The resources of large corporations and nation-states tend to dominate the resources available to individuals, so that in general, groups with many members can access newer technological capabilities more quickly than groups with few members.

For example, at its peak, the development of the atomic bomb used about 1/1000 of the American workforce and 1/500th of US GDP (Ord 2022, pg 6). In contrast, no non-state actor has ever built a nuclear weapon (as of writing!). Moreover, the organizations behind the Covid-19 mRNA vaccines,

Moderna, Pfizer, and BioNTech each employ thousands of individuals. While there have been scattered attempts by small groups or even individuals to design their own covid vaccines, these seem to either involve non-trivial groups of people with expert scientific talent, rely on less frontier vaccine techniques, or are less effective.

To the extent that most people do not want to harm others, groups that want to use novel technological capabilities to hurt people will tend to be smaller than groups that do not. The upshot of this is that large groups have a larger menu of technological options available to them for defense, since they will be able to muster the resources for new technologies that are closer to science (and more expensive).

For example, consider biological technology. A group that wishes to cause mass death and destruction would most like to engineer a new kind of virus, which spreads widely, is fatal, and for which there are no existing counter-measures. This is challenging for them, because frontier science increasingly requires large teams of individuals with specialized skills to work together, and even they struggle to produce technologies that work reliably enough to launch profitable organizations based on these technologies. Recruiting individuals with these skills is challenging, because most people do not want to cause death and destruction. In contrast, an organization seeking to deploy novel technologies defensively (for example, waste water monitoring, new forms of mRNA vaccines, and so on) can openly recruit, raise funds, and seek public feedback on their work from the global community of experts.

Indeed, we have good evidence that the scientific establishment reacts rapidly to biocatastrophes and attempts to deploy knowledge to mitigate the problem. In 2020, for example, roughly 1 in 20-25 papers published was related to Covid-19 in some way (from essentially none in 2019). In the end, the global deployment of novel technological capabilities (the mRNA vaccines) was the most effective strategy for mitigating the death toll of the virus. This suggests the rapid rollout of defensive technologies based on novel scientific discoveries, but requiring lots of funding and the coordination of a large group of people, could be used to mitigate future new technological perils.

Somewhat surprisingly, the above implies interventions that permanently accelerate the pace of scientific progress are net safety enhancing. To see why, consider the following simple model.

- We have a stock of regularities in nature that we use to make technologies
- Every year, this stock increases by *g*% due to the efforts of the scientific community
- Each of these regularities can be developed into a technology enabling some capability.

- Suppose share $s$ of these new capabilities are safe, and share $o$ are offensive.
- For simplicity, let's assume new capabilities are instantly available to very large organizations, such as the US government, who can afford them. For everyone else, it takes $T$ years to become available.

Let $A_t$ denote the set of technological capabilities available to large organizations in year $t$ and $a_t$ the set available to small organizations in year $t$. For $t > T$, the number of capabilities for large and small organization is:

$$A_t = (1 + g)^t A_0$$

$$a_t = (1 + g)^{t-T} A_0$$

As an aside, note this is a world where the technological capabilities available to everyone grow at a constant exponential rate, which is broadly consistent with actual economic growth. It is probably the case that sustaining $g$ is getting harder over time, but this isn't directly relevant to our model, so long as $g$ and $T$ don't change.

How does the number of safe capabilities, relative to the number of offensive ones, evolve over time? Well, if we assume offensive capabilities are only used by small organizations then the defensive-to-offensive ratio is:

$$[defensive\ capabilities]/[offensive\ capabilities] = (sA_t)/(oa_t)$$

Substituting in the definitions of $A_t$ and $a_t$ we obtain:

$$[defensive\ capabilities]/[offensive\ capabilities] = s(1 + g)^T/o$$

This is strictly increasing in $g$. In other words, the faster is scientific progress, the greater is the set of defensive capabilities, relative to offensive ones. Conversely, a slowdown in the rate of scientific progress (which is arguably underway!) reduces safety by "leveling the playing field" between large and small organizations. Note this result holds for all positive values of $s$ and $o$, including scenarios where science is much more likely to generate offensive than defensive technology (i.e., when $o >> s$).

Alternatively, suppose we are entering a period when artificial intelligence enables small organizations to access some fraction $x$ of new capabilities, which are normally only available to large organizations. We can capture this possibility as follows:

$$[safe\ capabilities]/[offensive\ capabilities] = (sA_t)/o(a_t + x(A_t - a_t))$$

Note here we have changed the denominator so that it includes all of $a_t$ as well as some fraction $x$ of the remaining capabilities $A_t - a_t$. Again, by substituting in the definitions of $A_t$ and $a_t$ we can obtain:

$$[safe\ capabilities]/[offensive\ capabilities] = s/(o(\frac{1-x}{(1+g)^T} + 1))$$

This is also increasing in $g$, so long as $x < 1$.

The important takeaway of this framework is:

1. Faster science increases the growth rate of both offensive and defensive technologies
2. Faster science increases the ratio of defensive to offensive technology

## A8.2 Scientific State Capacity

Section A8.1 considered the case that faster scientific progress is net safety enhancing. In this section I consider the argument that more competent and less dysfunctional scientific institutions may also be safety enhancing, via their ability to respond effectively to novel technological hazards. To the extent that our grant program accelerates science by contributing to healthier scientific institutions, this may be another venue through which a metascience program could be net safety enhancing.

A potentially related study is Omberg and Tabarrok (2022), which examines the efficacy of different methods of preparing for a biocatastrophe; specifically, the covid-19 pandemic. The study takes as its starting point the Global Health Security Index, which was completed in 2019, shortly before the onset of the pandemic. This index was designed by a large panel of experts to rate countries on their capacity to prevent and mitigate epidemics and pandemics. Omberg and Tabarrok examine how the index, and various sub-indices of it, are or are not correlated with various metrics of success in the covid-19 pandemic, mostly excess deaths per capita. The main conclusion is that almost none of the indices were correlated with covid-19 responses, whether those metrics related to disease prevention, detection, response, or the capacity of the health system.

There were a few exceptions though, and with the caveat that there is risk that these are false positives (which will happen when you regress a lot of variables on the same explanatory variables), these exceptions are intuitive and potentially informative.

1. Countries with higher trust in people (as measured by survey questions) tended to have lower excess deaths
2. Countries where people rated the importance of knowing about science in their daily lives had fewer excess deaths
3. Countries with more state capacity (measured by an index meant to capture a state's ability to implement its goals or policies), took less time to vaccinate 10% of the population, had more people vaccinated overall, and had fewer excess deaths.

None of these results are particularly robust. The first and second results disappear when a battery of controls is introduced. The results on state capacity and excess deaths, on the other hand, are only statistically significant in the presence of demographic and development indicators. Nonetheless, as they are intuitive, I think its appropriate to read this as modest evidence that a society that values science more, and which has more capable public institutions, will be better prepared to deal with novel technological perils.

To the extent metascience is a way for scientific institutions to improve their functioning, it seems plausible that secondary effects will be more public interest in science and better functioning government agencies (especially ones with knowledge most relevant to mitigating novel science-driven technological perils).

## A8.3 Modeling Better Science

To model the potential gains from more effective science, I modify my baseline model to allow for differences in the peril rate $d$ across the more science and less science scenarios. Whereas, in the baseline model, the more science scenario corresponded to the scientific status quo and the less science corresponded to a pause science scenario, now I think of the more science scenario as corresponding to a world where we have made science sufficiently efficient that we get the equivalent of an extra year of science over some indeterminate time period. This now has four effects:

5. We get the equivalent of an extra year of economic growth due to science (i.e., income in every period is increased by $G - g$
6. We get the equivalent of an increased in the growth rate of healthy life years from $\bar{s}$ to $s$.
7. We bring forward the time of perils by one year

8. We reduce the annual peril rate $d$ to $\bar{d}$

I furthermore assume this increase in the quality of science is temporary, so that we do not get a permanently higher economic growth rate, nor ongoing declines to mortality.

If we assume these take effect after $T$ periods, we can compute the change in utility by using equations (A2.1) and (A2.2), but replacing $d$ with $\bar{d}$ in the status quo - now renamed "better science" (BS) - scenario. I reproduce these two equations below:

$$V_{BS} = n(0)\left\{\sum_{t=0}^{t_1} (p(1 + s))^t(2 + tG) + \sum_{t=t_1+1}^{\infty} (p(1 + s))^t(1 - \bar{d})^{t-t_1}(2 + tG)\right\} + \sum_{t=0}^{\infty}(1 - p^t)\,\hat{n}(t)\,\hat{u}(t)$$

(A8.1)

$$V_{PS} = n(0)\left\{\sum_{t=0}^{t_1} (p(1 + s))^t(2 + tG) + \sum_{t=t_1+1}^{T} (p(1 + s))^t(1 - d)^{t-(t_1+1)}(2 + tG)\right\}$$
$$+ n(0)\left\{(1 + s)^T \sum_{t=T+1}^{\infty} p^t(1 + \bar{s})^{t-T}(1 - d)^{t-(t_1+1)}(2 + (t - 1)G + g)\right\} + \sum_{t=0}^{\infty}(1 - p^t)\,\hat{n}(t)\,\hat{u}(t)$$

(A8.2)

As in appendix A2, the difference between these two can be given as:

$$V_{SQ} - V_{PS} = n(0)\,\bar{\Delta}(t_1, t_2) + N_{T,BS}L(s, \bar{d})(U_{T+1} + GL(s, \bar{d})) - N_T\frac{1}{1-d}L(\bar{s}, d)(U_{T+1} - (G - g) + L(\bar{s}, d)G)$$

(A8.3)

Where we have replaced $d$ by $\bar{d}$ in the first set of terms in the brackets and where:

$$\bar{\Delta}(t_1, t_2) = \sum_{t=t_1+1}^{T} (p(1 + s))^t (2 + tG)\left[(1 - \bar{d})^{t-t_1} - (1 - d)^{t-(t_1+1)}\right]$$

$$N_{T,BS} = n(0)(p(1 + s))^T (1 - \bar{d})^{t_2}$$

(A8.4)

## A8.4 Modeling Extinction Risk with Better Science

As discussed in section 9.2, it is possible that a metascience grant program could reduce the annual peril rate, even as it accelerates the onset of perils. Let us suppose, as in section 9.0, that a metascience grant program has the effect of accelerating scientific progress by the equivalent of a year while reducing $d$ to $\bar{d}$. I'll now assume there is also an equal proportional decrease in $d_x$ to $\bar{d}_x$, such that $d/\bar{d} = d_x/\bar{d}_x$.

If we assume these take effect after $t_1$ periods, we can compute the change in utility by using equations (A7.3) and (A7.5), but replacing $d$ with $\bar{d}$ and $d_x$ with $\bar{d}_x$ in the more science scenario. I reproduce these two equations below:

$$V_0 = \sum_{t=0}^{\infty} (p(1 - \bar{d}_x))^t n(t)u(t) + \sum_{t=0}^{\infty} (p(1 - \bar{d}_x))^t (1 - p)E[V^*] \quad \text{(A8.5)}$$

$$V_0' = n'(0)u'(0) + \sum_{t=1}^{\infty} p^t (1 - d_x)^{t-1} n'(t)u'(t) + (1 - p)E[V^*] + p\sum_{t=0}^{\infty} (p(1 - d_x))^t (1 - p)E[V^*]$$

(A8.6)

If we again difference these, we obtain:

$$V_0 - V_0' = V_{SQ}(\bar{d}) - V_{PS}(d)$$

$$+ \sum_{t=0}^{\infty} (p(1 - \bar{d}_x))^t (1 - p)E[V^*] - (1 - p)E[V^*] - p\sum_{t=0}^{\infty} (p(1 - d_x))^t (1 - p)E[V^*]$$

(A8.7)

Using some geometric sums and algebra, we can write the above as:

$$V_0 - V_0' = V_{SQ}(\bar{d}) - V_{PS}(d) + (1 - p)E[V^*]\left\{\frac{1}{1-p(1-\bar{d}_x)} - \frac{1+pd_x}{1-p(1-d_x)}\right\} \quad \text{(A8.8)}$$

Note that if the final term is positive, then the impact of metascience on extinction risk is actually positive! This occurs when the additional year spent in the time of perils is outweighed by permanently lower extinction risk. This condition is satisfied when:

$$\frac{1}{1-p(1-\bar{d}_x)} > \frac{1+pd_x}{1-p(1-d_x)}$$

Cross multiply and we can obtain:

$$1 - p(1 - d_x) > 1 - p(1 - \bar{d}_x) + pd_x(1 - p(1 - \bar{d}_x))$$

This equation, in turn, has most terms drop out, and can be simplified to:

$$p(1 - \bar{d}_x)) > \bar{d}_x/d_x$$

In other words, for metascience to reduce the risks of existential reduction, despite accelerating it, we need the reduction in existential risk to be sufficiently strong. Define the proportional decrease in existential risk as $\lambda_x = \bar{d}_x/d_x$ and write:

$$p(1 - \lambda_x d_x)) > \lambda_x$$

We can then rearrange this to write:

$$\frac{p}{1-pd_x} > \lambda_x$$

As a final exercise, we can repeat the break-even exercise of section 4.3, but using equation (35) and again assuming $\overline{d}_x = 0.9974 d_x$. To compute this, substitute $E[V^*] = \lambda W$ into equation (35), set this equation equal to zero, and rearrange to obtain:

$$\frac{V_{SQ}(\overline{d}) - V_{PS}(d)}{W} \frac{1}{1-p} \left[ \frac{1+pd_x}{1-p(1-d_x)} - \frac{1}{1-p(1-\overline{d}_x)} \right]^{-1} = \lambda \qquad \text{(A8.9)}$$

# Appendix References